\documentclass[fleqn,usenatbib]{mnras}

\usepackage{newtxtext,newtxmath}

\usepackage[T1]{fontenc}

\DeclareRobustCommand{\VAN}[3]{#2}
\let\VANthebibliography\thebibliography
\def\thebibliography{\DeclareRobustCommand{\VAN}[3]{##3}\VANthebibliography}

\usepackage{graphicx}	
\usepackage{amsmath}	
\usepackage{orcidlink}
\usepackage{hyperref}
\usepackage{comment}
\usepackage{longtable}
\usepackage{natbib}
\usepackage{booktabs}
\usepackage[normalem]{ulem}
\usepackage{subcaption}
\usepackage{flafter}

\definecolor{blazeorange}{rgb}{1.0, 0.6, 0.2}
\definecolor{seagreen}{rgb}{0.18, 0.55, 0.34}
\definecolor{rufous}{rgb}{0.66, 0.11, 0.03}
\definecolor{royalfuchsia}{rgb}{0.79, 0.17, 0.57}
\definecolor{scarlet}{rgb}{1.0, 0.13, 0.0}
\definecolor{royalpurple}{rgb}{0.47, 0.32, 0.66}
\definecolor{darkblue}{rgb}{0, 0, 0.66}
\definecolor{violet}{rgb}{0.5,0.,0.5}

\newcommand\nabg[1]{{\color{violet}[NG]}}        

\title[GRB\,260310A: A Long GRB with a Complex Afterglow]{GRB\,260310A\,/\,SN\,2026fgk: A Multi-Wavelength Study of a Nearby Underluminous Long GRB and SN with a Complex Afterglow}

\author[Gill, Becerra et al. 2026]{
Ramandeep Gill\,\orcidlink{0000-0003-0516-2968},$^{1,37}$\thanks{E-mail: r.gill@irya.unam.mx (RG)}
Rosa~L.~Becerra\,\orcidlink{0000-0002-0216-3415},$^{2}$\thanks{E-mail: rbecerra@astro.unam.mx (RLB)}
Antonio de Ugarte Postigo\,\orcidlink{0000-0001-7717-5085},$^{3}$\thanks{E-mail: antonio.deugarte@lam.fr (AdUP)}
Christina C. Th\"one\, \orcidlink{0000-0002-7978-7648},$^{4}$\newauthor
Alan~M.~Watson\,\orcidlink{0000-0002-2008-6927},$^{2}$
Noémie~Globus\,\orcidlink{0000-0001-6148-6532},$^{5}$
Jean-Grégoire~Ducoin\,\orcidlink{0009-0008-7341-4825},$^{6}$
Peter~Veres\,\orcidlink{0000-0002-2149-9846},$^{7}$
Stanley~E.~Kurtz\,\orcidlink{0000-0003-4444-5602},$^{1}$\newauthor
Asuka Kuwata\,\orcidlink{0000-0002-6169-2720},$^{1}$
Antonio~Mart\'in-Carrillo\, \orcidlink{0000-0001-5108-0627},$^{8}$
Luca Izzo\,\orcidlink{0000-0001-9695-8472},$^{9}$
Christophe Adami\,\orcidlink{0000-0002-8904-3925}$^{3}$,\newauthor
Enrique~Moreno~M\'endez\,\orcidlink{0000-0002-5411-9352},$^{10}$
Nikos~Mandarakas\, \orcidlink{0000-0002-2567-2132},$^{3}$
Camila~Angulo-Valdez\,\orcidlink{0009-0002-6667-3294},$^{2}$
Stéphane~Basa\,\orcidlink{0000-0002-4291-333X},$^{3}$\newauthor
William H.~Lee\,\orcidlink{0000-0002-2467-5673},$^{2}$
Edilberto Aguilar-Ruiz\,\orcidlink{0000-0003-3502-4152},$^{1}$
Dalya~Akl\,\orcidlink{0009-0006-4358-9929},$^{11,12}$ 
Margo~F.~Aller\,\orcidlink{0000-0003-2483-2103},$^{13}$\newauthor
Miguel \'Angel Aloy\,\orcidlink{0000-0002-5552-7681}$^{14,15}$,
Jie~An\,\orcidlink{0009-0000-5068-3434},$^{16}$
Sarah~Antier\,\orcidlink{0000-0002-7686-3334},$^{17}$
Jean-Luc~Atteia\,\orcidlink{0000-0001-7346-5114},$^{18}$
Nathaniel R. Butler\,\orcidlink{0000-0002-9110-6673},$^{19}$\newauthor
Krittapas Chanchaiworawit\,\orcidlink{0000-0002-9650-4371},$^{20}$
Philipe V. De La Parra\,\orcidlink{0000-0001-5957-1412},$^{21}$
Damien~Dornic\,\orcidlink{0000-0001-5729-1468},$^{6}$
Francis~Fortin\,\orcidlink{0000-0003-3642-2267},$^{18}$\newauthor
Shaoyu~Fu\,\orcidlink{0009-0002-7730-3985},$^{22}$
Johan~P.~U.~Fynbo\,\orcidlink{0000-0002-8149-8298},$^{23,24}$
Lluis Galbany\,\orcidlink{0000-0002-1296-6887},$^{25,26}$
Leonardo~García-García\,\orcidlink{0000-0001-5125-1043},$^{5}$\newauthor
Stefan Geier\,\orcidlink{0000-0003-0516-2968},$^{27,28}$
Marion Guelfand\,\orcidlink{0009-0001-0357-3854},$^{6}$
Linbo He\,\orcidlink{0009-0004-7645-8218},$^{16}$
Shuaiqing~Jiang\,\orcidlink{0009-0001-8155-7905},$^{16}$
Emeric Le Floc'h\,\orcidlink{0000-0001-7421-4413},$^{29}$\newauthor
Massimiliano Lincetto\,\orcidlink{0000-0002-1460-3369},$^{6}$
Xing~Liu\,\orcidlink{0000-0002-4072-6899},$^{16}$
Gianluca~Lombardi\,\orcidlink{0000-0003-3412-0556},$^{27,28}$
Diego~L\'opez-C\'amara\,\orcidlink{0000-0001-9512-4177},$^{30}$\newauthor
Daniele~Bj{\o}rn~Malesani\,\orcidlink{0000-0002-7517-326X},$^{23,24,31}$
Francesco~Magnani\,\orcidlink{0009-0000-6101-7373},$^{5}$
Kanthanakorn Noysena\,\orcidlink{0000-0001-9109-8311},$^{20}$\newauthor
Margarita~Pereyra\,\orcidlink{0000-0001-6148-6532},$^{5,32}$
Ny~Avo~Rakotondrainibe\,\orcidlink{0009-0004-0263-7766},$^{3}$
Anthony~C.~S.~Readhead\,\orcidlink{0000-0001-9152-961X},$^{33,34}$\newauthor
Delphine Russeil\,\orcidlink{0000-0001-5400-7214},$^{3}$
Fredd~Sánchez-Álvarez\,\orcidlink{0009-0009-5612-3759},$^{2}$
Benjamin~Schneider\,\orcidlink{0000-0003-4876-7756},$^{3}$
Tirth D. Surti\,\orcidlink{0000-0002-6369-6266},$^{33}$\newauthor
Nial R. Tanvir\,\orcidlink{0000-0003-3274-6336},$^{35}$
Samaporn Tinyanont\, \orcidlink{0000-0002-1481-4676},$^{20}$
Dong~Xu\,\orcidlink{0000-0003-3257-9435},$^{16,36}$
Zipei Zhu\,\orcidlink{0000-0002-9022-1928}$^{16}$
\\
$^{1}$ Instituto de Radioastronomía y Astrof\'isica, Universidad Nacional Aut\'onoma de M\'exico, Antigua Carretera a P\'atzcuaro \# 8701,\\ Ex-Hda. San Jos\'e de la Huerta, Morelia, Michoac\'an, M\'exico C.P. 58089, M\'exico\\
$^{2}$ Instituto de Astronom\'ia, Universidad Nacional Aut\'onoma de M\'exico. A.P. 70-264, 04510. Ciudad de M\'exico, M\'exico.\\
$^{3}$ Aix Marseille Univ., CNRS, CNES, LAM, Marseille, France\\
$^{4}$ E. Kharadze Georgian National Astrophysical Observatory, Mt. Kanobili, Abastumani 0301, Adigeni, Georgia\\
$^{5}$ Instituto de Astronom{\'\i}a, Universidad Nacional Aut\'onoma de M\'exico, km 107 Carretera Tijuana-Ensenada, 22860 Ensenada, Baja California, México\\
$^{6}$ Aix Marseille University, CNRS, CPPM, Marseille, France\\
$^{7}$ Department of Space Science, University of Alabama in Huntsville, Huntsville, AL 35899, USA\\
$^{8}$ School of Physics and Centre for Space Research, University College Dublin, Belfield, Dublin 4, Ireland\\
$^{9}$ INAF, Osservatorio Astronomico di Capodimonte, Salita Moiariello 16, I-80121 Naples, Italy\\
$^{10}$ Facultad de Ciencias, Universidad Nacional Aut\'onoma de M\'exico, Apartado Postal 70-542, 04510 M\'exico, CDMX, M\'exico\\
$^{11}$ New York University Abu Dhabi, PO Box 129188, Saadiyat Island, Abu Dhabi, UAE\\
$^{12}$ Center for Astrophysics and Space Science (CASS), New York University Abu Dhabi, Saadiyat Island, PO Box 129188, Abu Dhabi, UAE \\
$^{13}$ Department of Astronomy, University of Michigan, 323 West Hall, 1085 S. University Avenue, Ann Arbor, MI 48109, USA\\
$^{14}$ Departament d'Astronom\'{\i}a i Astrof\'{\i}sica, Universitat de València, 46100 Burjassot, Spain\\
$^{15}$ Observatori Astronòmic, Universitat de València, 46980 Paterna, Spain\\
$^{16}$ National Astronomical Observatories, Chinese Academy of Sciences, Beijing 100101, China\\
$^{17}$ IJCLAB, Université Paris Saclay, Orsay, France\\
$^{18}$ IRAP, Université de Toulouse/CNRS/CNES, 9 avenue du colonel Roche, 31028 Toulouse, France\\
$^{19}$ School of Earth and Space Exploration, Arizona State University, Tempe, AZ 85287, USA\\
$^{20}$ National Astronomical Research Institute of Thailand, 260 Moo 4, Donkaew, Maerim, Chiang Mai 50180, Thailand\\
$^{21}$ CePIA, Astronomy Department, Universidad de Concepci\'on,  Casilla 160-C, Concepci\'on, Chile\\
$^{22}$ Department of Astronomy, School of Physics, Huazhong University of Science and Technology, Wuhan, 430074, People’s Republic of China\\
$^{23}$ Cosmic Dawn Center (DAWN), Denmark \\
$^{24}$ Niels Bohr Institute, University of Copenhagen, Jagtvej~155, 2200~Copenhagen~N, Denmark\\
$^{25}$ Institut d’Estudis Espacials de Catalunya (IEEC), E-08034 Barcelona, Spain\\
$^{26}$ Institute of Space Sciences (ICE, CSIC), Campus UAB, Carrer de Can Magrans, s/n, E-08193 Barcelona, Spain\\
$^{27}$ GRANTECAN S.A., Cuesta de San Jos\'e s/n, E-38712 Bre\~na Baja, La Palma, Spain\\
$^{28}$ Instituto de Astrofísica de Canarias, E-38205 La Laguna, Tenerife, Spain\\
$^{29}$ Université Paris-Saclay, Université Paris Cité, CEA, CNRS, AIM, 91191, Gif-sur-Yvette, France\\
$^{30}$ Instituto de Ciencias Nucleares, Universidad Nacional Aut\'onoma de M\'exico, Apartado Postal 70-264, 04510 M\'exico, CDMX, Mexico\\
$^{31}$ Department of Astrophysics/IMAPP, Radboud University Nijmegen, P.O.~Box 9010, Nijmegen, 6500~GL, The Netherlands\\
$^{32}$ Secretar\'ia de Ciencia, Humanidades, Tecnolog\'ia, e Innovaci\'on\\
$^{33}$ Owens Valley Radio Observatory, California Institute of Technology,  Pasadena, CA 91125, USA\\
$^{34}$ Institute of Astrophysics, Foundation for Research and Technology-Hellas, GR-70013 Heraklion, Greece\\
$^{35}$ School of Physics and Astronomy, University of Leicester, University Road, Leicester, LE1 7RH, UK.\\
$^{36}$ Altay Astronomical Observatory, Altay, Xinjiang 836500, China \\
$^{37}$ Astrophysics Research Center of the Open university (ARCO), The Open University of Israel, P.O Box 808, Ra'anana 43537, Israel
}
\date{Accepted XXX. Received YYY; in original form ZZZ}

\pubyear{\the\year{}}

\begin{document}
\label{firstpage}
\pagerange{\pageref{firstpage}-\pageref{lastpage}}
\maketitle
\clearpage


\begin{abstract}
We present a comprehensive multi-wavelength study of GRB\,260310A\,/\,SN\,2026fgk, a nearby ($z=0.153$), long-duration gamma-ray burst (GRB) with an exceptionally underluminous prompt $\gamma$-ray emission and a Comptonized spectrum. The burst occurred at the edge of a blue host galaxy at a projected distance of 15~kpc, which is one of the largest offsets reported for a long GRB. 
The bright optical afterglow, with dense coverage from COLIBRÍ, likely peaked at a few to several hours post-burst, followed by a shallow decay not expected from canonical afterglow models. Both the optical and X-ray light curves show a brief chromatic plateau from $4-7$\,days. We show that the subsequent rebrightening observed at $\sim20$\,days is best explained by the combined contribution of the associated Type Ic-BL supernova, identified in GTC spectra, and a late-time refreshed shock. The broadband optical to X-ray spectral energy distribution is well described by synchrotron emission from the forward shock, while the radio observations demand an additional emission component. We model the afterglow using (a) an on-axis uniform jet from a dirty fireball with late-time energy injection and (b) a misaligned jet with power-law angular structure, both having material emitting along our line-of-sight (LOS) moving with an initial Lorentz factor of $\Gamma_0\sim20-35$. We conclude that at more typical GRB distances ($z\gtrsim0.5$) the prompt $\gamma$-ray emission from this source would likely have escaped detection, whereas its optical afterglow would have remained observable, making the event appear as an orphan afterglow or a gamma-ray quiet fast X-ray transient.
\end{abstract}




\section{Introduction}
\label{sec:introduction}

Long-duration gamma-ray bursts (GRBs) are among the most luminous transient phenomena in the Universe. They are widely interpreted as the outcome of the core collapse of massive stars \citep[e.g.][]{Woosley1993,MacFadyen1999}, leading to the formation of a compact object and the launching of a relativistic jet \citep[see][for a review]{Kumar2015,Zhang2018}. In nearby events, the connection between long GRBs and broad-lined Type Ic supernovae has been firmly established, providing key insight into the final stages of massive stellar evolution and the diversity of engine-driven explosions \citep{Galama1998,Hjorth2003,Woosley2006,Cano2017}. Nevertheless, the observational properties of these events span a broad range in prompt energetics, afterglow luminosities, circumburst environments, and supernova signatures, implying that long GRBs likely originate from a heterogeneous population of massive progenitors and outflow configurations. 

In the standard picture of an ultra-relativistic uniform jet, having bulk Lorentz factor (LF) $\Gamma\sim10^2-10^3$, the prompt $\gamma$-ray emission is strongly beamed \citep{Rhoads-97,Sari-99} in the radial direction, causing the fluence to drop sharply for observers located outside of the jet core \citep{Granot+02,Eichler-Levinson-04}, and in some cases, to miss it altogether. Such off-axis events are expected to produce afterglows that rise gradually as the jet decelerates and its emission becomes visible to a wider range of angles \citep{Granot+02}. These transients may appear as so-called orphan afterglows \citep{Rhoads-03,Nakar2003,Perley+25} if the prompt gamma-ray emission falls below the detection threshold of high-energy instruments when our line-of-sight (LOS) is slightly outside the jet's sharp edge, or beamed away altogether when the jet is significantly misaligned. Alternatively, if the jet is heavily baryon loaded, as in a dirty fireball \citep{Rhoads-03}, causing it to only propagate at moderately relativistic LFs, i.e. $\Gamma\sim$ few tens, the very high opacity to photon-photon annihilation suppresses $\gamma$-ray emission \citep{Woods-Loeb-95,Baring-Harding-97,Lithwick-Sari-01} to low fluences that may only be detectable in nearby events.

The growing discovery space of optical and X-ray surveys such as the \textit{Einstein Probe} mission \citep{Yuan2022,Yuan2025} has revealed a population of fast-evolving transients lacking clear gamma-ray counterparts \citep{Ho2020,Yadav2025,Perley+25}, raising the question whether they represent intrinsically distinct phenomena or standard GRBs viewed under different geometric conditions, such as off-axis viewing angles \citep{Nakar2003}. In addition to their diverse temporal behaviour, several of these events have shown evidence for significant chromatic and spectral changes during their evolution. This highlights that multi-band color information can provide an independent and sensitive diagnostic of the underlying emission physics, complementing traditional light-curve analyses. In this context, nearby bursts offer a unique opportunity to investigate, within a single well-observed event, the detailed multi-wavelength evolution spanning from the first seconds to several weeks after the trigger, providing a complete picture of the prompt high-energy emission, the properties of the relativistic outflow and the evolution of the blast wave as it interacts with the surrounding medium.

In this paper, we present a comprehensive analysis of the multi-wavelength observations of the afterglow of GRB\,260310A/SN 2026fgk. This GRB was detected by \textit{Fermi}/GBM on 2026 March 10 and it provides an exceptional nearby laboratory to test our models. The identification of the optical counterpart of GRB\,260310A resulted from a rapid multi-facility effort involving GOTO, ATLAS, ZTF, and LAST observations \citep{ONeill2026,Tonry2018,Hinds+26,GCN43974}, which established the  association of this transient source with the GRB and also identified its blue host galaxy at a redshift of $z=0.153$ \citep{GCN43977,GCN43984}. Despite the long duration of the burst and subsequent emergence of a supernova, the prompt gamma-ray energetics place it at the low-$E_{\rm iso}$ end of the long-GRB population with $E_{\rm iso}=(4.51\pm0.47)\times10^{50}$ erg, lower than $98.6\%$ of the long GRBs in our \textit{Fermi}/GBM comparison sample. This low energy combined with a complex multi-frequency afterglow evolution indicates that the observed behaviour may be influenced by geometric effects and/or additional complexity in the outflow and surrounding medium.

The paper is organised as follows. In Section~\ref{sec:observations}, we describe the photometric and spectroscopic data sets, including observations from gamma-rays, X-rays, optical and radio. The properties of the host galaxy are presented in Section\,\ref{sec:hg}. In Section~\ref{sec:analysis} we analyse the temporal and spectral evolution of the prompt emission and afterglow, including the afterglow chromatic evolution, supernova emergence, and the radio image properties. Section\,\ref{sec:bulk-Gamma-constraints} presents an explanation for the origin of the prompt emission and derives constraints on the coasting LF using a non-dissipative fireball model. The theoretical modelling of the broadband afterglow using two distinct jet models is presented in Section\,\ref{sec:afterglow-modeling}. Finally, we discuss our findings in Section~\ref{sec:discussion} and summarise our conclusions in Section~\ref{sec:summary}.

Throughout this work, we adopt a flat $\Lambda$CDM cosmology with $H_0 = 67.7$~$\mathrm{km\ s^{-1}\ Mpc^{-1}}$ and $\Omega_m = 0.31$ \citep{Planck2020}. 

\section{Multi-Waveband Observations}
\label{sec:observations}

\begin{figure*}
    \centering
    \includegraphics[width=0.80\linewidth]{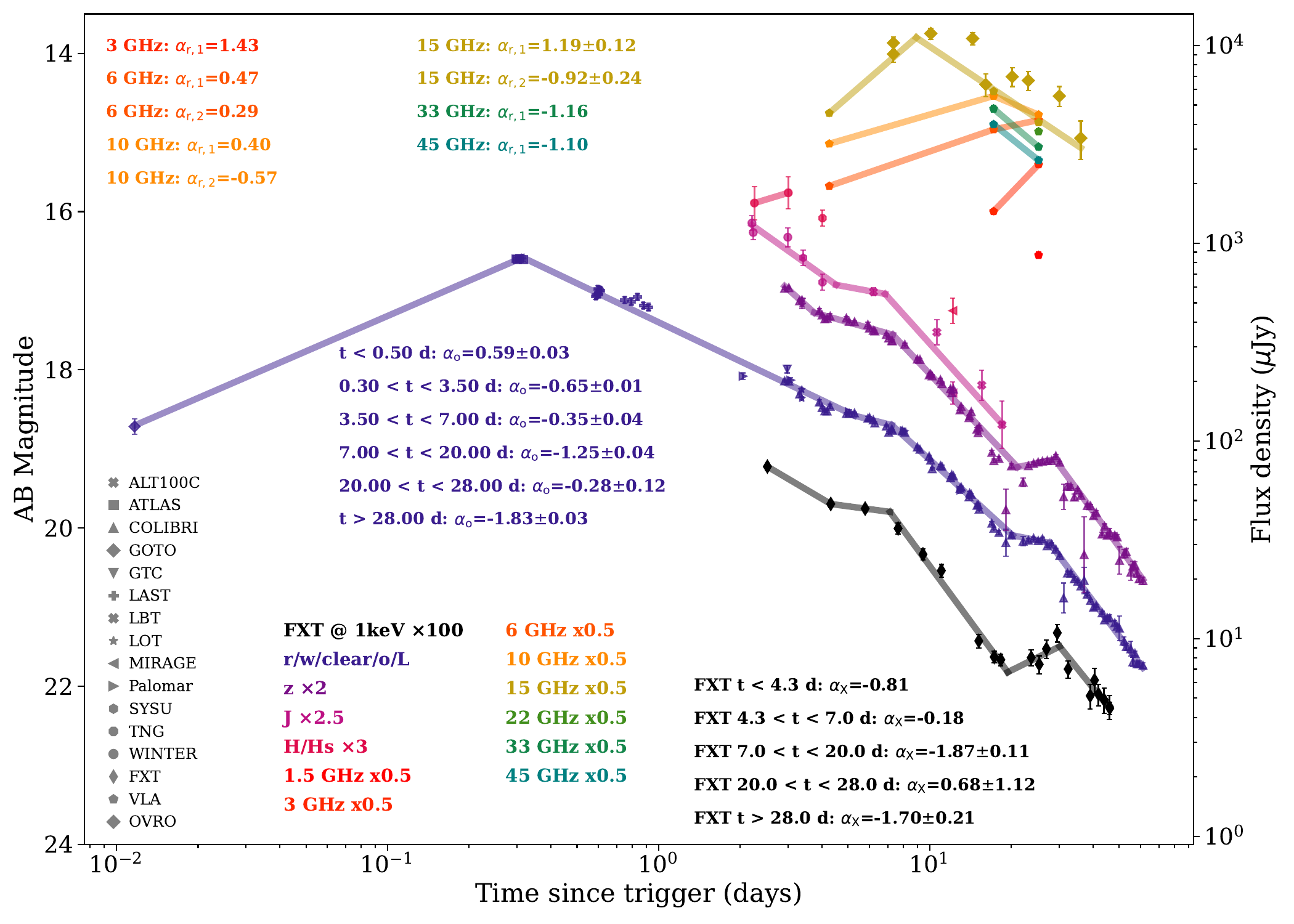}
     \includegraphics[width=0.80\linewidth]{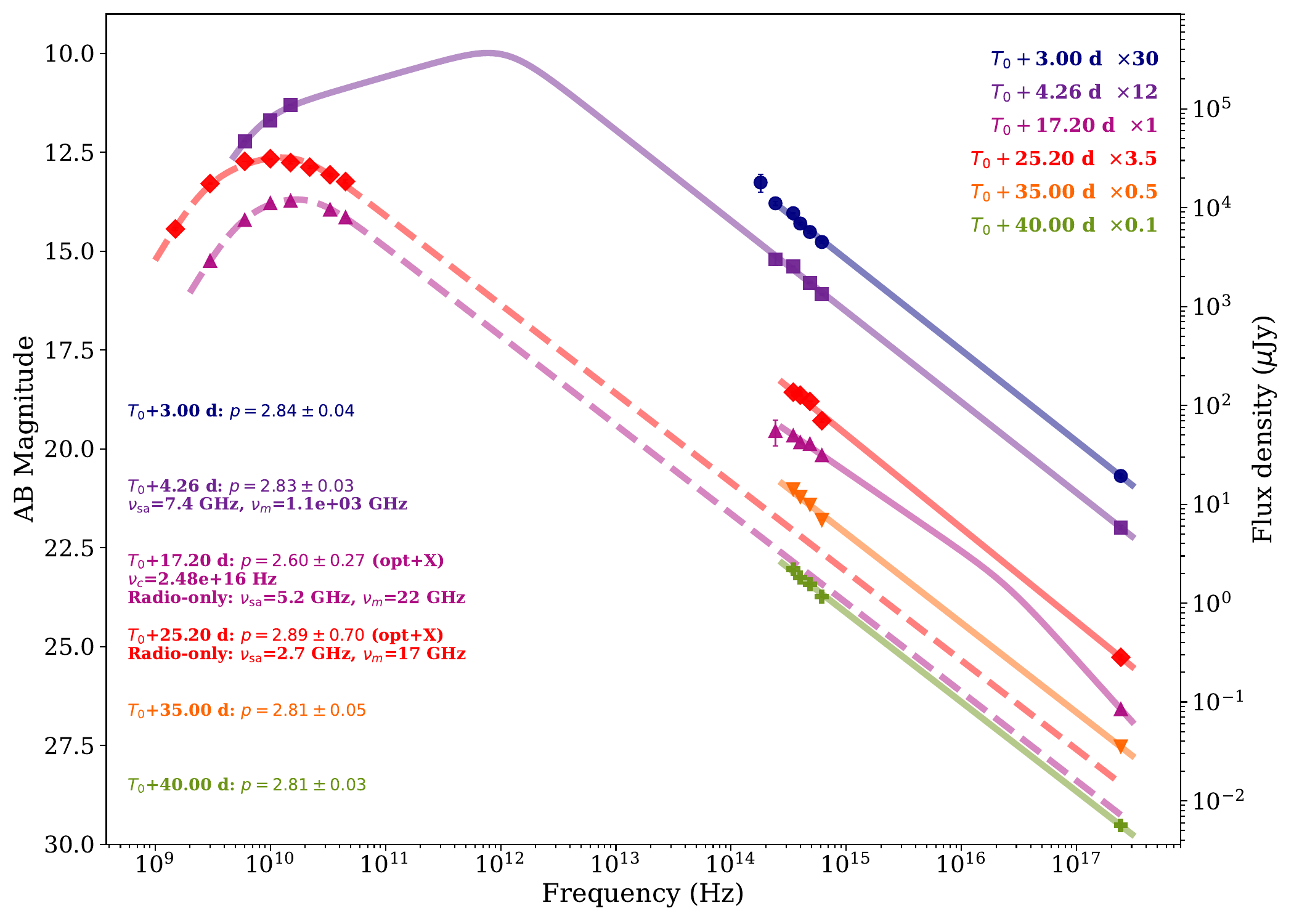}
     \caption{
    Top: Multi-wavelength light curves of GRB~260310A / SN\,2026fgk spanning radio, optical/nIR, and X-ray observations. AB magnitudes are shown on the left axis and the corresponding flux densities on the right axis. The optical photometry has been corrected for Galactic extinction and, in optical bands, for the contribution of the host galaxy. Dashed lines represent broken power-law fits to the temporal evolution, where $F_{\nu}\propto t^{\alpha}$, with the derived decay indices $\alpha$ indicated for the different time intervals and frequency bands. For clarity, the photometry has been scaled.
     Bottom: Spectral energy distributions (SEDs) at different epochs, spanning radio to X-ray frequencies. The parameter $p$ denotes the electron energy distribution power-law index. For epochs with radio coverage ($T_0+4.26$, 17.20, and 25.20~d), we model the SEDs using either a single smooth synchrotron spectrum ($T_0+4.26$~d) or a two-component model ($T_0+17.20$ and $T_0+25.20$~d) to account for the observed curvature between the radio and higher-frequency data. For epochs without radio data, the SEDs are well described by a single power-law segment, $F_{\nu} \propto \nu^{\beta}$ (solid lines). The characteristic break frequencies ($\nu_{\rm sa}$, $\nu_m$, $\nu_c$) are indicated in the figure. For clarity, the SEDs have been shifted vertically.
     }  
     \label{fig:multi-waveband-LC-Spectra}
     \end{figure*}

We performed an extensive multi-wavelength follow-up campaign of GRB\,260310A\,/\,SN\,2026fgk spanning gamma-rays, X-rays, optical/near-infrared, and radio frequencies. Figure~\ref{fig:multi-waveband-LC-Spectra} summarizes the temporal and spectral evolution of the source across the full photometric campaign. The top panel of Figure~\ref{fig:multi-waveband-LC-Spectra} presents the broadband light curves, whereas the bottom panel shows representative spectral energy distributions (SEDs) at key epochs. In this section, we summarize the observational datasets used throughout this work, while the detailed temporal, spectral, and broadband analysis is presented in Section~\ref{sec:analysis}. 

\subsection{\textit{Fermi}/GBM Detection}
\label{sec:fermi}

The {\it Fermi} Gamma-ray Burst Monitor (GBM, \citealt{Meegan+09}) consists of 12 NaI detectors, sensitive in the $\sim$8 to $\sim$1000 keV range, and two BGO detectors, sensitive in the 0.2 to 40~MeV range. GRB\,260310A triggered  GBM at $T_0=\mbox{04:57:10}$ UT on 10 Mar 2026 by the  Gamma-ray Burst Monitor (GBM) instrument on board the {\itshape Fermi}/Gamma‑ray Space Telescope \citep{GCN43951} with a position centred at is RA, Dec = 14:14, 78:42 (J2000), with a statistical uncertainty of 4.3~deg \citep{GCN43951}.

The NaI GBM detectors {\tt na} and {\tt nb} point within 60~deg to the source and are suitable for spectral analysis together with BGO detector {\tt b1}. Detector {\tt n9} has a larger angle to the source, and it is not used for spectral analysis. However, it still retains significant flux, and it is used for temporal analysis. 

\begin{figure}
    \centering
    \includegraphics[width=0.95\linewidth]{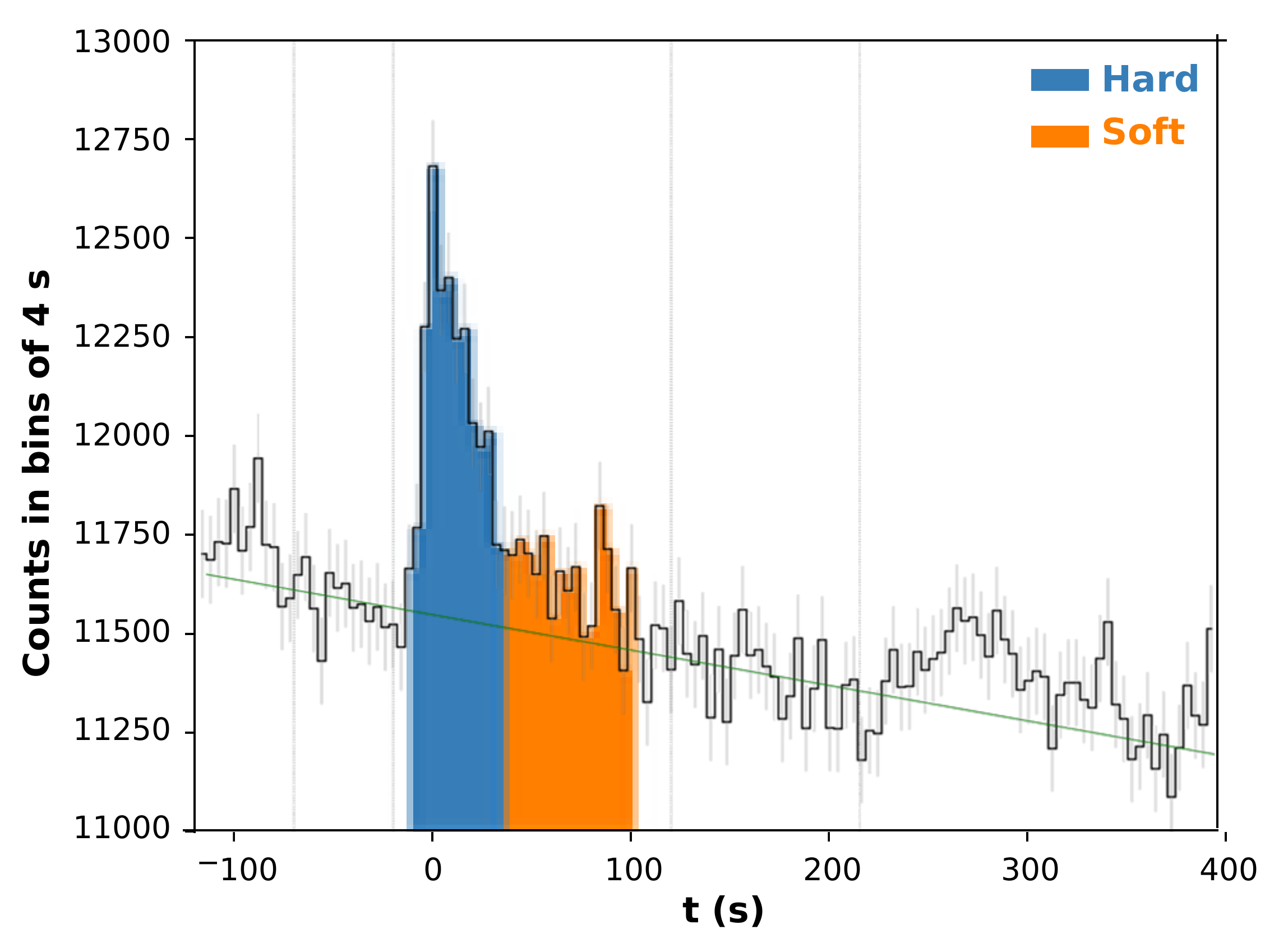}\\ 
    \caption{\textit{Fermi}/GBM Light curve of GRB~260310A with a 4~s resolution, 10-1000~keV energy range and summed over detectors {\tt n9}, {\tt na} and {\tt nb}. Time is measured with respect to the trigger time. Dotted horizontal lines mark the background selections, while the green line is a first degree polynomial fit to the background. The excess flux above background from $T_0+220$~s to $T_0+350$~s is not related to the GRB. }
    \label{fig:lcfermi}
\end{figure}

Figure~\ref{fig:lcfermi} shows the light curve of the GRB. The prompt emission is clearly visible as a single pulse. The duration, measured as the interval between the 5th and 95th percent of the cumulative flux in the 50-300~keV energy range is $T_{90}=60.1\pm2.4$~s~\citep{GCN43975}. The trigger occurred when the spacecraft just exited the South Atlantic Anomaly (SAA) region (at $T_0-130$ s). This is a region of the orbit with high particle activity and during the passage the instruments are turned off. 

Given the proximity of the orbit to the SAA, we carefully inspect the light curve to identify emission that belong to the GRB. The dominant pulse from $T_0-8$ to $T_0+41$~s (blue in the figure) is the main emission containing most of the flux. The episode from about $T_0+40$~s to $T_0+100$~s (orange in the figure) is relatively weak, and it is only discernible below 50~keV. It does not show appreciable flux in the 50-300~keV interval and hence it does not significantly contribute to the $T_{90}$ duration measure. However, the sky location of this emission is consistent with the location of the GRB, and therefore, it is considered part of the GRB. We also perform a localization of the emission above background from $T_0+220$~s to $T_0+350$~s and find that this emission is inconsistent with the GRB location. This emission is likely from particle activity. In summary, the GRB consists of a clear emission episode from $T_0-8$ to $T_0+41$~s followed by a weaker and softer emission from $T_0+41$~s to $T_0+98$~s.

GRB 260310A was also detected by AstroSat CZT \citep{Chattopadhyay+19} with a duration of $T_{90} = 30^{+2}_{-8}$~s in the 100-500~keV energy range \citep{GCN43958}. 

\subsection{X-ray campaign: \textit{Einstein Probe} Follow-up}
\label{sec:ep}

The \textit{Einstein Probe} Follow-up X-ray Telescope (FXT) \citep{Yuan2025,Yuan2022} performed X-ray follow-up of GRB\,260310A / SN\,2026fgk. \cite{GCN43994} reported a bright, fading X-ray counterpart spatially consistent with GRB\,260310A in an early observation. 

Table \ref{tab:xray_photometry} and the upper panel of Figure \ref{fig:multi-waveband-LC-Spectra} show the full follow-up campaign spanning 16 epochs from $T_0+2.5$ to $T_0+46.1$~days. 
The light curve shows an initial shallow decline, followed by a steepening at about $T_0+8$~days first reported by \cite{GCN44095}, and a rebrightening starting around $T_0+20$~days that lasts for approximately 10~days.
During the first observation, the source displayed an unabsorbed flux of approximately $5.1\times10^{-12}$~erg~s$^{-1}$~cm$^{-2}$ in the 0.5--10~keV energy range, gradually fading to about $5.0\times10^{-13}$~erg~s$^{-1}$~cm$^{-2}$ at the latest epochs. The corresponding flux density at 1~keV also shows an overall decay throughout the campaign. The temporal evolution and its interpretation are discussed in Section~\ref{sec:closure}.

No strong spectral evolution is observed between the optical and X-rays bands during the campaign (see the bottom panel of Figure~\ref{fig:multi-waveband-LC-Spectra}). 


\subsection{Optical Photometry}
\label{sec:opticallog}

\subsubsection{AT\,2026fgk: Discovery, Classification, and Association with GRB 260310A}

The discovery of GRB\,260310A and its association with GRB 260310A was a multi-step process with important contributions from several groups. It illustrates the collaborative aspect of modern transient astronomy and the benefits of rapid data sharing.

\cite{ONeill2026}, using the Gravitational-wave Optical Transient Observer \citep[GOTO;][]{Steeghs+2022,Dyer2024}, discovered the  optical transient AT\,2026fgk a position of RA, Dec = 14:37:16.141 +71:50:30.32 (J2000) and reported a magnitude of $L = 18.84 \pm 0.06$ at 2026 March 10 05:14:30~UTC, or $T_0 + 0.29$~hours. 
The GOTO project searches and characterizes transients in difference images, so this magnitude has any host contribution already removed \citep{Lyman+2006}. \cite{gcn43991} later confirmed these early results using observations with DDOTI from $T_0+0.2$ to $T_0 + 4.6$ hours.

\cite{Hinds+26} reported that GOTO, ATLAS, and ZTF photometry showed that GRB\,260310A had a light curve that rose and faded rapidly, and that it had a color of $g-r\approx +0.4$. They characterized it as a \emph{red fast optical transient}.

\cite{GCN43974}, using the Large Array Survey Telescope \citep[LAST; ][]{Ofek2023a,Ofek2023b}, observed the field of GRB\,260310A starting at $T_0 + 11.67$ hours. They report that their automated pipeline found no credible counterpart. Nevertheless, these authors were the first to suggest that AT\,2026fgk was the optical counterpart of GRB~260310A, noting that it coincided spacially and temporally with the GBM source, having been detected at $T_0 + 0.29$ hours and 7.0~deg from the \textit{Fermi}/GBM localization center (see Figure~\ref{fig:loc}), and had a light curve that first rose and then fell. Furthermore, they reported serendipitous LAST observations of the field at $T_0 - 2.15$ hours that showed no detection to a $5\sigma$ limiting AB magnitude of 20.74 in a clear filter, confirming its fast rise.

\subsubsection{COLIBRÍ Observations}

\begin{figure}
\centering
    \includegraphics[clip, width=0.95\linewidth]{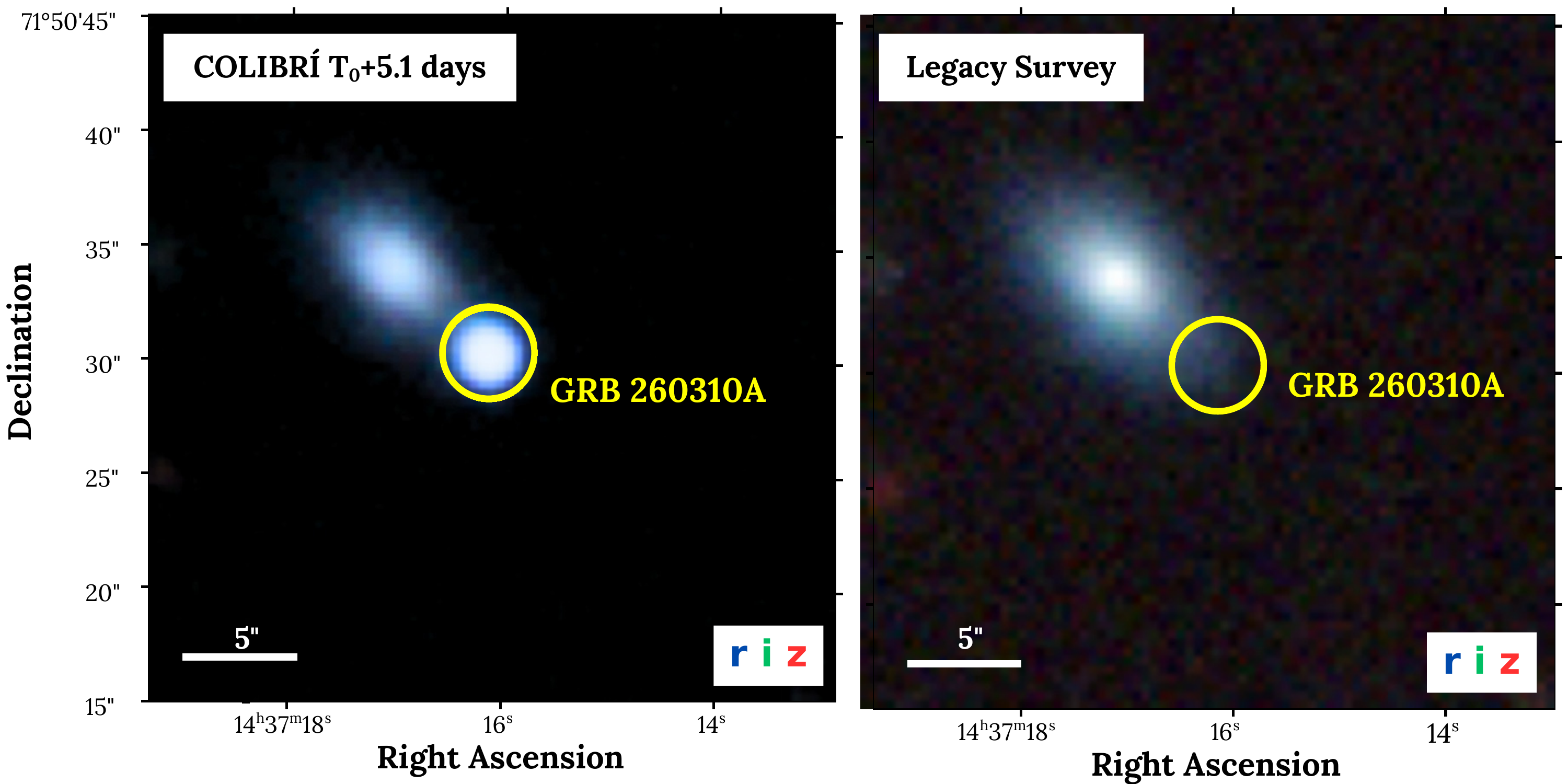}
    \caption{False colour zoomed images of the afterglow field. The left panel shows the COLIBRÍ image obtained at $T_0+5.1$~days. The right panel shows archival Legacy Survey image of the same field, where no source is visible at the afterglow position. The location of GRB\,260310A is marked by the yellow circle in each panel.}
\label{fig:loc}
\end{figure}

We observed the field of GRB\,260310A with the DDRAGO wide-field imager on the COLIBRÍ telescope. COLIBRÍ\footnote{\url{https://www.colibri-obs.org/}} is a Franco-Mexican fast, robotic 1.3~m telescope operated by the Observatorio Astronómico Nacional (OAN) in the Sierra de San Pedro Mártir, Baja California \citep{Basa2022,Basa2026}. DDRAGO is a two-channel imager, with the blue channel operating in $gri$ and the red channel in $zy$ \citep{Langarica2024}. Our observations were carried out in the $g$, $r$, $i$, $z$, and $y$ filters, closely matching the SDSS/Pan-STARRS photometric system \citep[see details in][]{Watson2025}. Fig.~\ref{fig:loc} shows our images at $T_0+5.1$\,days of GRB\,260310A, compared with the same region in the Legacy Survey \citep{Dey+2019}.

The COLIBRÍ ASU pipeline (Butler, in preparation) automatically reduces, aligns, removes sky, coadds, and calibrates the raw images, and produces a list of potential candidates with photometry in the direct image and in the image after subtraction of the corresponding Pan-STARRS DR1 image. Our photometry of GRB\,260310A is obtained by PSF-fitting after subtracting the template image. We performed a manual local recalibration using bright stars within 3~arcmin of GRB\,260310A, which reduced the calibration dispersion to about 0.01~mag. This additional uncertainty is included in the photometry. As we use the same template images consistently and the transient is in a smooth outer part of the host galaxy, uncertainties in the templates are likely to result in shifts in our photometry at levels equivalent to 24--25 mag, but are not likely to increase the statistical noise.

We followed-up the transient over multiple epochs from 3 to 62~days after the trigger, some of the preliminary photometry was reported by \citep{GCN43980,GCN44001,GCN44125}. 
Table~\ref{tab:new_optical_photometry} gives our photometry with COLIBRÍ. Our magnitudes are in the natural AB system without the application of any colour terms and without correction for the Galactic extinction in the direction of the burst. These corrections are however applied before analysis and are given in Table~\ref{tab:extinction}, following \citet{Schlafly+2011}.

\begin{table*}
\centering
\caption{Spectroscopic observing log. T-T$_0$ refers to the time since the GRB, Inst./tel. indicate the instrument and telescope used for the observation, Slit is the width of the slit used for spectroscopy, Pos. Angle the position angle of the slit, measured from North to East, with Par. indicating that the slit was placed in the parallactic angle, along the elevation, and Perp. indicating that it was placed perpendicular to the major axis of the host galaxy to minimize the host contribution. GRISM indicates the dispersing element used for the observation and Exposure time the number of exposures and their duration for each GRISM observation.
}

\label{tab:LogSpectra}
\begin{tabular}{ccccccc}
\hline
\hline
$T-T_0$	& Instrument/Telescope & Slit	& Pos. Angle & Seeing & GRISM & Exposure Time \\
(day)& 	& $^{\prime\prime}$&(deg N-E)&$^{\prime\prime}$& &       (s)	  \\
\hline
3.00 	& OSIRIS+/GTC& 1.0	& 47.0 & 3.1 &R1000B+R1000R& $3\times600+3\times600$  \\
5.97    & MISTRAL/1.93m & 1.9& Perp. & 1.5--2.0  & Blue (700)          &   $2\times1800$           \\ 
6.02 	& OSIRIS+/GTC& 1.0	& Par. & 1.1 & R1000B	   & $3\times600$ \\
9.96    & MISTRAL/1.93m & 1.9& Perp. & 1.5--2.0  & Blue (700)          & $2 \times 1800 + 1\times 1500$                \\ 
13.94   & MISTRAL/1.93m & 1.9& Perp. & 1.5--2.0  & Blue (700)         &  $4 \times 1800 + 1\times 1500$            \\ 
16.99	& OSIRIS+/GTC& 1.0	& Par. & 0.9 & R1000B	   & $3\times900$ \\
22.88   & MISTRAL/1.93m & 1.9& Perp. & 1.5--2.0  & Blue (700)          & $4 \times 1800$              \\ 
26.92	& OSIRIS+/GTC& 1.0	& Par. & 1.3 &R1000B+R1000R& $3\times900+3\times900$  \\
34.93	& OSIRIS+/GTC& 1.0	& Par. & 1.0 & R1000B	   & $3\times900$ \\
\hline
\end{tabular}
\end{table*}

\subsubsection{Additional Optical and NIR Photometry}

For fitting the light curve at later times, we use the COLIBRÍ photometry described above. 
However, to constrain our models at early times, we use photometry from GOTO \citep{ONeill2026} and LAST \citep{GCN43974}, and 
the ATLAS forced-photometry service \citep{Tonry2018, Smith2020, Shingles2021}, which gives observations in the ATLAS $o$ filter at four epochs between $T_0 +7$ and $T_0 + 8$ hours. 
The GOTO, ATLAS, and LAST photometry is shown in Table \ref{tab:new_optical_photometry}. We note that the GOTO, ATLAS, LAST, and COLIBRÍ photometry have all been corrected for the contribution of the host galaxy.

We note that \cite{Yang+2026} constrain the early light curve using the DDOTI magnitude for the interval $T_0 + 0.2$ to 4.6 hours, taken from the GCN Circular by \cite{gcn43991}, with an assigned uncertainty of 0.3 mag. However, we further note that the original authors label the magnitude as “tentative” and do not give an uncertainty. The magnitude is also apparently averaged over a considerable interval. Given these considerations, it is unclear to what extent this measurement can robustly support the conclusions subsequently drawn by \citeauthor{Yang+2026}

Table \ref{tab:new_J_photometry} shows photometry in the $J$ band from GCNs and new photometry from the ALT100C 100-cm telescope of the JinShan project at Altay, Xinjiang, China. This is used as an independent check of our models and to derive SEDs. We note that no correction for host-galaxy contribution has been applied to the $J$ band photometry, as no sufficiently deep template images are available. In particular, archival data from the 2MASS survey \citep{Skrutskie2006} are too shallow to reliably subtract the host contribution, which may introduce an additional systematic uncertainty in the measured fluxes.

Table \ref{tab:new_extraphotometry} shows additional photometry from the Nordic Optical Telescope (NOT), the Thai Robotic Telescope at Sierra Remote Observatories (TRT-SRO), and the ALT100B 100-cm telescope of the JinShan project at Altay, Xinjiang, China. These observations are not used for fitting, as they do not provide additional constraints, but some are used to derive SEDs.



\subsection{Optical Spectroscopy}
\label{sec:spectroscopy}
\subsubsection{OSIRIS+ at the 10.4~m GTC}
Five spectroscopic observations were made with the OSIRIS+ instrument on the 10.4~m Gran Telescopio Canarias (GTC) telescope, at the Roque de los Muchachos Observatory (La Palma, Spain). The data were processed using standard procedures using self developed scripts under an {\sc iraf} environment. These reductions included bias correction, response correction using flat fields, wavelength correction based on HgAr, Ar, Ne and Xe lamps. Flux calibration was performed using as reference spectrophotometric standards Ross~640 and Hiltner~600. The spectra were then corrected for slit losses taking into account the recorded seeing in the acquisition images, the slit width, and the position angle. Finally, the spectral flux was scaled using photometry to correct for additional effects due to non-photometric conditions.

The first spectrum, obtained 3.00~days after the burst, had the slit aligned with the core of the host galaxy, to study the spectrum of the galaxy in addition to the GRB light. Unfortunately, during that observation the seeing was very bad, at $\sim3^{\prime\prime}$, limiting the information that can be extracted. The rest of observation were performed with a parallactic position angle to limit slit losses.

The spectra show prominent emission lines due to the host galaxy. We identify the Balmer lines from H$\alpha$ to H$\delta$, [\ion{O}{II}], [\ion{O}{III}] and [\ion{S}{II}] all at a common redshift of $0.1532\pm0.0002$. [\ion{N}{II}] emission line is right at the edge of the telluric A-band, but with adopted fitting procedures we are able to measure its flux as well.   

The second epoch, obtained 6~days after the burst, has a good signal-to-noise ratio and is afterglow dominated. We used this spectrum to search for absorption features within the spectral range. We do not detect any of the two H and K \ion{Ca}{II} features down to a $3\sigma$ limit of $< 0.18$~{\AA}. A similar limit is obtained for \ion{Ca}{I}\,$\lambda$4227. The only feature that is clearly detected in absorption is the \ion{Na}{I} D $\lambda\lambda5890/5896$ doublet with a combined observed equivalent width of $0.35\pm0.07$ {\AA} (0.15 and 0.23) with a redshift of $0.1526\pm0.0002$, which we interpret as the most accurate redshift for the GRB.

We can compare the detection limits obtained for the \ion{Ca}{II} lines with those of a large sample of GRB spectra by calculating the Line Strength Parameter (LSP) as described by \citet{deUgartePostigo2012}. This parameter compares the strength of the spectral features with the distribution of their strength seen in a sample of 69 GRB afterglow spectra. A value of zero would imply that the line of sight has typical features, whereas a positive number imply strong features and a negative one weak features. The detection limits obtained for \ion{Ca}{II} imply an LSP $<-2.46$ which implies that the spectral features are weaker than all but one GRB sight-lines of that sample.

\begin{figure*}
    \includegraphics[clip, width=1\linewidth]{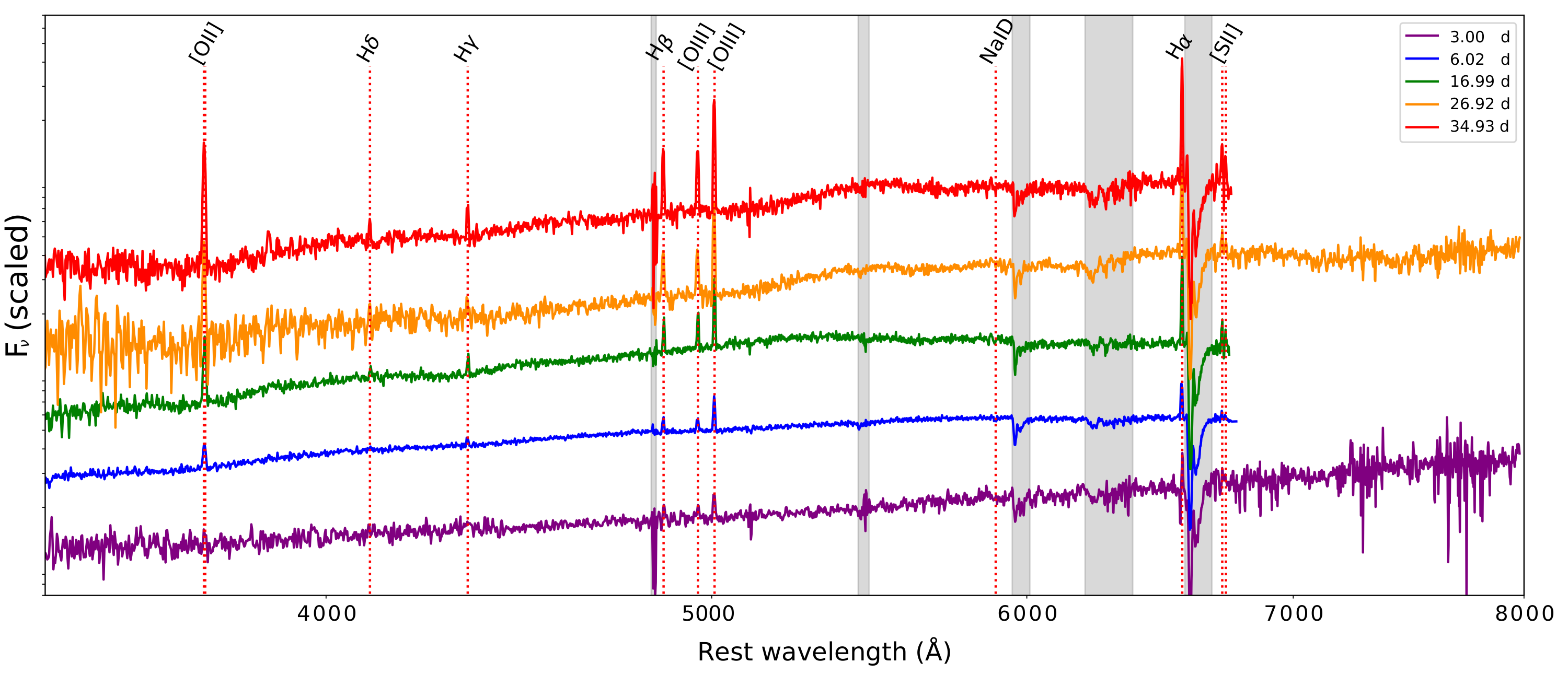}
    \caption{Spectral sequence of GRB\,260310A obtained with OSIRIS+ at the 10.4m GTC. Vertical gray bands indicate spectral regions affected by telluric absorption or sky line residuals. Several prominent emission lines and the \ion{Na}{I}D absorption are indicated with dotted red lines. 
\label{fig:spectra}}
\end{figure*}

From the third epoch on, the continuum started to show prominent undulations that we associated with SN features that allowed us to identify an underlying broad lined SN Ic \citet{deUgartePostigo2026}, which we refer to as SN\,2026fgk \citep[also see][]{OConnor2026}. In the subsequent epochs the spectral shape evolves with time as the afterglow contribution recedes and the SN features, produced by the ejecta, decline in velocity. The spectra are shown in Fig.~\ref{fig:spectra}.

\subsubsection{MISTRAL / Observatoire de Haut-Provence}
\label{sec:mistral}
We obtained four optical spectra of GRB\,260310A at $T_0+5.97$, $T_0+9.96$, $T_0+13.94$, and $T_0+22.88$~days after the \textit{Fermi}/GBM trigger (see Table~\ref{tab:LogSpectra}) using the MISTRAL instrument on the 1.93~m telescope at the Observatoire de Haut-Provence (France). The slit was positioned perpendicular to the host main axis in order to minimize the host contribution to the GRB emission. Observations were made under a seeing of the order of 1.5 to 2~\arcsec.

The spectra were normalised and smoothed for visualization purposes, and are presented in Figure~\ref{fig:ohp} (online supplementary material). The spectra were normalised and smoothed for visualization purposes and are presented in Figure~\ref{fig:ohp}. The observations at $T_0+5.97$ and $T_0+9.96$ bracket the break at $T_0+8$~days shown in Fig.~\ref{fig:multi-waveband-LC-Spectra}, while the spectrum obtained at $T_0+22.88$~days corresponds to the beginning of the plateau phase at $T_0+20$--$30$~days in Fig.~\ref{fig:multi-waveband-LC-Spectra}.

At all epochs, we detect narrow emission lines consistent with a host galaxy at $z = 0.153$, including \ion{O}{II}, \ion{H}{$\beta$}, \ion{O}{III}, \ion{H}{$\alpha$}, \ion{N}{II}, and \ion{S}{II}. The positions of these lines are indicated in Figure~\ref{fig:ohp}.

The earliest spectrum is well described by a relatively featureless continuum, consistent with a dominant afterglow component. A noticeable characteristic is however a small [\ion{N}{II}]@6548A present in the GRB spectrum and not visible in all the GTC spectra (see Fig.~\ref{fig:spectra}). This [\ion{N}{II}] emission is mainly visible in $T_0+22.88$~days spectrum.


\subsection{Radio campaign}
\label{sec:radiolog}

GRB\,260310A was extensively monitored at radio wavelengths with the Karl G.~Jansky Very Large Array (VLA) \citep{Perley2011}, revealing one of the brightest nearby GRB afterglows observed in recent years. The first broadband VLA campaign was performed at $T_0+4.26$~d, covering central frequencies of 6, 10, and 15~GHz \citep{GCN44045}. A bright source was clearly detected at all bands, with flux densities of $3894\pm8$, $6368\pm13$, and $9104\pm29~\mu$Jy. These early detections indicate a rapidly rising radio spectrum, with $F_{\nu}\propto \nu^{0.9}$ between 6 and 15~GHz (see the bottom panel in Fig.~\ref{fig:multi-waveband-LC-Spectra}). 

A second VLA epoch was obtained at $T_0+17.2$~d with nearly continuous spectral coverage from 3 to 45~GHz. The source remained bright, reaching $2.90\pm0.04$~mJy at 3~GHz and peaking at $11.73\pm0.16$~mJy near 15~GHz, before declining toward higher frequencies to $7.98\pm0.27$~mJy at 45~GHz \citep{GCN44160}. 

Monitoring of GRB\,260310A at 15~GHz was performed with the OVRO 40~m telescope, providing a light curve between $T_0+7$ and $T_0+36$~days. There is a nearby unrelated radio source within the 40 m telescope beam located 1.29~\arcmin from the GRB position at RA(J2000) = 14:36:59.770, Dec(J2000) = +71:50:42.77, identified in the VLASS 3.1 quick-look cutouts (via CIRADA) and cross-matched with WISEA J143659.59+715042.9. This source has a 15~GHz flux density of $\sim3.8$~mJyand is located near the half-power point of the OVRO beam when pointed at GRB\,260310A, so small pointing offsets introduce significant uncertainties in the detected flux density of the confusing source. The OVRO 15~GHz flux density of GRB\,260310A, is reported in Table~\ref{tab:radio_photometry}, where, to account for the residual uncertainty associated with the nearby contaminating source, we added a systematic uncertainty of 1~mJy in quadrature to the statistical errors of the OVRO measurements.

The complete radio dataset used in this work is listed in Table~\ref{tab:radio_photometry}.

The high signal-to-noise VLA detections reported in this campaign motivate an assessment of whether spatially resolved radio signatures could be detectable. We examine the expected radio image evolution and flux-centroid motion in Section~\ref{sec:radiocentroid}.

Additional radio observations were carried out by the Arcminute Microkelvin Imager Large-Array (AMI-LA) \citep{GCN44005}, the Submillimeter Array (SMA) \citep{GCN44134} and the NOrthern Extended Millimeter Array (NOEMA) \citep{GCN44057}.

\begin{table*}
\centering
\caption{Summary of SED fitting results. (1) stellar mass in log scale, (2) SFR in log scale, (3) amount of dust attenuation in the V band, (4) metallicity, (5) age of the main stellar population in the galaxy, (6) attenuation curve slope for stellar continuum, (7) AGN fraction, (8) mass fraction of the late burst population.}
\label{Table:SEDresults}

\begin{tabular}{cccccccc}
\hline
log$_{10}$(M$_{\star}$/[M$_{\odot}$]) &  log$_{10}$(SFR/[M$_{\odot}$ yr$^{-1}$]) & A$_V$ (mag) & $Z$ & t$_0$ (Myr) & $\delta$ & $f_{\rm agn}$ & $f_{\rm burst}$ \\
\hline
$10.19_{-0.18}^{+0.13}$ & $0.14_{-0.40}^{+0.20}$ & $0.28 \pm 0.12$ & $< 0.010 $ & $6718 \pm 2711$ & $-0.51 \pm 0.40$ & $< 0.15$ & $< 0.09$ \\
\hline
\end{tabular}
\end{table*}

\section{Host Galaxy \& Environment}
\label{sec:hg}

GRB\,260310A is spatially coincident with a bright blue galaxy identified in the Legacy Survey \citep{Dey+2019}. This source lies only $5.6\arcsec$ from the transient position and has magnitudes of $g = 18.67 \pm 0.01$ and $r = 18.67 \pm 0.01$ (not corrected for Galactic extinction). The galaxy is classified as a Sérsic-type and exhibits a relatively large half-light radius of $R_{\rm half} \approx 2.84\arcsec$, corresponding to $\sim7.8$~kpc, indicative of a well-resolved and potentially massive system (see Figure~\ref{fig:loc}). Using the formalism of \citet{Bloom2002} and using the values presented by \citet{Becerra2023}, we estimate a probability of chance coincidence of $P_{\rm cc} \approx 3.0 \times 10^{-3}$, suggesting that this galaxy is the likely host candidate.

The offset of the optical afterglow of $5.6\arcsec$ from the center of the host galaxy corresponds to a projected separation of 14.99~kpc \citep[also see][]{OConnor2026}, or 1.92 half light radii (r$_{50}$) of the galaxy. This offset is among the largest reported for long-duration (Type II) GRBs \citep{Blanchard2016}, but not unprecedented, especially for large host galaxies \citep[see e.g. GRB 171205A][]{Thoene2024}. Despite the large offset, the transient position still lies within the visible extension of the host, at the edge of a luminous region of the galaxy (see Figure~\ref{fig:loc}). The host itself is a blue, star-forming galaxy \citep{GCN43977,GCN43984} and its morphology suggests a moderately face-on spiral galaxy, with the GRB occurring in a star-forming region in one of the outer spiral arms.

Using the archive CFHT, Pan-STARRS, and WISE photometry described in Appendix~\ref{app:hgphoto}, we performed the SED fitting of this galaxy using CIGALE\footnote{\url{http://cigale.lam.fr}} (Code Investigating GALaxy Emission; \citealt{Boquien2019}), adopting a parameter space similar to \citet{Corre2018}. We have adopted a delayed star formation rate ($\rm SFR \propto t/\tau_0^2 \cdot e^{-t/\tau_0}$), on top of which a possible recent burst of star formation is enabled. The initial mass function of \citet{Chabrier03} is adopted with the stellar synthesis models of \citet{BC03}. The stellar emission absorbed by dust and re-emitted in the IR is modelled using IR templates from \citet{Dale14} and using $\alpha_{IR} \in [1,3]$. The host galaxy SED fit, which resulted in a reduced $\chi^2 = 0.18$, is shown in Fig.~\ref{fig:host} (online supplementary material) and the derived properties of the host galaxy are displayed in Table~\ref{Table:SEDresults}. The SED fit indeed confirms this galaxy to be a massive, mildly star-forming galaxy with a log stellar mass of 10.19$^{+0.13}_{-0.18}$ M$_\odot$, again similar to the host of GRB\,171205A with log (M*/M$_\odot$) = 10.29 M$_\odot$ \citep{Thoene2024}.

\begin{figure}
	\includegraphics[width=\columnwidth]{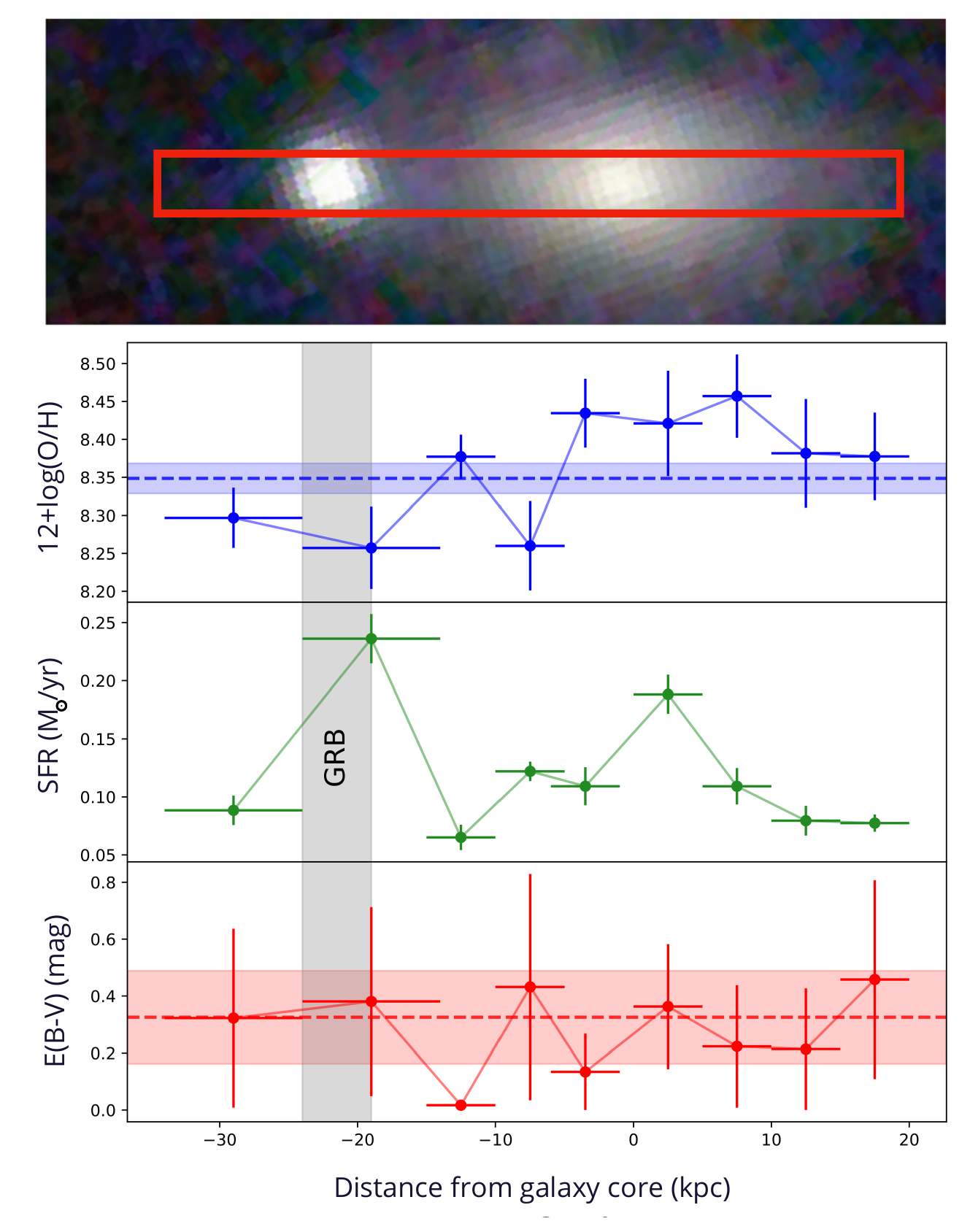}
    \caption{The image indicates the location of the slit on the host galaxy. We extracted bins of 10 pixels (2.54~\arcsec) for the outer two regions on the left of the panel (including the GRB site) to increase the S/N, and 5 pixels (1.27~\arcsec) for the rest of the slit. 
    Top: The first panel shows the metallicity, determined from the O3N2 parameter \citep{Marino13}, which reaches a minimum around the location of the GRB and peaks around the galaxy core. Middle: It shows the star formation rate which is maximum at the location of the GRB, implying that it must be located in the neighbourhood of a prominent star formation region. 
    Bottom: We show the extinction which is relatively uniform across the galaxy and overall rather low. The extinction has also been used to correct the emission line fluxes in each bin. The dashed horizontal lines and shaded regions mark the metallicity, extinction and corresponding errors determined from the integrated spectrum of the host along the slit.}
\label{fig:hostresolved}
\end{figure}

Using the first OSIRIS+/GTC spectrum at $T_0 + 3$ days, we perform a study of the host galaxy in emission. The slit was aligned along the major axis of the host galaxy, covering its core and also the location of the GRB. We performed multiple extractions across this slit that allows us to measure the host properties in different parts of the galaxy, including the site of the GRB. We measure the flux of [OIII], H$\beta$, H$\alpha$ and [NII] to determine (1) emission line metallicities using the O3N2 parameter in the \cite{Marino13} calibration, (2) the extinction from the Balmer decrement (H$\alpha$/H$\beta$) and the absolute SFR using the H$\alpha$ flux. All emission lines have been corrected for a Galactic extinction of E(B--V)$=$0.018 and the intrinsic host extinction determined in each bin. The results are shown in  Figure~\ref{fig:hostresolved}. 
The GRB is located in a region of lower than average metallicity and a high local star formation rate, indicating its position in a possibly young HII region. The extinction is low across the entire galaxy. These properties are typical for the site of long GRB, produced by the collapse of massive stars. We also determine properties of the summed galaxy spectrum, which gives an average metallicity of 12+log(O/H)$=$8.35$\pm$0.02 and E(B--V)=0.33$\pm$0.16 and a total SFR of 1.13$\pm$0.14 M$_\odot$yr$^{-1}$.

\begin{figure}
	\includegraphics[clip, width=\columnwidth]{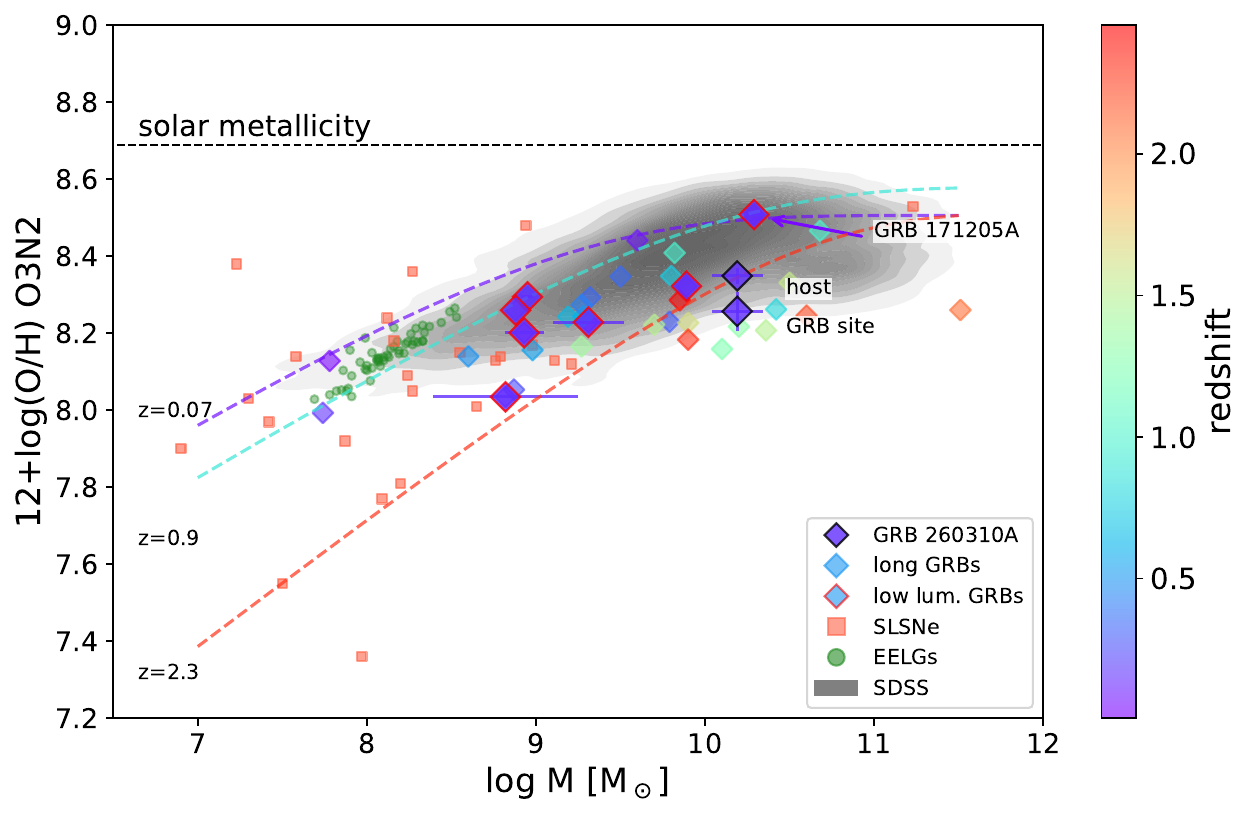}
    	\includegraphics[clip, width=\columnwidth]{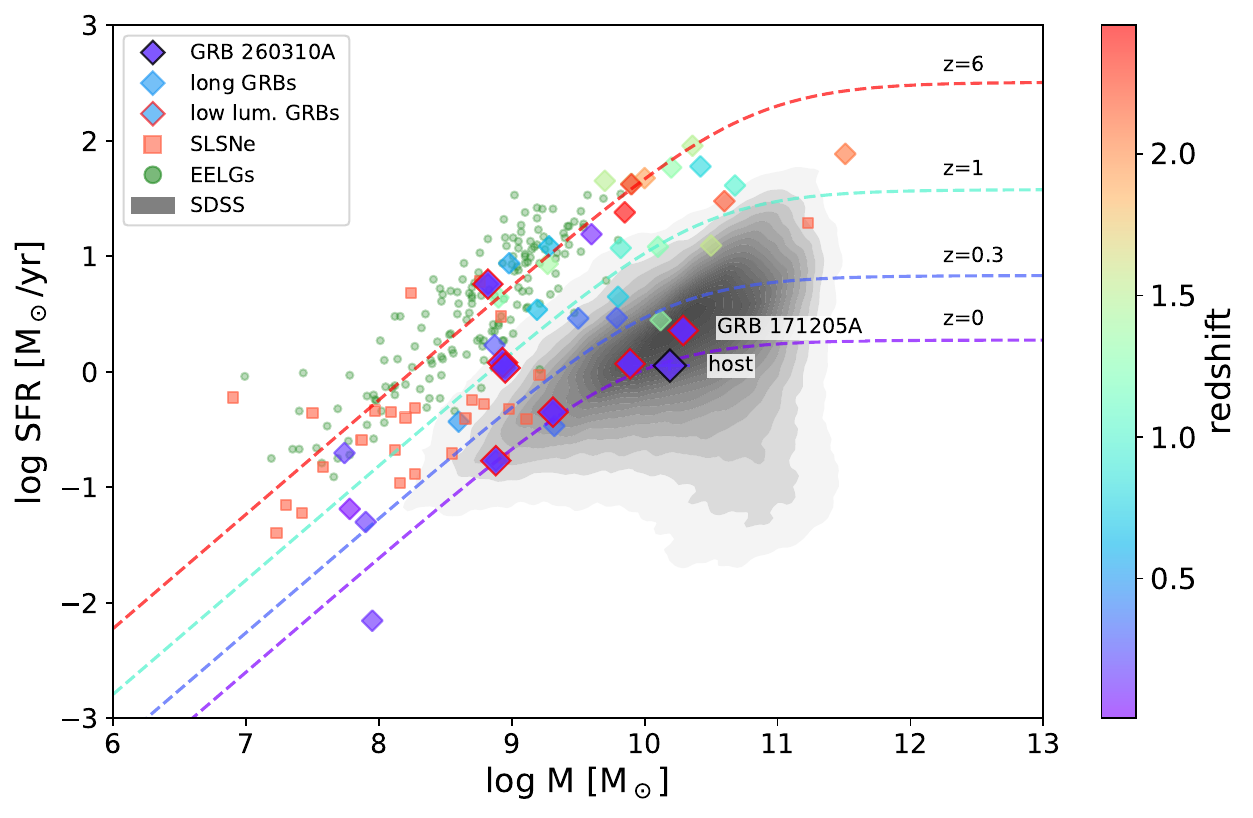}
    \caption{Comparison of the host of GRB 260310A with other long GRB hosts as well as other transient host samples: Superluminous supernova Type I hosts \citep{Leloudas15, Perley16}, extreme emission line galaxies \citep{Amorin15} and SDSS star-forming galaxies using DR 16. GRB hosts data were obtained from the samples of \citealt{Kruhler15, Han10}, with additional GRBs from \citealt{DellaValle06, Christensen08, Kelly13, Schulze14, Thoene14, Izzo17, Heintz18, Cano17, deUgarte18, Melandri19, deUgarte20, Thoene2024}, stellar masses from the GHostS database (http:\/\/www.grbhosts.org) and \citealt{Palmerio19}. Top panel: mass-metallicity relation indicating the metallicity at the GRB site and the global host metallicity. M-Z relations at different redshifts are from \citealt{Zahid13} and \citealt{Wuyts2014} and transformed into the O3N2 metallicity calibration established by \citealt{Marino13}. Bottom panel: mass-SFR relation, taking the global SFR of the host. The star-formation main sequence relations at different redshifts are taken from \citealt{Popesso2022, Lee2015}. Error bars of the comparison samples are omitted for clarity. In both plots we also include the host of GRB 171205A, mentioned in the text, which had a very similar mass to GRB 260310A at an even lower redshift, as well as other low redshift low luminosity GRB host (red edged symbols) of GRB\,980425, GRB\,031203, GRB\,100316D, GRB\,120422A, GRB\,171205A, GRB\,060505, GRB\,161219B.}
\label{fig:hostcomparison}
\end{figure}

Comparing the properties of the host and GRB site to other long GRB environments and field galaxies, the host is a rather average star-forming galaxy at z$\sim$1, but with a lower metallicity. In the M-Z relation (see Fig.\ref{fig:hostcomparison}, top panel), the host falls among the average metallicity values for other GRB hosts while the site is clearly low for its redshift. However, long GRB hosts in general, including GRB 260310A, have higher metallicities and also higher masses than e.g. superluminous supernova hosts or extreme emission line galaxies, which has been noted before. At its redshift, the host has one of the highest stellar masses determined for long GRB hosts. In the M-SFR relation (Fig~\ref{fig:hostcomparison}, bottom panel), the host is in the bulk of SDSS star-forming galaxies, right on top of the SFR main sequence at z$=$0. Most long GRB hosts lie somewhat above the SFR main sequence for their corresponding redshift. However, this is likely a consequence of the rather large host mass of GRB 260310A, where properties vary more across the host galaxy compared to dwarf galaxies, and not an indication for a different environment, especially considering the high SFR rate at the actual GRB position compared to other locations in the host. 

\section{Analysis}
\label{sec:analysis}
\subsection{Prompt Emission \& Amati Relation}
\label{sec:promptanalysis}

\begin{figure}
    \includegraphics[width=0.95\linewidth]{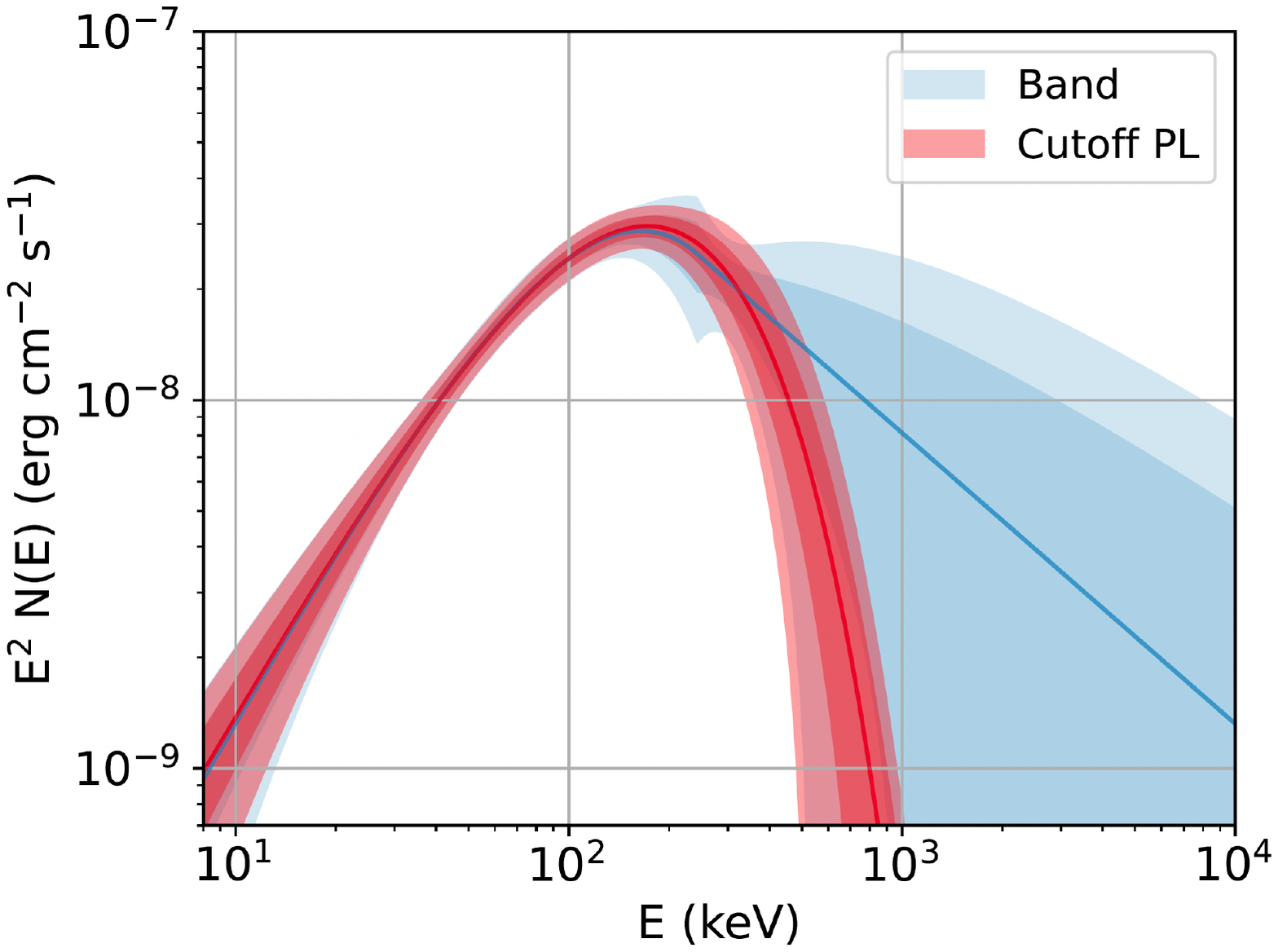} \\
    \includegraphics[width=0.95\linewidth]{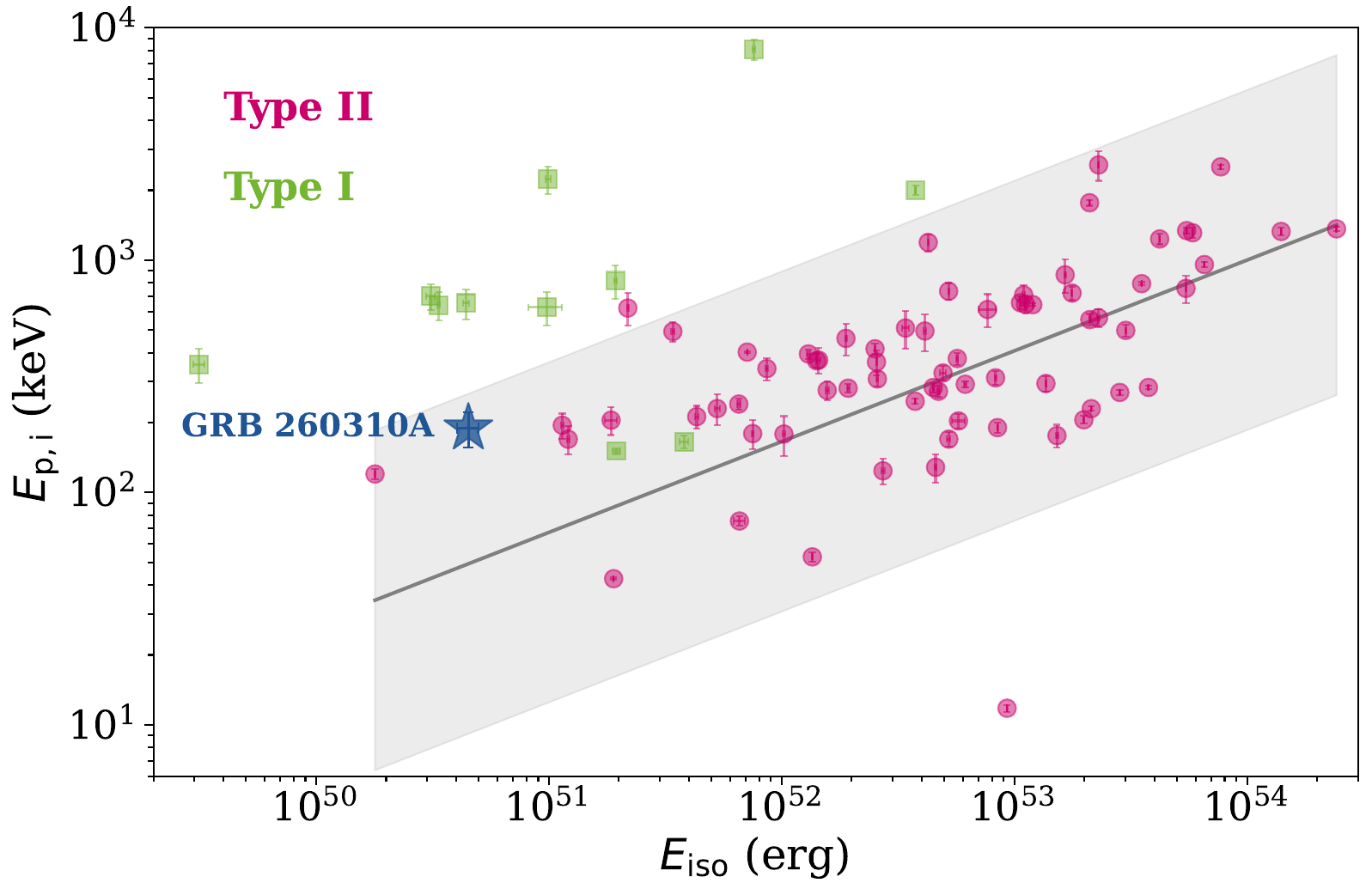}\\
    \caption{Top: Comparison of the cut-off power-law and Band function description of the entire spectrum. The two shaded regions around the best fitting curves mark the 1 and 2$\sigma$ confidence regions. The spectral fits are shown in Fig.\,\ref{fig:prompt-spec-fit} (online supplementary material).
    Bottom: Location of GRB~260310A in the $E_{\rm iso}$–$E_{\rm p,i}$ (Amati) plane. The blue star marks GRB~260310A. The comparison sample consists of \textit{Fermi}/GBM GRBs with measured redshifts. The solid grey line represents an orthogonal distance regression fit to the Type II (collapsar) population in log–log space, using only events with $E_{\rm peak}$ fractional uncertainties below 10\%, and the shaded region indicates the corresponding $2\sigma$ dispersion. GRB~260310A lies in the region of Type II population.}
    \label{fig:prompt-Amati}
\end{figure}

We analyse the prompt emission pulse over two intervals, as shown in Fig.\,\ref{fig:lcfermi}, where the earlier (blue) one is energetically hard while the later (orange) one is soft. The spectra in these two intervals separately and over the entire burst (see top panel of Fig.\,\ref{fig:prompt-Amati}) are best fit by a power-law function with an exponential cutoff (or cutoff PL; see model description in Appendix\,\ref{sec:prompt-models}). The best fit values are given in Table~\ref{tab:prompt-spec-params} and the model fit in count space is shown in the left panel of Fig.\,\ref{fig:prompt-spec-fit} (online supplementary material). We note that in the first interval the photon index is relatively hard, but given the uncertainties it is not an outlier in the broader population. The isotropic-equivalent energy release in the 1-10,000~keV energy range is $E_{\rm iso}=(3.6\pm 0.3)\times 10^{50}$ erg and the timescale over which the central 90\,per\,cent of the fluence is accumulated is $T_{90}\simeq60$\,s (50-300\,keV; \citealt{GCN43975}).

We also fit the spectrum with the commonly used Band function (see model description in Appendix\,\ref{sec:prompt-models} and the spectral fit in the right panel of Fig.\,\ref{fig:prompt-spec-fit}). It does not provide a statistically significant improvement to the fits, although it also gives an acceptable description of the spectrum. For the $T_0 - 8$ to $T_0 + 98$~s interval, the best-fitting Band-function parameters are $\alpha_B = -0.34 \pm 0.26$, $\beta_B = -2.79 \pm 0.81$, and $E_{\rm peak} = 164 \pm 28$~keV. This model yields a flux of $(6.67 \pm 0.70)\times10^{-8}$~erg~cm$^{-2}$ s$^{-1}$ and an isotropic-equivalent energy in the 1--10,000~keV range of $E_{\rm iso} = (4.51 \pm 0.47)\times10^{50}$~erg.

We also estimate the timescale of the shortest coherent variations in the light curve, namely the minimum variability timescale (MVT). Using the method of \citet{Golkhou+15variability} \citep[see also][]{Bala+26mvt}, we find $T_{\rm MVT} = 4.3 \pm 1.8$~s.

Using the isotropic-equivalent energy of $E_{\rm iso} = (4.5 \pm 0.5)\times10^{50}$~erg and a cosmological rest-frame peak energy of $E_{\rm p,i} = (198 \pm 6)$~keV derived before, we put GRB\,260310A in the context of other \textit{Fermi}/GBM GRBs. The bottom panel of Fig.~\ref{fig:prompt-Amati} shows the location of GRB\,260310A as a blue star in the Amati $E_{\rm iso}$–$E_{\rm p,i}$ plane \citep{Amati2008,Amati2013} and compares it to the sample of \textit{Fermi}/GBM GRBs with measured redshifts \citep{vonKienlin2020,Gruber2014,vonKienlin2014,Bhat2016}. For each burst, $E_{\rm iso}$ is computed from the reported fluence assuming the cosmology described in Section~\ref{sec:introduction}, and $E_{\rm p,i}$ is obtained by correcting the observed peak energy to the rest frame. The sample is restricted to events with well-constrained spectral parameters, requiring fractional uncertainties below 20\% in both fluence and $E_{\rm peak}$. Long GRBs (Type II; $T_{90} > 2$~s) are shown in magenta, while short GRBs (Type I; $T_{90} < 2$~s) are shown in green. The solid line represents an orthogonal distance regression fit to the long-GRB population, performed in log–log space using only bursts with $E_{\rm peak}$ uncertainties below 10\,per\,cent, and the shaded region indicates the corresponding $2\sigma$ dispersion. GRB~260310A lies in the region of collapsar-driven (Type II) events and is broadly consistent with the Amati relation. Moreover, the nature and subsequent classification of the event is unambiguously reinforced by the emergence of an associated supernova component discussed in later sections.

\subsection{The Afterglow}
\label{sec:closure}
\subsubsection{Temporal \& Spectral Analysis}
\label{sec:qualitative}

We present the multi-wavelength light curve and spectral energy distributions (SEDs) at different epochs in Fig.~\ref{fig:multi-waveband-LC-Spectra}. We model both using multiple broken power-law segments to extract the temporal ($\alpha$) and spectral ($\beta$) indices, with the flux density given by $F_{\nu}\propto T^{\alpha}\nu^{\beta}$.

The optical emission exhibits a structured temporal evolution that can be divided into several phases. At early times ($T-T_0 < 0.50$~d), the light curve shows a rise ($\alpha_{\rm o} = 0.59 \pm 0.03$). The emission then transitions into a decay phase extending to $T-T_0\sim 3.5$~days ($\alpha_{\rm o} = -0.65\pm 0.01$) and from $T-T_0\approx3.5$ to $T-T_0\approx7$~days a shallower decay of ($\alpha_{\rm o} = -0.35\pm 0.04$), consistent with a plateau phase. After $\sim 7$~days, the light curve steepens significantly ($\alpha_{\rm o} = -1.25 \pm 0.04$) and shows a supernova bump around $T-T_0\gtrsim20$\,days before declining more rapidly ($\alpha_{\rm o} = -1.83 \pm 0.03$) at $T\gtrsim28$\,days. 

The X-ray light curve follows a broadly similar evolution since the first observation carried out at $T-T_0\approx2.5$\,days but with some differences. It shows an initial decay phase ($0.1 < T-T_0 < 4.3$ d) that is characterized by $\alpha_{\rm X} \approx -0.81$, and then followed by a flattening between $4.3$ and $7.0$~days ($\alpha_{\rm X} \approx -0.18$), which is temporally coincident with the optical plateau. The X-ray emission then undergoes a steep decline between $7$ and $20$~days ($\alpha_{\rm X} = -1.87 \pm 0.11$)  consistent with that reported by \citet{GCN44095}, before showing tentative evidence for a late-time rebrightening, consistent with that reported initially by \citet{GCN44278} and \cite{GCN44375}.

At radio frequencies, the light curves display a markedly different behaviour. The emission rises or remains nearly flat at early times, with frequency-dependent temporal indices (e.g. $\alpha \sim 0.4$ at 10~GHz and steeper at lower frequencies), and peaks around $\sim 10$~days. After the peak, the radio emission decays with $\alpha \sim -1.0$ to $-1.2$ at high frequencies (33–45~GHz), while lower frequencies evolve more gradually, indicating spectral evolution and the passage of characteristic synchrotron frequencies through the observing bands.

The SEDs at multiple epochs, spanning radio to X-ray frequencies (see the bottom panel in Fig.~\ref{fig:multi-waveband-LC-Spectra}), are interpreted within the synchrotron framework of \citet{Granot-Sari-02}. At $T_0 + 4.26$~days, where full multi-wavelength coverage is available, the SED is well described by a single synchrotron component with smooth breaks, allowing us to constrain the characteristic break frequencies ($\nu_{\rm sa}$, $\nu_m$, and $\nu_c$).

At later epochs with radio coverage ($T_0+17.20$ and $T_0+25.20$~days), a single component synchrotron model cannot reproduce the observations, and shows a mismatch between the radio and higher-frequency emission. We therefore adopt a two component model, where the optical-to-X-ray emission and the radio emission are fitted separately, using a synchrotron spectrum with smooth breaks. We find that this approach provides a significantly better description of the data and suggests a more complex emission structure at these times. For epochs without radio data, the SEDs are well described by a single power-law segment, $F_{\nu} \propto \nu^{\beta}$, consistent with a single spectral regime of the synchrotron spectrum. 

Overall, the multi-band temporal evolution reveals clear chromatic features that will be discussed in Section~\ref{sec:chromatic}.

\subsubsection{Interpretation with a Spherical Blast Wave}

Next, we apply the standard afterglow closure relations for a spherical outflow \citep[e.g.][]{Granot-Sari-02} to check for consistency with an on-axis uniform jet, and to also infer the radial density profile of the external medium as well as the energy power-law index of the electrons accelerated at the shock front. 

The density of the external medium into which the relativistic blast wave propagates is typically parameterized as a power law in radius, $n_{\rm ext}(R)\propto R^{-k}$, where $k=0$ signals an interstellar medium (ISM) that is mostly valid for compact binary mergers. Alternatively, in a collapsar-driven GRB the massive star progenitor alters the external medium density profile by driving powerful winds pre-core-collapse, in which case $k=2$ but more generally $0<k\leq2$. The interaction of the blast wave with the external medium produces an external forward shock that accelerates electrons from the swept up medium into a power-law energy distribution, with comoving number density (in the frame of the blast wave) given by $n_e(\gamma)\propto\gamma^{-p}$ for electron LFs $\gamma>\gamma_m$, where $\gamma_m$ is the LF of the minimal energy electrons. These electrons cool by emitting synchrotron radiation and the closure relations allow to constrain both $k$ and $p$.

The afterglow optical emission is typically observed at frequencies $\nu_m<\nu<\nu_c$ (or power-law segment (PLS) G; see \citet{Granot-Sari-02}) in a slow-cooling synchrotron spectrum. Here $\nu_m$ is the characteristic synchrotron frequency of electrons with LF $\gamma_m$ and $\nu_c$ is the cooling break frequency corresponding to those electrons that are cooling on the dynamical time \citep{Sari+98}. Our optical light curve data lacks a clear peak, but there must be one between the first optical data point at $T\sim10^{-2}$\,days and the second at $T\sim0.3$\,days. Assuming that the peak in the optical light curve at $T-T_0\lesssim0.3$\,days is due to the deceleration of the blast wave, the pre- and post-deceleration temporal indices for PLS G are $\alpha_{\rm pre}=3-k(p+5)/4$ and $\alpha_{\rm post}=[k(3p-5)-12(p-1)]/4(4-k)$ \citep[e.g.][]{Beniamini+22}, where the spectral index is $\beta=(1-p)/2$. Lacking early-time optical data it is difficult to ascertain the true temporal index below the peak, however, when assuming that the temporal index from our power-law fit of $\alpha_{\rm pre}\simeq0.59$ is close to its true value and $\alpha_{\rm post}\simeq-0.65$, we find $k\simeq1.5$ and $p=1.47$. The value of $p$ is inconsistent with the expectation of $2<p<3$ where diffusive shock acceleration theory argues for $p=2.2$. This inconsistency stems from the shallow post-deceleration slope of the optical light curve which is arguably affected by the energy injection episode (sec.\,\ref{sec:uniform-jet-with-Einj}) or the jet angular structure (sec.\,\ref{sec:misaligned-PLJ}). In the case of energy injection the light curve is expected to return to its pre-injection slope when energy injection ceases, if the emission is still arising from the same PLS of the synchrotron spectrum. Under that assumption the temporal slope of $\alpha=-1.87$ (from the X-ray light curve since the optical light curve may be affected by the SN emission) at $8\lesssim T-T_0({\rm days})\lesssim 20$ yields $k=1.17$ and $p=3.2$. Both of these values are now within their respective ranges, although marginally for $p$. Alternatively, we find that the mean spectral index is $\beta\sim-0.9$ which gives $p\sim2.8$ and yields $k\simeq1.24$ from the pre-deceleration temporal slope that is not affected by energy injection or jet angular structure. This simple analysis argues for neither an ISM nor a wind external medium, but something in between.

\begin{figure}
    \centering
    \includegraphics[width=1.\linewidth]{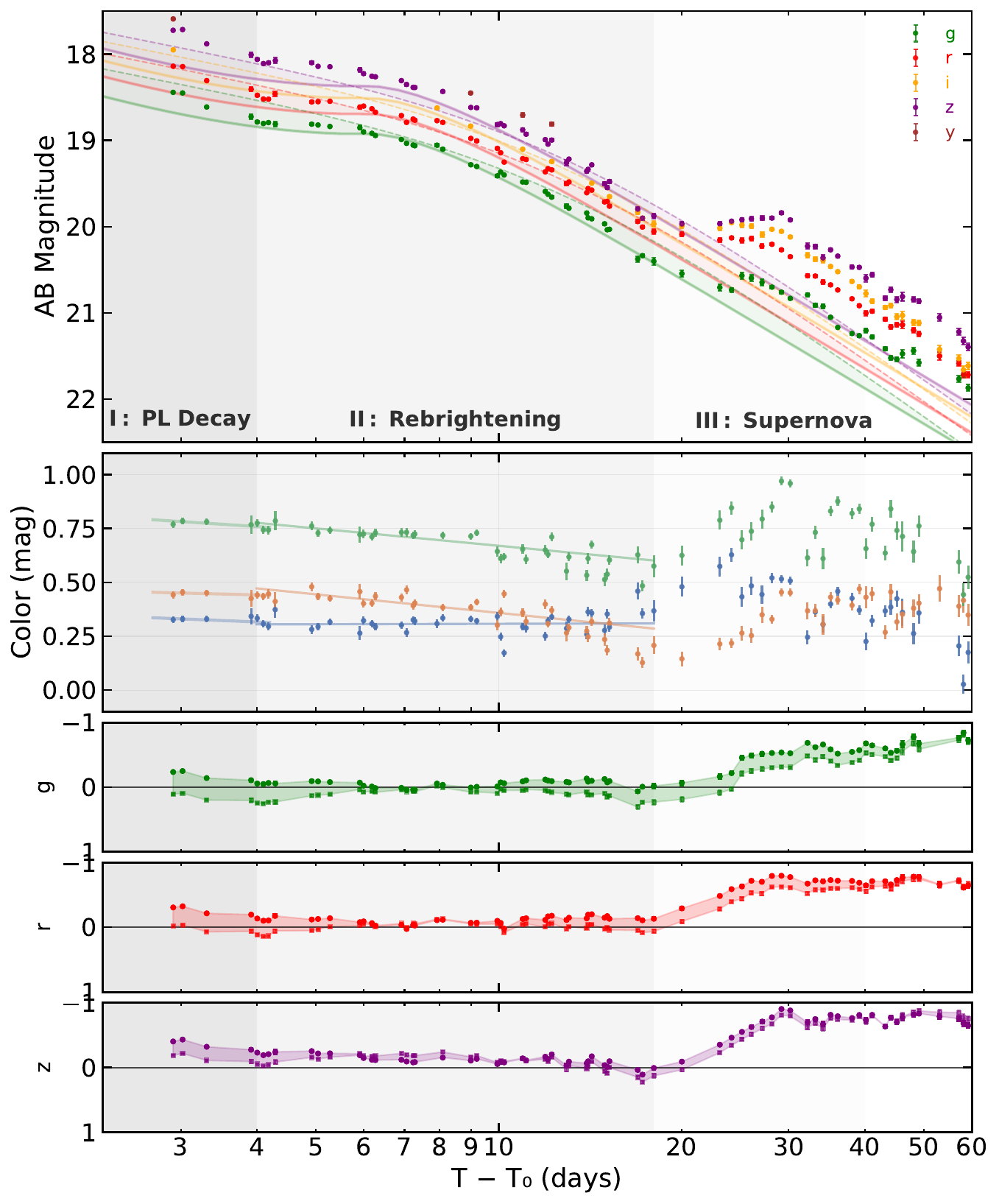}
    \caption{ COLIBRÍ Multi-band optical light curves, color evolution, and model residuals. Top: Light curves in the $g$, $r$, $i$, $z$, and $y$ bands as a function of time since $T_0$. All bands exhibit a common temporal structure consisting of an initial decay (Phase I), followed by a rebrightening/flattening phase (Phase II), and a late-time excess (Phase III). The overplotted curves show a global fit using the on-axis energy injection model (solid lines, Section~\ref{sec:uniform-jet-with-Einj}) and the misaligned structured jet model (dashed lines, Section~\ref{sec:misaligned-PLJ}), capturing the dominant achromatic afterglow evolution, with clear deviations emerging at the onset of the rebrightening phase, particularly in the redder bands. Middle: Evolution of optical colors $g-r$, $r-z$, and $g-z$. While $g-r$ remains approximately constant during the rebrightening phase, both $r-z$ and $g-z$ exhibit a systematic blueward evolution with slopes given by Eq.~\ref{eq:color-slopes}. At late times ($T-T_0 \gtrsim 18$~days), all colors redden, signaling the emergence of the supernova. Bottom: Residuals between the best-fit afterglow models and the observed fluxes in the $g$, $r$, and $z$ bands, for two models: circles indicate the energy injection model, and squares indicate the structured misaligned afterglow model. Shaded regions highlight the difference between model predictions. One can see that both models reproduce the bluer bands more accurately than the $z$ band. 
    }
    \label{fig:color_evolution}
\end{figure}

\subsection{Chromatic Evolution}
\label{sec:chromatic}

Figure~\ref{fig:color_evolution} presents the multi-band light curve from COLIBRÍ and the corresponding colour evolution of GRB\,260310A. The top panel shows the light curves in the $g, r, i, z,$ and $y$ bands. All bands exhibit a broadly similar temporal behaviour, characterized by an initial decay (Phase I), followed by a clear flattening and shallower decay which is the rebrightening phase (Phase II), and finally a late-time excess dominated by the emergence of a new component (Phase III). 

The colored curves show a global fit performed simultaneously to the $g$, $r$, and $z$ bands using the energy injection model described in Section~\ref{sec:afterglow-modeling}. This fit captures the dominant achromatic component of the light-curve evolution. However, deviations from the model are clearly visible, particularly at the onset of the rebrightening phase, where chromatic evolution becomes significant.

The panel below the light curves shows the evolution of optical colors: $g-r$, $r-z$, and $g-z$. A key feature is the differential behavior between colors. The $g-r$ color remains approximately constant throughout Phase II. In contrast, both $r-z$ and $g-z$ exhibit a systematic decrease with time, with measured slopes 
$d(r-z)/d\log_{10} T = -0.29 \pm 0.04,\; d(g-z)/d\log_{10} T = -0.27 \pm 0.03$. 
That is both the $r-z$ and $g-z$ colors become bluer with time suggesting that either there is an increasing flux in $g$ and $r$ or a decreasing flux in $z$.

The constancy of $g-r$ compared with the bluing in $r-z$ 
is suggestive of a spectral break 
located between the $r$ and $z$ band during Phase II and which evolves with time. In Section~\ref{sec:afterglow-modeling}, this behavior will be interpreted as the passage of the characteristic synchrotron frequency $\nu_m$ through the optical bands, from which we will be able to infer the strength of energy injection. 
However, this interpretation relies on the assumption that a single emission component dominates during the rebrightening phase. If multiple components contribute—for example, emission from distinct shocked regions—then the observed color evolution may instead reflect the changing relative weights and temporal evolution of these components, rather than the motion of a single spectral break.

The late-time evolution ($T-T_0 \gtrsim 18$~days) shows a clear reddening of the optical colors together with an increasing flux excess relative to the afterglow model, particularly in the redder bands. This behavior cannot be naturally explained by a single synchrotron component, and instead indicates the emergence of an additional emission component with a softer spectral energy distribution. Such a component is naturally interpreted as thermal emission from an expanding supernova photosphere, whose spectrum is approximately blackbody-like and evolves to lower temperatures with time. This leads to a progressive reddening of the observed colors and a deviation from the single power-law spectrum expected from synchrotron emission. The combination of (i) the redward color evolution, (ii) the wavelength-dependent flux excess, and (iii) the timing relative to the afterglow evolution provides strong evidence for the emergence of an underlying supernova associated with GRB~260310A, independently of its identification in our GTC spectra.

Finally, the bottom three panels show the residuals for each photometric band. The shaded regions indicate the range between the two model predictions, highlighting their differences across time. During Phase II, the residuals are small, indicating that the model accurately reproduces the early-time decay. In Phase III, systematic deviations begin to emerge and a clear excess develops, particularly in the redder bands. This behavior is consistent with the emergence of an additional emission component that is not captured by the afterglow models.

\subsection{Supernova component}
\label{sec:sn}

\begin{figure}
	\includegraphics[clip, width=1\linewidth]{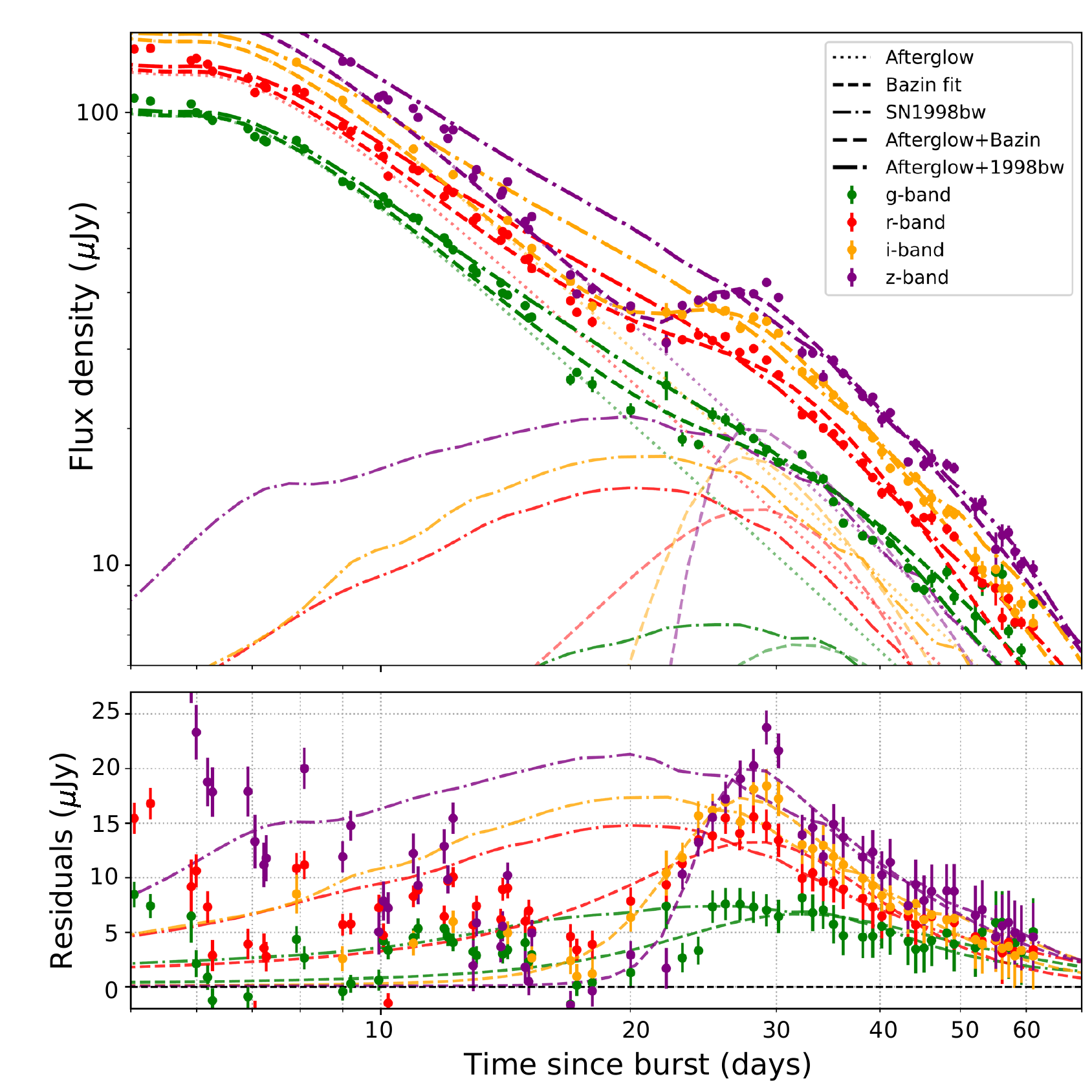}
    \caption{The late evolution of the light curve shows a prominent bump that could be associated with the emergence of an associated SN. The top panel shows the light curve evolution in the different bands and the bottom panel shows the light curve after the afterglow emission (modeled in Sect.~\ref{sec:afterglow-modeling}) has been subtracted. A phenomenological fit to these data is shown.}
\label{fig:sn-lc}
\end{figure}

The chromatic evolution discussed in Sect.~\ref{sec:chromatic} suggests that the early optical behaviour during the rebrightening phase is primarily driven by afterglow-related processes, while the late-time reddening indicates the emergence of a supernova component. However, the decomposition of the optical emission is not unique. In particular, the simultaneous appearance of a late X-ray rebrightening raises the possibility that part of the late optical excess may arise from an additional achromatic afterglow component, such as a refreshed shock, superposed on the emerging SN emission. In this section we therefore explore two alternative interpretations for the late optical evolution: first, a scenario in which the bump is attributed predominantly to the SN itself, and second, a model including both a late refreshed-shock contribution and an underlying GRB-SN.

The late evolution of the optical light curves is dominated by what we interpret as an emerging SN \citep{Guelfand2026}. The optical light curve, shown in Figs.~\ref{fig:color_evolution} and \ref{fig:sn-lc}, displays a rapid decay followed by a bump, which is most prominent in the reddest bands. By subtracting the on-axis uniform jet model with energy injection, described in section~\ref{sec:afterglow-modeling}, we are left with an initial excess starting from around 5 days after the burst followed by a more prominent bump. This later bump resembles a supernova light curve, but it has a later and sharper rise than is typically seen for the supernovae associated with GRBs, it also seems to peak at similar times for all bands. 

When we attempt to fit this late bump with SN1998bw templates \citep{Galama1998} using the k (amplitude) and s (temporal) stretching factors \citep{Zeh2004}. The parameters that deliver the best fits are: k = 0.25, 0.3, 0.4, 0.8 in the \textit{g}, \textit{r}, \textit{i}, \textit{z}, respectively and s = 1.6, 1.1, 1.0, 0.9 in \textit{g}, \textit{r}, \textit{i}, \textit{z}, respectively. These significantly different stretch parameters in the different bands are not expected for GRB-SN. Even if  we get an acceptable fit at late times, the supernova rise at around 20 days is not well reproduced and the 1998bw template before the bump is significantly above the observations.  In particular, our photometry shows, in all bands, a dip with respect to the SN1998bw templates between 15 and 25 that is not expected. The rise beyond 20 days is much later and sharper than can be described by using SN1998bw templates with simple stretching factors (see Fig.~\ref{fig:sn-lc}).  

A better fit of the late bump is obtained using a phenomenological Bazin function \citep{Bazin2009} applied to the data after two weeks after the burst. These fits are shown as dashed lines in Fig.~\ref{fig:sn-lc}. The different bands peak at a similar time, around 27 days after the burst, corresponding to 23.4 days after the burst in rest frame, which can be compared with the 14 to 20 days for SN 1998bw, where the bluer bands peak earlier than the redder bands. For 1998bw, the bluer light curves peak earlier than the red ones, which is different to what we see here. We also note that this fit does not allow for a detectable SN component at early times, in contrast to the detection of SN features starting 6 days after the burst and also hinted by the colour evolution already starting to appear at that time.

\begin{figure}
	\includegraphics[clip, width=1\linewidth]{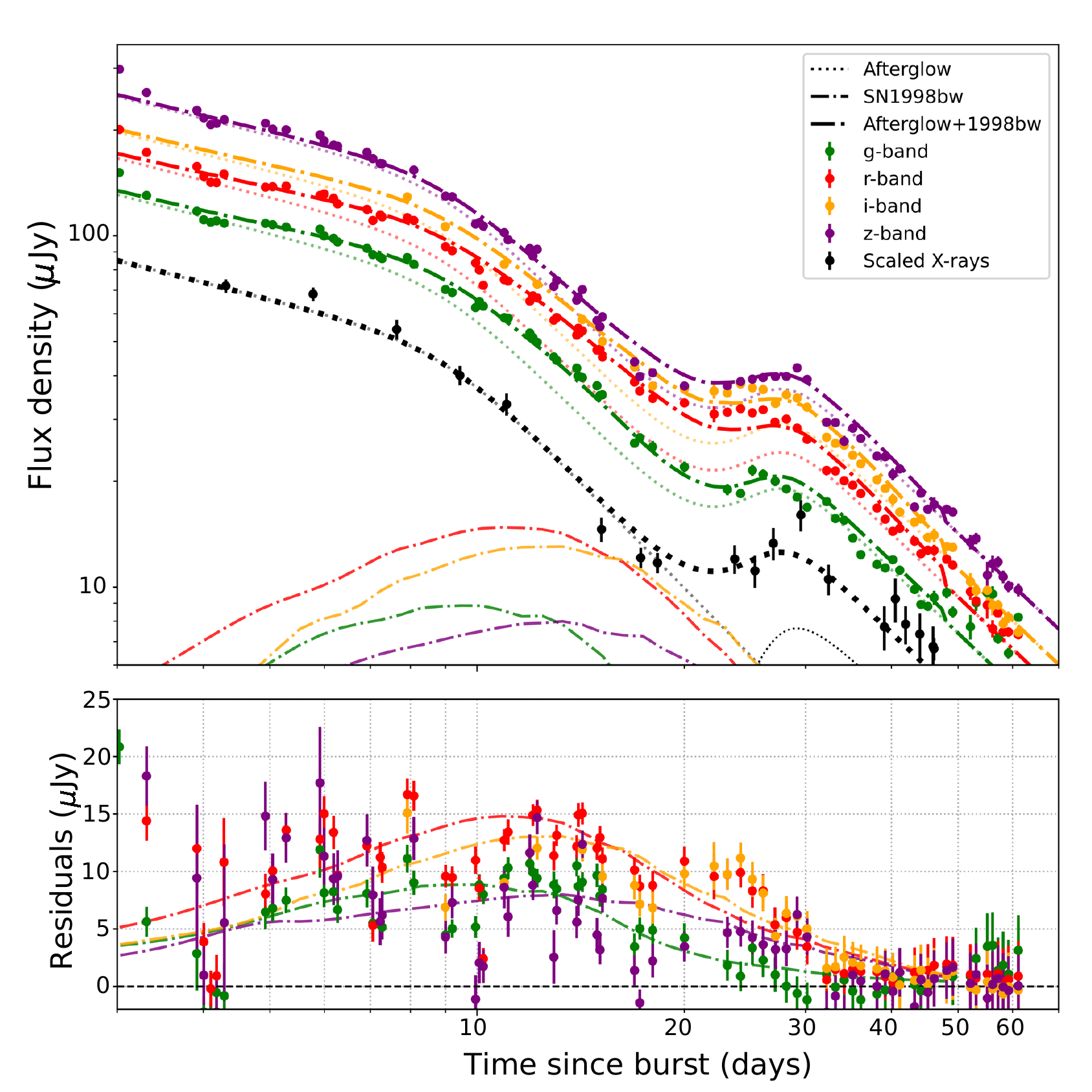}
    \caption{ Light curve evolution assuming a late refreshed shock plus supernova. The top panel shows the light curve evolution in the different bands and the bottom panel shows the light curve after the afterglow emission (including the late bump revealed in X-rays) has been subtracted. The residuals are well reproduced by faint and fast SN1998bw templates.}
\label{fig:sn-lc2}
\end{figure}

Another factor that should be considered while studying the late optical bump is that the X-ray emission shows a late rebrightening at a similar time as we see the late optical bump (see Fig.~\ref{fig:multi-waveband-LC-Spectra}). If this late X-ray component has a spectrum that extends to the optical bands, it could affect the overall shape of our light curve that we are measuring. To test this, we reproduce the X-ray bump by a broken powerlaw in addition to the broadband model, increasing the flux density achromatically in all bands in the same way as in the X-rays. This can be physically understood as a further late refreshed shock. 

We include this achromatic component to the afterglow model in all bands, and by subtracting it from the light curves, we get the residuals shown in Fig.~\ref{fig:sn-lc2}. These residuals are now well reproduced with SN1998bw templates. In this case we get similar stretching factors in all bands, with an amplitude factor k = 0.3 and a temporal stretching factor of s = 0.6, which are consistent with the relation that has been identified between these two parameters \citep{Cano2014}. Moreover, considering the excess light due to these SN templates, we are able to accurately reproduce the colour evolution shown in the second panel of Fig.~\ref{fig:color_evolution}.

\begin{figure}
\centering
	\includegraphics[clip, width=\linewidth]{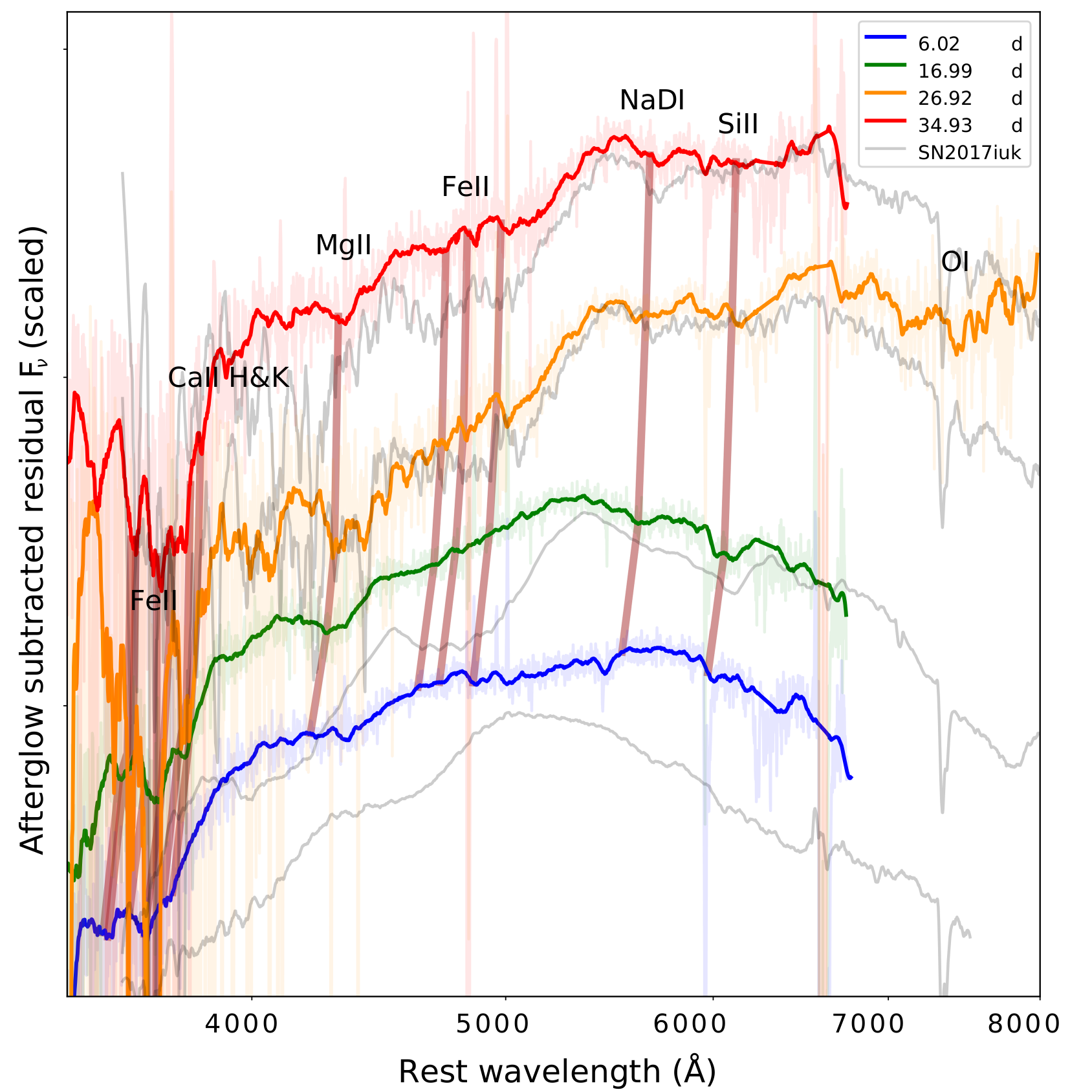}
    \caption{ Spectral evolution of the supernova, after the GTC/OSIRIS+ spectra had the afterglow component subtracted. Several absorption features are identified with varying velocities. For comparison, we display, in gray, the spectra of SN2017iuk associated to GRB\,171205A at similar epochs.}
\label{fig:sn-spectra}
\end{figure}

In a similar way as we did with the supernova light curve, we subtract the afterglow model from the GTC spectra to obtain cleaner supernova spectra. We note that the spectral shape that we obtain from the SN spectra compares better with existing GRB-SN templates when assuming the additional refreshed shock. The supernova spectral analysis does not include the first spectrum, since it was afterglow dominated. The residuals, dominated by supernova emission, are shown in Fig.~\ref{fig:sn-spectra}. In that figure, we can see that the spectra, starting on day 6, resemble those of a GRB-SN as was already seen and reported from the unsubtracted spectra \citep{deUgartePostigo2026}. In particular, we identify and indicate the location of several of the most prominent features usually seen in type Ic broad-line supernovae associated with long GRBs. For comparison, we also plot the spectra of SN\,2017iuk the type Ic broad lined supernova associated with GRB\,171205A \citep{Izzo2019}, at similar epochs. This was also a faint SN located on the outskirts of a nearby galaxy \citep{Thoene2024}. 

\begin{figure}
	\includegraphics[clip, width=1\linewidth]{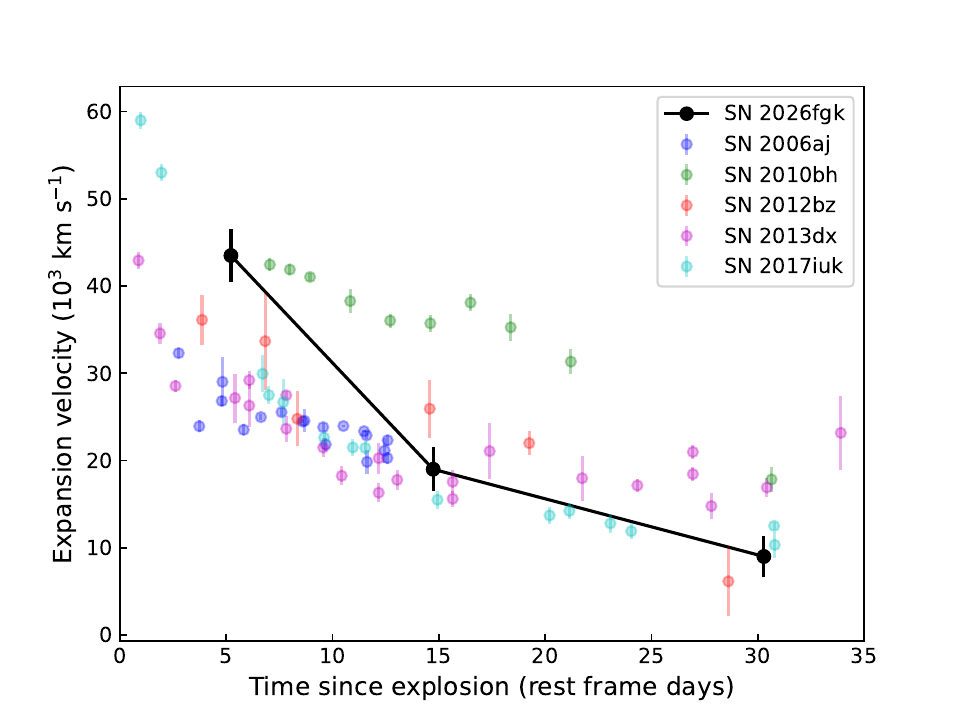}
    \caption{Expansion velocity evolution of the ejecta features seen in the supernova spectra of SN\,2026fgk (Fig.~\ref{fig:sn-spectra}) in comparison with a sample of GRB-SNe from \citet{finneran2025b}.}
\label{fig:sn-vel}
\end{figure}

To study the expansion velocity and its evolution, we followed the methodology described in detail in \cite{finneran2025b}. Fig.~\ref{fig:sn-vel} shows the expansion velocity evolution of SN\,2026fgk compared to a sample of GRB/SNe measured uniformly with the same method by \cite{finneran2025b}. The evolution resembles that of SN\,2017iuk and is overall consistent with the velocities seen in other broad-lined Ic SNe associated with GRBs. Similar estimates of the SN expansion velocity are obtained by \citet{OConnor2026}.

\subsection{Radio Image Size}
To obtain the best possible angular resolution, and hence the best possible constraints on source size, we reprocessed the second and third VLA epoch data, to image the highest frequency (44~GHz) data.  The data were imaged with Uniform, Natural, and Briggs weighting, to provide the best angular resolution, the best sensitivity and the optimal compromise between the two, respectively. The results were consistent within the three weighting schemes; here we report the results from the Briggs-weighted imaging; its higher signal-to-noise ratio more than compensates for the slightly larger synthesized beam size, compared to the uniformly-weighted data. Although the source is unresolved by the VLA observations, a 2D-Gaussian fit to the image suggests an upper limit to the source size of  $20 \times 12$ milli-arcsec.  Our peak flux densities of  $8.05 \pm 0.18$ mJy/b (1st epoch) and  $5.12 \pm 0.18$ mJy/b (2nd epoch) are consistent with the values reported by \citet{GCN44160} and \citet{GCN44235}. 

\section{Constraints on $\Gamma_0$ and Origin of Prompt Emission}
\label{sec:bulk-Gamma-constraints}
The prompt emission spectrum of the initial hard pulse is well described by a cutoff power law or Comptonized spectrum, with a photon index of $\alpha=-0.06\pm0.21$ and observed peak energy of $E_{\rm pk}\simeq198.5\pm16.8$\,keV. This yields an energy of $E_{\rm pk,z} = (1+z)E_{\rm pk}\sim230$\,keV in the cosmological rest frame of the GRB and an $E_{\rm iso}=(3.12\pm0.21)\times10^{50}$\,erg emitted over a duration of $T_{\rm pulse}\simeq50$\,s. The origin of this spectrum must lie in a different radiative process other than the non-thermal one that produces the typical Band-like prompt GRB emission.

A Comptonized spectrum can be produced in a non-dissipative outflow where the radiation advected with the optically thick flow is released at the photosphere, as in the fireball model \citep{Goodman-86,Paczynski-86,Piran+93,Kumar2015}. The bulk LF of the optically thick fireball grows linearly with radius, $\Gamma(R)=R/R_0$, and saturates to $\Gamma_s\equiv\Gamma(R_s)=\eta$ at the saturation radius $R_s=\eta R_0$, where $\eta = E_0/M_0c^2$ is the energy per baryon, $E_0$ is the total energy of the fireball, $M_0$ is the baryon load, and $R_0\sim10^7$\,cm is the typical launching radius. The temperature at the base of the flow is given by $T_0 = (L_0/4\pi\sigma_{\rm SB}R_0^2)^{1/4}$ modulo an order unity correction, where $L_0$ is the luminosity of the thermal radiation field and $\sigma_{\rm SB}$ is the Stefan-Boltzmann constant. As the radiation dominated fireball expands under its own pressure it is adiabatically cooled with its comoving\footnote{Quantities with a prime are expressed in the fluid-frame.} temperature declining linearly with radius, $T'(R)/T_0=R_0/R$. Its luminosity is given by $L=4\pi R^2c\Gamma^2\sigma_{\rm SB}T'^4=L_0$ and the peak energy in the engine-frame by $E_{\rm pk}(1+z) \simeq \Gamma E_{\rm pk}' \simeq 3k_BT'\Gamma=3k_BT_0=E_{\rm pk,0}$, where $E_{\rm pk,0}$ is the initial spectral peak energy and $k_B$ is the Boltzmann constant. The flow continues to expand until the radiation escapes at the photosphere that occurs when the Thomson optical depth, $\tau_T=n_e'\sigma_TR/2\Gamma$, to electron scattering with Thomson cross-section $\sigma_T$ reaches unity, which yields the photospheric radius $R_{\rm ph} = \sigma_TL_0/8\pi m_pc^3\eta\Gamma^2$. If the photosphere is reached before the saturation radius, $R_{\rm ph}<R_s$, then the spectral peak energy of the quasi-thermal photospheric emission is $E_{\rm pk}(R_{\rm ph})=E_{\rm pk,0}/(1+z)$. By equating $L$ with the mean isotropic-equivalent $\gamma$-ray luminosity of $L_{\rm iso}\simeq (1+z)E_{\rm iso}/T_{\rm pulse}\simeq7.2\times10^{48}\,[(1+z)/1.153]E_{\rm iso,50.5}T_{\rm pulse,1.7}^{-1}\,{\rm erg\,s^{-1}}$, we find
\begin{equation}
    E_{\rm pk} = 198\,\left(\frac{1+z}{1.153}\right)^{-3/4}E_{\rm iso,50.5}^{1/4}T_{\rm pulse,1.7}^{-1/4}R_{0,8.11}^{-1/2}\,{\rm keV}\,
\end{equation}
that matches the observed peak energy for a launching radius of $R_0\simeq1.2\times10^8$\,cm. Once the fireball becomes optically thin it stops accelerating. Therefore, the maximum LF is attained when $R_{\rm ph}=R_s$ that yields \citep{Meszaros-Rees-00}, 
\begin{equation}
    \Gamma_{\max} = \frac{R_{\rm ph}}{R_0}\simeq75\,\left(\frac{1+z}{1.153}\right)^{1/4}E_{\rm iso,50.5}^{1/4}T_{\rm pulse,1.7}^{-1/4}R_{0,8.11}^{-1/4}\,.
\end{equation}

If on the other hand, the fireball becomes optically thin well after it starts to coast, i.e. $R_{\rm ph}>R_s$, then the temperature at the photosphere is lower, $T_{\rm ph}=T_0(R_{\rm ph}/R_s)^{-2/3}$, due to further adiabatic cooling of the radiation field. This would produce a burst with luminosity $L_{\rm ph}=L_0(R_{\rm ph}/R_s)^{-2/3}$ and spectral peak energy $E_{\rm pk}=(1+z)^{-1}E_{\rm pk,0}(R_{\rm ph}/R_s)^{-2/3}$. From the mean observed $\gamma$-ray luminosity, we can now constrain the initial fireball luminosity, spectral peak energy, and photospheric radius, 
\begin{eqnarray}
    L_0 &\simeq& 3.9\times10^{49}\,\left(\frac{1+z}{1.153}\right)^3E_{\rm iso,50.5}^3T_{\rm pulse,1.7}^{-3}R_{0,7}^{-2}\eta_{2.06}^{-8}\,{\rm erg\,s^{-1}} \\
    E_{\rm pk} &\simeq& 200\,\left(\frac{1+z}{1.153}\right)^{-9/4}E_{\rm iso,50.5}^{-5/4}T_{\rm pulse,1.7}^{5/4}R_{0,7}\eta_{2.06}^6\,{\rm keV} \\
    R_{\rm ph} &\simeq& 1.5\times10^{10}\,\left(\frac{1+z}{1.153}\right)^3E_{\rm iso,50.5}^3T_{\rm pulse,1.7}^{-3}R_{0,7}^{-2}\eta_{2.06}^{-11}\,{\rm cm}\,,
\end{eqnarray}
and find an agreement with the observed peak energy if the emitting material was moving at ultrarelativistic speeds with $\Gamma_0=\eta=116$. If the engine operated for $\sim T_{\rm pulse}/(1+z)$, the total energy of the fireball must be 
\begin{eqnarray}
    E_0 &\sim& L_0T_{\rm pulse}/(1+z) \nonumber \\
    &\simeq& 1.7\times10^{51}\,\left(\frac{1+z}{1.153}\right)^2E_{\rm iso,50.5}^3T_{\rm pulse,1.7}^{-2}R_{0,7}^{-2}\eta_{2.06}^{-8}\,{\rm erg}\,,
\end{eqnarray}
which is sufficient to power the afterglow as shown below.

Due to emission arising from different radii corresponding to different polar angles away from the LOS but still within the $1/\Gamma$ beaming cone, the spectrum deviates from a true black body. \citet{Lundman+13} show that if the bulk LF has angular structure, with $\Gamma(\theta)\propto\theta^{-\kappa}$, then the photon index below the spectral peak of a quasi-thermal spectrum is $\alpha=-(1/4)(1+3/\kappa)$. Then to match the observed photon index of $\alpha=-0.06\pm0.21$ the jet has to be unphysically steep with $\kappa\gg1$ and still it would only reach its asymptotic photon index of $\alpha=-1/4$. However, if some dissipation occurs below the photosphere then the low-energy spectrum can be made softer \citep{Vurm-Beloborodov-16} from the hard thermal spectrum to match the observed one.

\section{Afterglow Modeling}
\label{sec:afterglow-modeling}
We present two different scenarios to explain the multi-waveband afterglow light curve of GRB\,260310A. In both cases we assume the standard afterglow theory \citep{Rees-Meszaros-92,Meszaros-Rees-93,Sari-Piran-95} that considers the deceleration of the ejecta, with isotropic-equivalent kinetic energy $E_{\rm k,iso}$ and initial bulk LF $\Gamma_0$, as it sweeps up the external medium in its path. We consider a general radial density profile for the external medium, $n_{\rm ext} = n_0(R/R_0)^{-k}=AR^{-k}$, with the normalization given at a fixed radius $R_0=10^{18}$\,cm. The interaction of the ejecta with the external medium produces two external shocks, where a forward shock propagates ahead of the ejecta and shock heats the swept up medium. A reverse shock propagates through the ejecta and decelerates it by extracting its kinetic energy, most of which is then delivered to the shocked swept up medium near the characteristic deceleration radius. The afterglow emission in this scenario is then attributed to the synchrotron radiation of shock-accelerated electrons \citep{Sari+98,Granot-Sari-02}. These electrons receive a fraction $\epsilon_e$ of the total internal energy density of the shocked material behind the shock, while a fraction $\epsilon_B$ of it goes into amplifying or generating in-situ microscopic-scale magnetic fields. In the thin-shell case \citep{Sari-Piran-95} the reverse shock is unable to decelerate the ejecta and remains mostly Newtonian, such that its contribution to the afterglow flux remains subdominant and the observed flux is obtained entirely from the forward shock. We assume this case in what follows and ignore any contribution of the reverse shock. 

\subsection{On-Axis Uniform Jet with Energy Injection}
\label{sec:uniform-jet-with-Einj}

We first consider a simple model of an on-axis uniform jet with an energy injection episode. The latter is needed to explain the shallow decay of the optical light curve post-peak and the plateau seen in both the optical and X-ray light curves around $T\sim5$\,days. When the observer's LOS is sufficiently far away from the jet's edge at $\theta=\theta_j$, so that $\theta_{\rm obs}+\Gamma^{-1}<\theta_j$ at all relevant timescales, then the observer has no knowledge of the collimated nature of the outflow and it can be treated as if it was a spherical flow. Besides, there is no strong indication of an achromatic jet break that occurs when the above condition is violated, which therefore supports this simple model of a spherical shell.

As shown in Fig.\,\ref{fig:multi-waveband-LC-Spectra}, the peak of the optical light curve is at $T-T_0\lesssim0.3$\,days. The origin of the peak can be explained by two different effects: (i) blast wave deceleration when $\nu_m<\nu_o<\nu_c$ or (ii) crossing of the injection frequency, $\nu_m$, across the optical band at $T>T_{\rm dec}$. Typically, the X-ray light curve serves as a good indicator to constrain the deceleration radius and time, since the X-rays are always above $\nu_m$, and therefore always show a decaying light curve during the self-similar blast wave evolution at $T>T_{\rm dec}$. In the present case there are no X-ray observations in our dataset at earlier times. Hence the confusion between the two different origins of the light curve peak. When assuming the first case, we know that the deceleration time of the blastwave is given by
\begin{eqnarray}
    T_{\rm dec} &=& (1+z)\left[\frac{(3-k)E_{\rm k,iso}}{2^{5-k}\pi A c^{5-k}\Gamma_0^{2(4-k)}}\right]^{1/(3-k)} \\
    &\simeq& 0.23\,E_{\rm k,iso,53}^{1/2}n_{0,-2}^{-1/2}\Gamma_{0,1.5}^{-3}\,{\rm days}\quad (k=1)\,,
\end{eqnarray}
where we have assumed $k=1$, as inferred from the closure relations in sec.\,\ref{sec:closure}, for which $A=n_0R_0$. It is atypical to have such a late deceleration time that requires the coasting bulk LF to be much smaller than the typical value of $\Gamma_0\sim10^2-10^3$. The above estimate of the deceleration time is supported by our afterglow modelling shown below.

\subsubsection{Energy Injection}
Some form of energy injection into the initial blast wave that refreshes the shock is required in this model to explain the shallow light curve decay post-peak and plateau in the optical and X-rays at $T\sim5$\,days. This can occur in two distinct ways \citep[e.g.][]{Angulo2026}: (i) a collision between two mass shells ejected by the central engine at two different times only separated by a small time interval \citep{Kumar-Piran-00,Sari-Meszaros-00,Zhang-Meszaros-02,Vlasis+11,Moreno-Mendez+15}, and (ii) a continuous and gradual energy injection by either due to the ejecta comprising a radial velocity stratification, with progressively slower moving inner shells trailing behind the faster outer shell \citep{Rees-Meszaros-98,Sari-Meszaros-00}, or where a rapidly spinning central engine e.g. a millisecond magnetar, continuously injects energy into the blast wave via a magneto-hydrodynamical (MHD) wind as it spins down \citep{Dai-Lu-98,Zhang-Meszaros-01,Zhang-Meszaros-02}. Here we model the energy injection using scenario (ii) that is also simpler to model in comparison to the first. This kind of energy injection has been used in many earlier works \citep[e.g.][]{Laskar+18,Schroeder+24,deWet+24,Schroeder+25,Angulo2026} to explain rebrightening features in afterglow light curves, with the assumption that the trailing ejecta in the radially stratified case only catches up with the faster moving blast wave at the time of the rebrightening.

We model this scenario by injecting energy over a narrow radial width $\Delta R=R_{\rm end}-R_{\rm inj}$, where injection commences at the injection radius $R_{\rm inj}$ and ceases at $R_{\rm end} = (1 + \Delta R/R_{\rm inj})R_{\rm inj}$. During this time a total amount of energy $E_{\rm inj}\equiv f_{\rm inj}E_{\rm k,iso}$, that is a fraction $f_{\rm inj}$ of the initial kinetic energy of the blast wave, is injected as a power-law in radius, with index $s_{\rm inj}$, at the rate \citep{Angulo2026}
\begin{equation}
    \frac{dE}{dR} = \frac{(1+s_{\rm inj})}{\Delta R} E_{\rm inj} \left(\frac{R-R_{\rm inj}}{\Delta R}\right)^{s_{\rm inj}}\propto R^{s_{\rm inj}}\,,
    \label{eq:energy_inj}
\end{equation}
which causes the energy of the blast wave to grow as $E(R)\propto R^{1+{s_{\rm inj}}}$. When the injected energy dominates over the initial energy of the blast wave, energy conservation in an adiabatic blast wave guarantees that $E(R)\sim\Gamma^2M_{\rm sw}c^2\propto\Gamma^2R^{3-k}$, where $M_{\rm sw}\propto n_{\rm ext}(R)R^3\propto R^{3-k}$ is the swept up mass. Since the blast wave energy increases as a power law over the radius $\Delta R$, the bulk LF also grows as a power law, with $\Gamma\propto R^{(s_{\rm inj}+k-2)/2}$. This scaling is the asymptotic limit and applies strictly when $f_{\rm inj}\gg1$, otherwise an intermediate regime is realized. Since $R\propto \Gamma^2T$, this yields the scaling for apparent time $T\propto R^{3-(s_{\rm inj}+k)}$ that further yields the temporal scalings 
\begin{equation}
\label{eq:Gamma_Energy_scaling}
\Gamma\propto T^{(s_{\rm inj}+k-2)/2[3-(s_{\rm inj}+k)]}\,\quad {\rm and} \quad E\propto T^{(1+s_{\rm inj})/[3-(s_{\rm inj}+k)]}\,.
\end{equation}
The temporal scalings for a constant energy adiabatic blast wave can be obtained when $s_{\rm inj}=-1$. 

To solve the dynamical evolution of the blast wave, including this form of energy injection, we use the numerical code of \citet{Gill-Granot-23} that yields equal-arrival-time-surface integrated afterglow emission in the observer frame. It makes the explicit assumption that the shock-microphysical parameters remain constant throughout the evolution of the blast wave. However, nothing guarantees that this assumption must be obeyed during the energy injection processes. Therefore, the obtained model solutions must be interpreted with the knowledge of this caveat.

\begin{figure}
    \centering
    \includegraphics[width=0.48\textwidth]{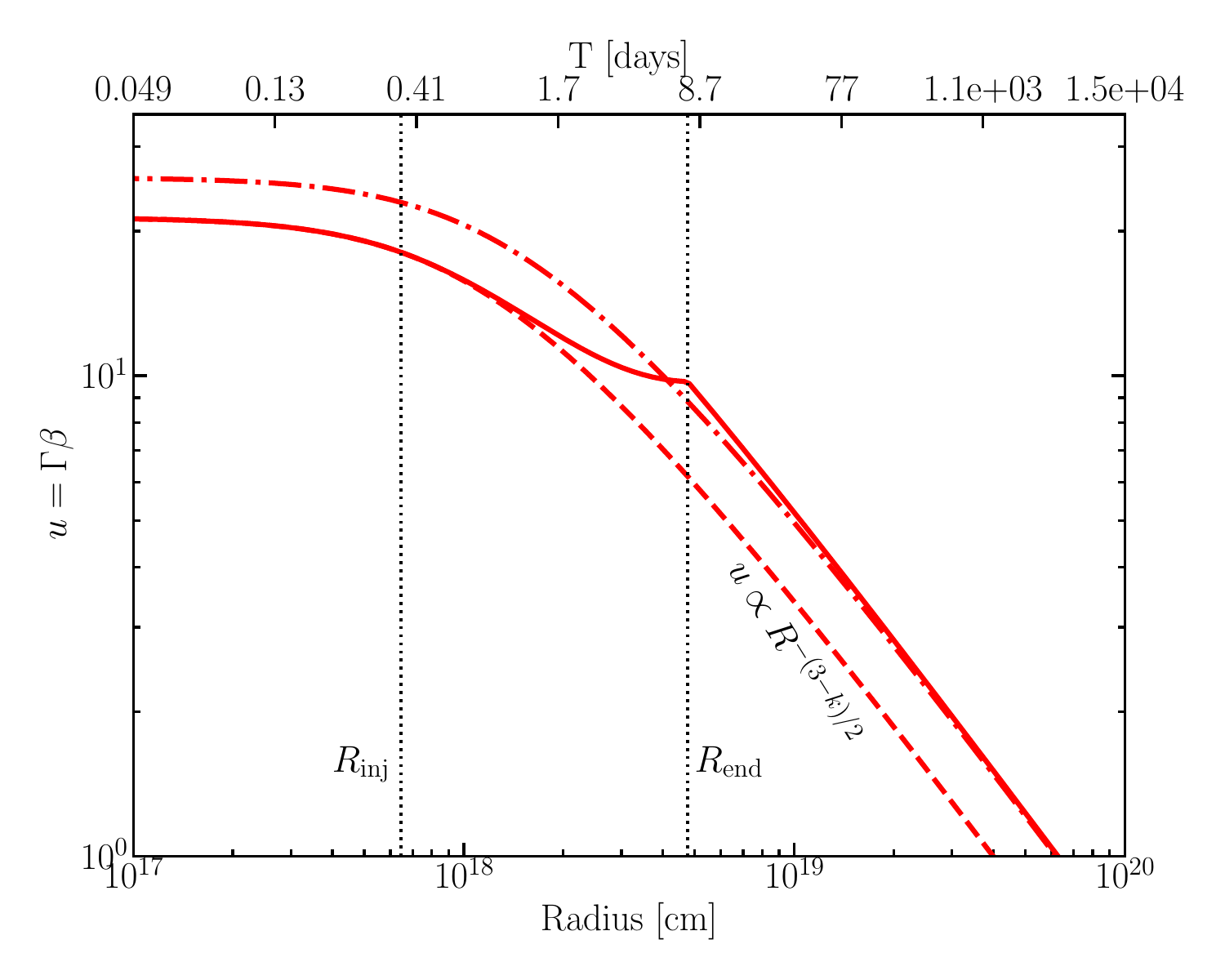}
    \caption{Radial evolution of the proper speed, $\Gamma\beta$, of the spherical shell, with energy injection commencing at $R=R_{\rm inj}$ and ending at $R=R_{\rm end}$. The shell has initial isotropic-equivalent kinetic energy $E_{\rm k,iso}$ and bulk LF $\Gamma_0$. The top x-axis shows the arrival time of photons emitted along the LOS from this shell. The dashed curve shows the proper speed of the shell in the absence of energy injection. The top x-axis only corresponds to this case for $R<R_{\rm inj}$. The dot-dashed curve shows the dynamical evolution of the shell that has energy $(1+f_{\rm inj})E_{\rm k,iso}$ and initial bulk LF $f_{\rm inj}\Gamma_0$. The top x-axis only corresponds to this case at $R>R_{\rm end}$.}
    \label{fig:Gamma-R-Sph-Einj}
\end{figure}

The dynamical evolution of a blast wave with initial isotropic-equivalent kinetic energy, $E_{\rm k,iso}$, and coasting LF, $\Gamma_0$, is shown in Fig.\,\ref{fig:Gamma-R-Sph-Einj}, both with (solid curve) and without (dashed curve) energy injection. The solid curve starts to deviate at $R>R_{\rm inj}$ as energy is gradually injected and resumes the same decay trend at $R>R_{\rm end}$ as that of the dashed curve. The dot-dashed curve shows the evolution of a blast wave with kinetic energy $(1+f_{\rm inj})E_{\rm k,iso}$ and coasting bulk LF $f_{\rm inj}\Gamma_0$. This last condition is obtained by demanding that the baryon mass in both shells, one in which energy is injected at a later radius and the other that starts with the larger energy initially, remains the same.

\begin{figure*}
    \centering
    \includegraphics[width=0.48\linewidth]{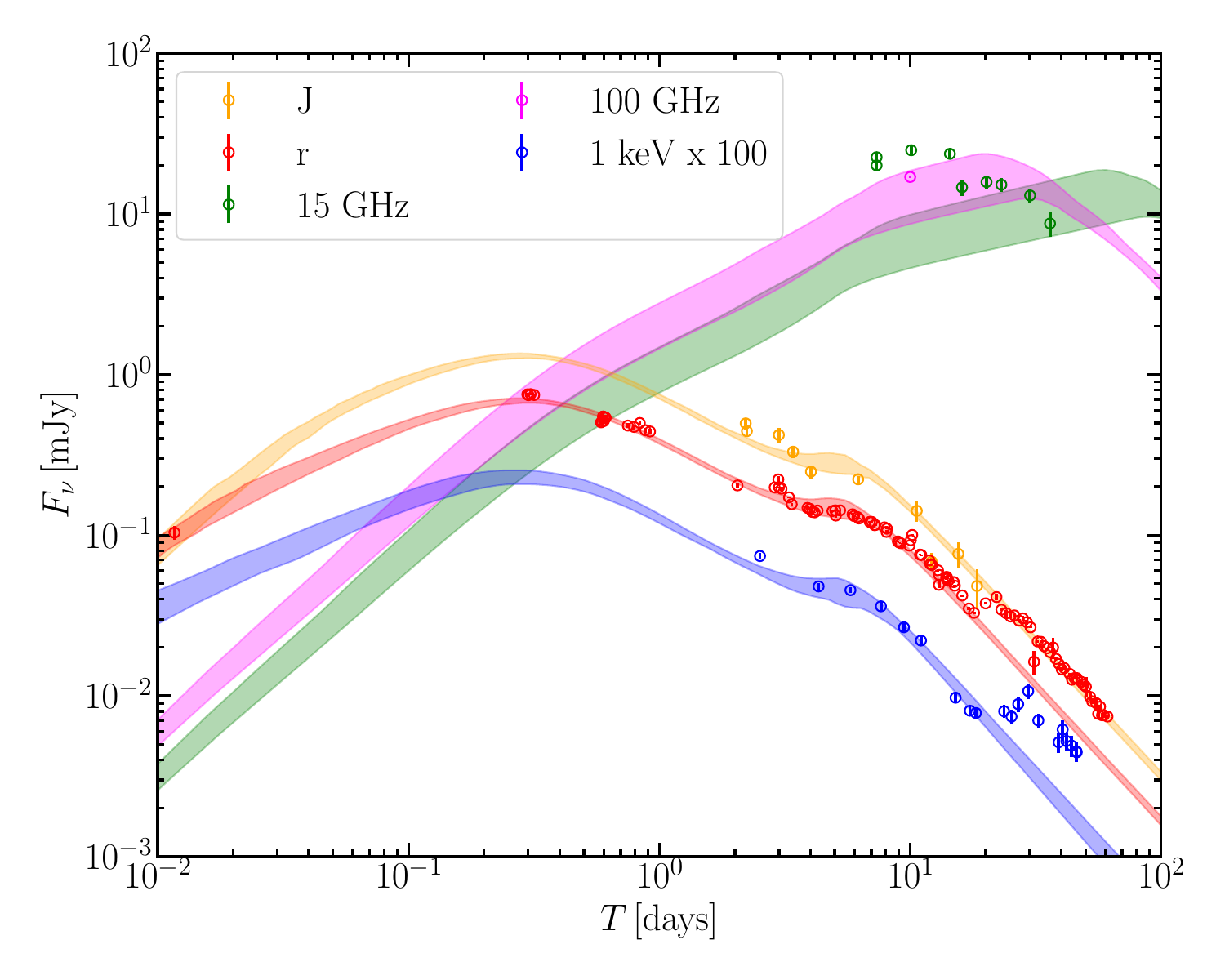}
    \includegraphics[width=0.48\linewidth]{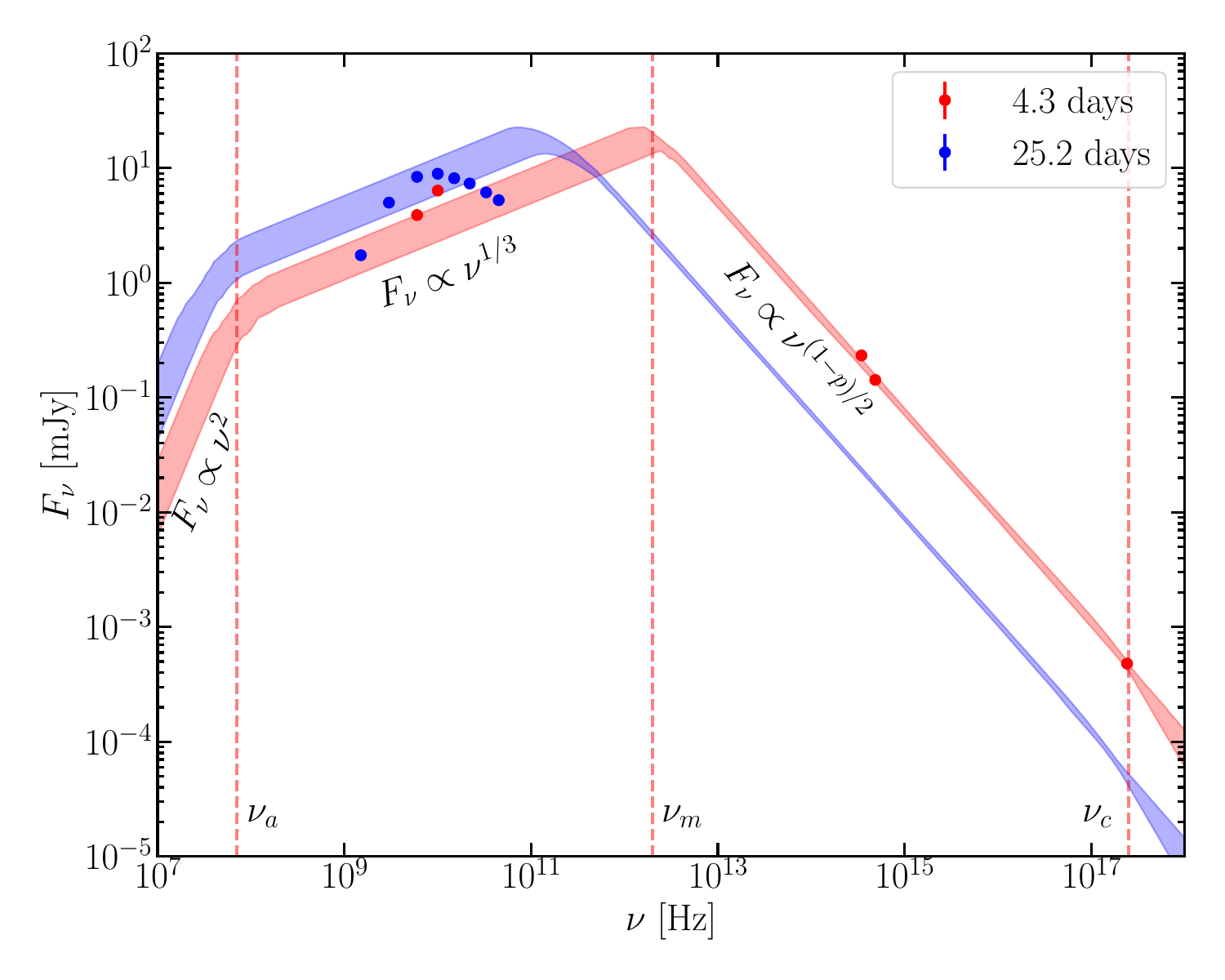}
    \caption{(\textbf{Left}) Afterglow model fit to the multi-waveband observations from an on-axis uniform jet with energy injection. Only the r-band and X-ray data were used for the fit. The J-band observations are shown for comparison with the light curve obtained from the model. The r-band light curve shows an excess at $T\gtrsim20$\,days that arises from a combination of emission from the SN and a refreshed shock. The best-fit model parameters are shown in Table\,\ref{tab:best-fit-params} and the parameter posterior distributions are presented in Fig.\,\ref{fig:fit-posteriors-Sph-Einj} (online supplementary material). 
    (\textbf{Right}) Comparison of model spectra with observations at two different epochs. The radio spectrum is not explained by the forward shock emission and must require an additional spectral component.
    }
    \label{fig:fit-spherical-shell}
\end{figure*}

\begin{table*}
    \centering
    \begin{tabular}{c|c|c|c|c|c|c|c|c|c|c|c}
    \multicolumn{11}{|c|}{On-Axis Uniform Jet with Energy Injection} \\
    \hline
       $E_{\rm k,iso}$ [erg] & $\Gamma_0$ & $n_0\,[{\rm cm^{-3}}]$ & $k$ & $p$ & $\epsilon_e$ & $\epsilon_B$ & $f_{\rm inj}$ & $s_{\rm inj}$ & $R_{\rm inj}$ [cm] & $\Delta R/R_{\rm inj}$\\
       \hline
       $8.5\times10^{52}$ & $21.4$ & $8.7\times10^{-3}$ & $1.16$ & $2.82$ & $0.23$ & $2.8\times10^{-3}$ & $1.21$ & $1.23$ & $6.4\times10^{17}$ & $6.37$ \\
       \hline \\
       \\
       \multicolumn{12}{|c|}{Misaligned Power-Law Jet} \\
       \hline
       $E_{\rm tot}$ [erg] & $E_{\rm k,iso,c}$ [erg] & $\Gamma_{0,c}$ & $n_0\,[{\rm cm^{-3}}]$ & $k$ & $p$ & $\epsilon_e$ & $\epsilon_B$ & $\theta_c$ [deg] & $\theta_{\rm obs}$ [deg] & $a$ & $b$ \\
       \hline
       $9.5\times10^{50}$ & $6.9\times10^{53}$ & 363 & $0.056$ & $0.15$ & $2.4$ & $0.22$ & $1.5\times10^{-3}$ & $1.54$ & $6.9$ & $2.42$ & $1.54$ \\
       \hline
    \end{tabular}
    \caption{(\textbf{Top}) Best-fit model parameters for the on-axis uniform jet model with energy injection. 
    (\textbf{Bottom}) Best-fit model parameters for the misaligned power-law jet model. $E_{\rm tot}$ is the total jet kinetic energy after integrating over the jet angular profile.}
    \label{tab:best-fit-params}
\end{table*}

\subsubsection{MCMC Fit to the Afterglow}
We fit the optical and X-ray observations at $10^{-2}\leq T-T_0\,(\rm days)\leq 20$ with our afterglow model using Monte Carlo Markov Chains, and derive model parameter constraints with maximum likelihood estimation using the public package \texttt{emcee} \citep{emcee}. The posterior distributions for the model parameters are shown in Fig.\,\ref{fig:fit-posteriors-Sph-Einj} (online supplementary material). The radio data is excluded from the fit since (a) it is only available during and after the energy injection phase and (b) the radio emission can be very sensitive to the exact energy injection physics, and therefore will not be easily reproduced in a simple model used here and ultimately bias the fit. Instead, we compare the radio emission produced by the forward shock in our model with observations to check for consistency. 

The left panel of Fig.\,\ref{fig:fit-spherical-shell} shows our model fit to the multi-waveband light curves using the 1$\sigma$ shaded region sampled from the parameter posterior distributions. The best-fit parameters for our model are presented in Table\,\ref{tab:best-fit-params}. The optical and X-ray light curves are described well by this model, where it is able to reproduce the brief plateau seen in the two optical bands as well as in X-rays at $T\sim5$\,days. Some color evolution is evident beyond the sharp break in the optical light curve at $T\sim7$\,days, after which time the r-band flux deviates mildly from the model prediction but the J-band flux and that in X-rays is in fact consistent with the model. At $T\gtrsim20$\,days the r-band light curve shows an excess. This is consistent with emission from the associated Ic-BL SN (see sec.\,\ref{sec:sn} for more details). The right panel of Fig.\,\ref{fig:fit-spherical-shell} shows the spectrum at two different epochs, one during the energy injection phase and the second after energy injection ceases. At both epochs the X-ray band is very close to the broad cooling break, and therefore both the optical and X-ray light curves maintain similar shapes.  

This model is unable to account for the radio emission and under-produces it just when energy injection ceases. The radio spectrum at $T=25.2$\,days is sharply peaked, likely due to $\nu_a$ and $\nu_m$ being close to each other, as also obtained in the smoothly broken power-law fits in Fig.\,\ref{fig:multi-waveband-LC-Spectra}. In comparison, the model spectrum shows the two characteristic frequencies to be widely separated, producing instead a power-law spectral component with $F_\nu\propto\nu^{1/3}$. These discrepancies argue for the radio emission to be coming from another emission component. In the recent work of \citet{Christy+2026} the radio emission is explained by a reverse shock component, where they also find that the radio emission from the forward shock does not match observations. Alternatively, it is conceivable that the shock-microphysical parameters evolve during energy injection that may lead to a better match between the model and observations. However, since the optical and X-rays are well reproduced with a fixed set of shock-microphysical parameters over the entire afterglow evolution, this possibility appears to be less favourable. 

The MCMC fit finds a standard isotropic-equivalent kinetic energy of $E_{\rm k,iso}=8.5\times10^{52}$\,erg, but a much smaller than typical coasting bulk LF of $\Gamma_0\simeq21$. The external medium density has a radial profile with $k\simeq1$, which is neither an ISM nor a wind, and its normalization, $n_0 \simeq 8.7\times10^{-3}\,{\rm cm^{-3}}$ at $R_0=10^{18}$\,cm, agrees well with the location of the GRB at the outskirts of its host galaxy. While the energy of the blast wave and the external medium density are more standard, the initial bulk LF is driven to low values by the unusual late peak of the optical light curve found by the model fit. Similarly low coasting LFs have been inferred in other GRBs, e.g. AT2019pim \citep{Perley+25}, AT2021lfa \citep{Li+25}, where most such cases only show \textit{orphan} afterglows, i.e. without any prompt $\gamma$-ray counterpart, and are generally interpreted as having \textit{dirty} fireballs \citep{Dermer+99,Huang+02,Rhoads-03}. In a dirty fireball, the coasting LF becomes much lower than that obtained in a \textit{clean} fireball due to entrainment of excessive baryons while exiting from the stellar envelope. In the present case, the baryon load of the fireball is $M_0=E_{\rm k,iso}/\Gamma_0c^2=2.2\times10^{-3}M_\odot$ as compared to the typical value of $M_0\simeq5.6\times10^{-5}E_{\rm k,iso,52}\Gamma_{0,2}^{-1}\,M_\odot$ when assuming standard GRB parameters. 

Energy injection starts rather quickly, with $R_{\rm inj}\simeq R_{\rm dec}$, as soon as the outer mass shell starts to decelerate. The energy of the blast wave grows almost quadratically with radius over $\Delta R/R_{\rm inj}\simeq6.4$, with an increase in its total energy by a factor of $(1+f_{\rm inj})=2.2$. When the ejecta has a radial velocity gradient so that the mass above a given bulk LF scales as $M(>\Gamma)\propto \Gamma^{-s}$, the energy of the blast wave grows as $E(>\Gamma)\propto\Gamma^{1-s}\propto T^{-(3-k)(1-s)/(7+s-2k)}$. By comparing this scaling with that derived above in the more general case of energy injection adopted in this work, we find that the two are compatible when $s = (4-k+s_{\rm inj})/(2-k-s_{\rm inj})<0$. Therefore, the energy injection solution obtained here is incompatible with a radially stratified ejecta since the latter assumes $s>0$.

Our model assumes a spherical shell, but in reality the flow must be collimated into a jet half-opening angle of $\theta_j$. This would produce an achromatic steepening in the light curve, referred to as a \textit{jet break}, when the beaming cone of emission from the material at the edge of the jet includes the observer's LOS as the blast wave decelerates. Although it is explained naturally in our spherical shell model, the achromatic steepening at $T\sim6$\,days could be indicative of a jet break. When assuming a top-hat jet, the condition for a jet break is expressed more generally as $\Gamma(T)(\theta_j-\theta_{\rm obs})=1$. Taking $\theta_{\rm obs}=0$ yields a jet break time of $T_j = (\Gamma_0\theta_j)^{2(4-k)/(3-k)} T_{\rm dec} \simeq 7.6[(1+z)/1.153]\,E_{\rm k,iso,52.93}^{0.53}n_{0,-2.06}^{-0.53}\theta_{j,-1}^{3.07}$\,days for a fiducial $\theta_j=0.1$\,rad. For this typical fiducial jet opening angle, the true jet kinetic energy of such an outflow would be 
\begin{equation}
    E_k\approx\frac{1}{2}\theta_j^2E_{\rm k,iso}\simeq4.2\times10^{50}\theta_{j,-1}^2E_{\rm k,iso,52.93}\,{\rm erg}\,,
\end{equation}
which is consistent with the estimate of the total energy release obtained in Sec.\,\ref{sec:bulk-Gamma-constraints}. In the absence of lateral expansion, the post jet-break light curve is expected to be steeper with a change in the temporal index of $\Delta\alpha=(3-k)/(4-k)\simeq2/3$ for $k=1$. Since the X-rays should be unaffected by any contribution from the SN, we take the pre-jet-break temporal index from the X-ray light curve as $\alpha_{\rm pre}\sim-0.18$. Then the expected post-jet-break temporal index is $\alpha_{\rm post}=\alpha_{\rm pre}-\Delta\alpha\simeq-0.85$ that does not match the observed index of $\alpha_{\rm post}\simeq-1.87$. Alternatively, if the jet expands sideways then the expected post jet-break index is $\alpha_{\rm post}=-p\simeq-2.8$ that again does not match the observed index. Therefore, the achromatic steepening at $T\sim6$\,days is incompatible with a jet break.

\subsubsection{Chromatic Rebrightening from Energy Injection}
We now investigate whether the observed chromatic evolution during the rebrightening phase can be explained within the energy injection scenario described above. This emission is characterized by modified blast-wave dynamics during the energy injection phase, which may in principle also be accompanied by changes in the shock microphysical parameters, as well as additional emission components from different shocked regions, {\it e.g.}, a reverse shock. Here, however, we consider a simplified first-order phenomenological correction to the underlying model, in which the emission comes from the external forward shock, by considering an additional emission component that may contribute to the observed flux during the rebrightening phase. 
This will be used to demonstrate that the observed color evolution can naturally arise from energy injection through the passage of a synchrotron spectral break across the optical bands in the additional emission component, which will then modulate the colors in the total flux. During this phase, the forward-shock dynamics are modified by the continuous increase in blast-wave energy, leading to a different temporal evolution of the characteristic synchrotron frequency in the additional component, $\tilde\nu_m$, compared to the standard constant-energy solution.

As can be seen in the middle panel of Fig.~\ref{fig:color_evolution}, a key feature is the differential behavior between colors: $g-r$ remains approximately constant throughout Phase II while both $r-z$ and $g-z$ show a systematic decrease with time, with measured slopes. 
\begin{equation}
\frac{d(r-z)}{d\log_{10} T} = -0.29 \pm 0.04,
\qquad
\frac{d(g-z)}{d\log_{10} T} = -0.27 \pm 0.03.
\label{eq:color-slopes}
\end{equation}
The observed color evolution during the energy injection phase can be explained if the characteristic synchrotron frequency of the minimum-energy electrons lies between the optical bands, specifically $\nu_z < \tilde\nu_m < \nu_r$. In the $z$-band, the observed photons lie below the characteristic synchrotron frequency $\tilde\nu_m$, where the spectrum approaches $F_\nu \propto \nu^{1/3}$,  and is largely insensitive to the electron index $p$. In contrast, the $r$ and $g$ bands 
are affected by the spectrum above $\tilde\nu_m$, where it follows $F_\nu \propto \nu^{-(p-1)/2}$ and directly reflects the shock acceleration physics.


Using the scalings derived in Eq.~\ref{eq:Gamma_Energy_scaling}, we can derive the temporal evolution of the characteristic synchrotron frequency during the energy injection phase. The minimum Lorentz factor of the accelerated electrons scales as $\gamma_m\propto\Gamma$, while the post-shock magnetic field scales as $B\propto\Gamma n^{1/2}\propto\Gamma R^{-k/2}$. The characteristic synchrotron frequency therefore evolves as
\begin{equation}
\tilde\nu_m \propto \Gamma \gamma_m^2 B
\propto \Gamma^4 R^{-k/2},
\end{equation}
where this expression describes only the temporal scaling of $\tilde\nu_m$; its absolute normalization additionally depends on the blast-wave energy and the shock microphysical parameters.  
Using the radial dependence of $\Gamma$ and the scaling of radius with apparent time derived above Eq.~\ref{eq:Gamma_Energy_scaling}, we obtain $\tilde\nu_m \propto R^{2s_{\rm inj}+\frac{3}{2}k-4}$ and $R \propto T^{1/[3-(s_{\rm inj}+k)]}$ that yields
\begin{equation}
\frac{d\ln \tilde\nu_m}{d\ln T}
=
\frac{2s_{\rm inj}+\frac{3}{2}k-4}{3-(s_{\rm inj}+k)}
\label{eq:num_evol}
\end{equation}
and which gives the standard result without energy injection when $s_{\rm inj}=-1$.

In this model, the dominant contribution to the color evolution comes from the motion of $\tilde\nu_m$ relative to the observing bands. This gives the approximation
\begin{equation}
B\equiv
\frac{d(\mathrm{color})}{d\log_{10} T}
\approx
-2.5\,\ln(10)\,
(\beta_{\rm high} - \beta_{\rm low})
\frac{d\ln \tilde\nu_m}{d\ln T}.
\end{equation}

When $\tilde\nu_m$ lies between two observing bands, the emission in each band receives contributions from electrons both below and above the characteristic frequency. However, the relative weighting of these contributions differs between bands, leading to an effective color evolution that reflects the motion of $\tilde\nu_m$ across the observing window. In the standard synchrotron spectrum, the flux density scales as $F_\nu \propto \nu^{1/3}$ below $\tilde\nu_m$, corresponding to a spectral index $\beta_{\rm low} = 1/3$, and as $F_\nu \propto \nu^{-(p-1)/2}$ above $\tilde\nu_m$, corresponding to $\beta_{\rm high} = -(p-1)/2$. Therefore, the difference in spectral slopes between the two bands is $\beta_{\rm high} - \beta_{\rm low} \approx -(p-1)/2 - 1/3$, 
where this expression represents the difference between the asymptotic spectral slopes, and therefore provides an approximate description of the color evolution when the observing bands straddle $\tilde\nu_m$.

The measured slopes are given by Eq.~\ref{eq:color-slopes}, yielding an average value $B = -0.28 \pm 0.02$. Using $k=1.1$ from our broadband modeling (Table~\ref{tab:best-fit-params}) and also making the simplifying assumption by adopting the same electron power-law index as found for the forward shock emission, {\it i.e.} $p=2.77$, we find 
\begin{equation}
\frac{d\ln \tilde\nu_m}{d\ln T}
=
\frac{B}{-7.01}
=
-0.040 \pm 0.003.
\end{equation}
which gives
$s_{\rm inj} \approx 1.16$.
This value is consistent with the result $s_{\rm inj}=1.23$ obtained from the full MCMC modeling. The apparent agreement therefore suggests that the level of energy injection inferred from the broadband light-curve modeling is of the same order as that required to reproduce the observed color evolution, under the assumption that it is driven by the passage of a synchrotron spectral break through the optical bands. We emphasize, however, that the MCMC fit was not constructed to reproduce the detailed chromatic evolution during the rebrightening phase, and that, in principle, the two approaches need not trace the same effective $s_{\rm inj}$ if different emission components contribute to the light curve and color evolution.

\begin{figure*}
    \centering
    \includegraphics[width=0.48\linewidth]{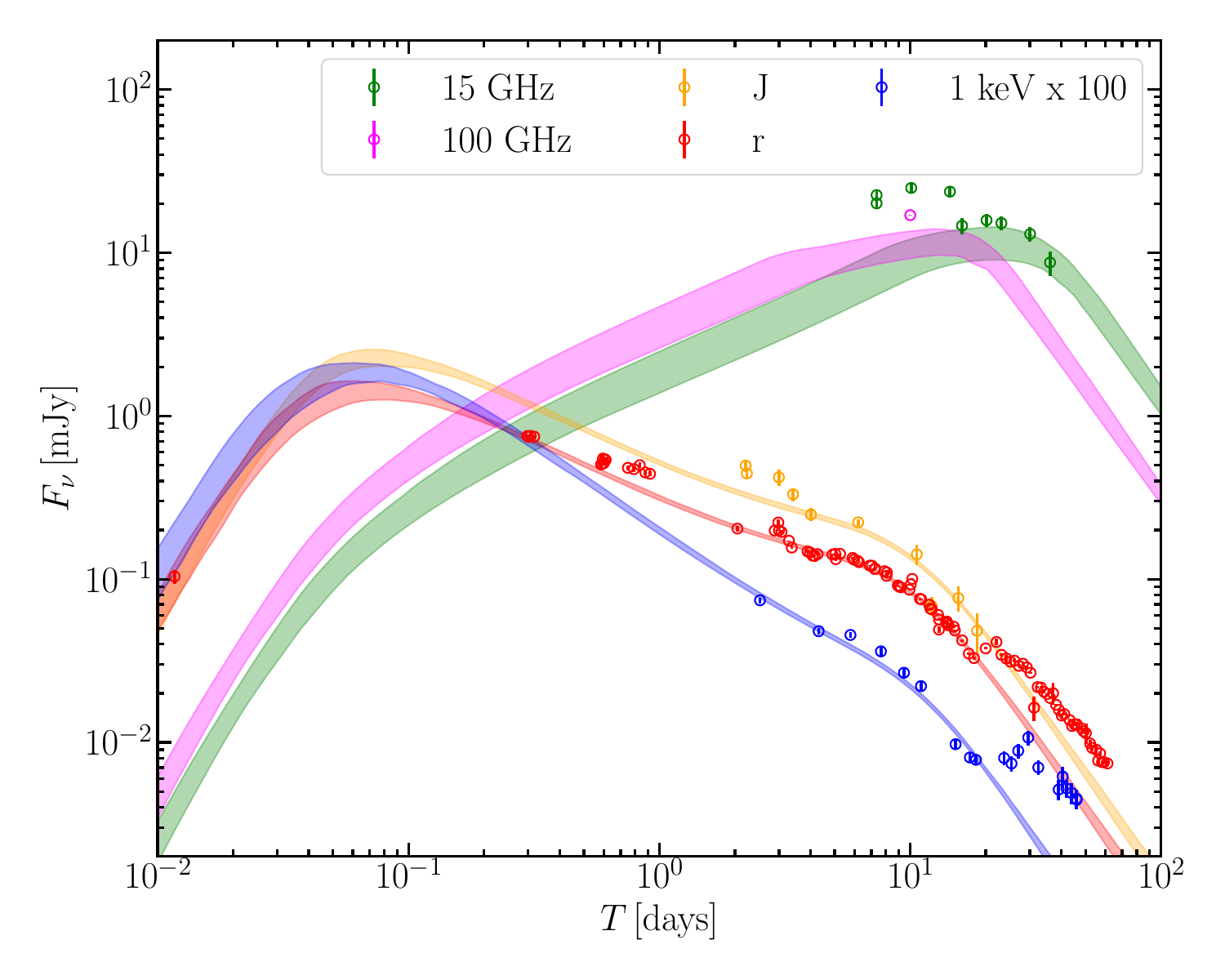}
    \includegraphics[width=0.48\linewidth]{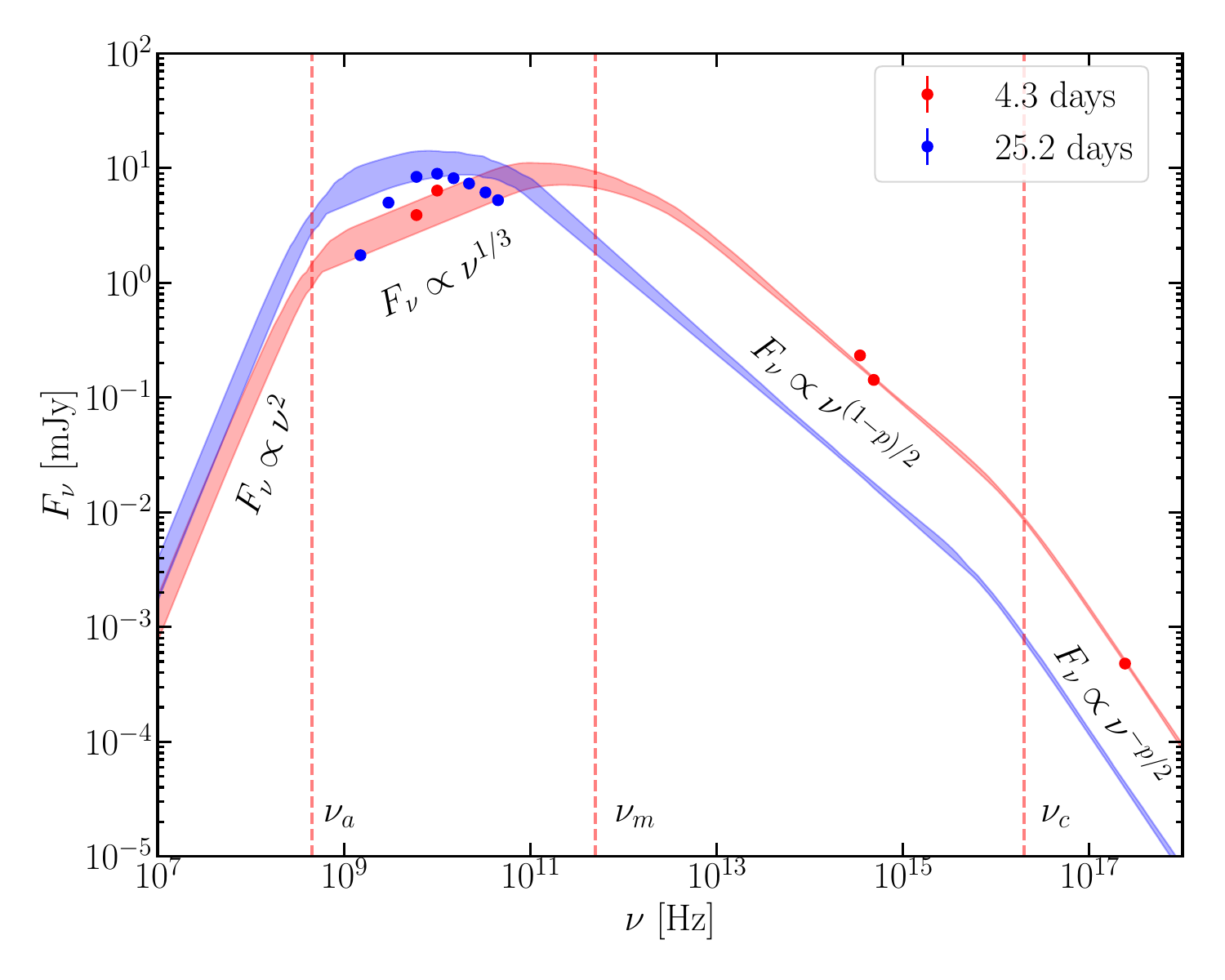}
    \caption{(\textbf{Left}) Afterglow model fit to the multi-waveband observations from a misaligned power-law angular structured jet. Only the r-band and X-ray data were used for the fit. The J-band observations are shown for comparison with the light curve obtained from the model. The best-fit model parameters are shown in Table\,\ref{tab:best-fit-params} and the parameter posterior distributions are presented in Fig.\,\ref{fig:fit-posteriors-PLJ} (online supplementary material). Shaded regions show the $1\sigma$ spread in the light curves randomly sampled from the parameter posterior distributions. 
    (\textbf{Right}) Comparison of model spectra with observation at two different epochs.
    }
    \label{fig:fit-misaligned-jet}
\end{figure*}

\subsection{Misaligned Jet with Angular Structure}
\label{sec:misaligned-PLJ}
Here we consider the model of a misaligned jet with angular structure \citep[e.g.][]{Kumar-Granot-03,Gill-Granot-18b,Beniamini+20, Beniamini+22}. Both its isotropic-equivalent kinetic energy, $E_{\rm k,iso}(\theta)$, and initial bulk LF, $\Gamma_0(\theta)$, decline as a power law with polar angle $\theta$ measured from the jet symmetry axis,
\begin{eqnarray}
    E_{\rm k,iso}(\theta)&=&4\pi\frac{dE_k(\theta)}{d\Omega}=E_{\rm k,iso,c}\Theta^{-a} \\
    \Gamma_0(\theta) &=& 1 + (\Gamma_c-1)\Theta^{-b} \\
    \Theta(\theta) &=& \sqrt{1 + \left(\frac{\theta}{\theta_c}\right)^2}\,,
\end{eqnarray}
where $dE_k/d\Omega$ is the kinetic energy per unit solid angle, and $E_{\rm k,iso,c}$ and $\Gamma_c$ are the normalizations obtained at the jet symmetry axis ($\theta=0$). Both profiles share the same core angle, $\theta_c$, but have different decay power-law indices. 

The evolution of the jet follows locally spherical dynamics \cite[see, e.g.,][]{Gill-Granot-18b} where material at every angle propagates as if it is part of a spherical flow with its local $E_{\rm k,iso}(\theta)$ and $\Gamma_0(\theta)$. Our model does not include the effect of lateral spreading that is expected to influence the emission substantially only when the flow becomes trans-Newtonian \citep[e.g.][]{Granot-Piran-12,Govreen-Segal-Nakar-24}. The flux density at the observer is obtained by performing an integration over the Equal Arrival Time Surfaces (EATS) \citep[e.g.][]{Granot+99,Gill-Granot-18b}. The optical and X-ray afterglow light curves show a brief plateau around $T-T_0\sim5$\,days followed by a break at $T-T_0\sim10$\,days. To account for that, our structured jet model would need to be supplemented with a similar energy injection process as implemented earlier. This will likely add 4 additional parameters on top of the 11 parameters used in the model without energy injection, which may lead to the model being under-constrained and not yield robust results. Therefore, we will dispense with energy injection and obtain a more robust fit below.

\subsubsection{MCMC Fit to the Afterglow}
We use the above model to obtain in Fig.\,\ref{fig:fit-posteriors-PLJ} (online supplementary material) the parameter posterior distributions from our MCMC fit to the multi-waveband afterglow observations. The light curve fit is shown in the left panel of Fig.\,\ref{fig:fit-misaligned-jet} and the best-fit model parameters are given in Table\,\ref{tab:best-fit-params}. We find that a misaligned jet with power-law angular structure does a reasonably good job at fitting the observations at $T-T_0\gtrsim1$\,day. The model lacks the ability to account for the plateau, but it is able to describe the broad break in both the optical and X-ray light curves. Similar to our earlier model, this model too fails to reproduce the radio emission and again points to the existence of an additional emission component, likely arising from the reverse shock \citep{Christy+2026} during the energy injection process. The optical and X-ray model light curves show different temporal trends, indicative of the emission in the two bands coming from different power-law segments of the synchrotron spectrum. Indeed, this can be inferred from the model spectra shown at two different epochs in the right panel of Fig.\,\ref{fig:fit-misaligned-jet}, which confirms that the optical band remains below the cooling break frequency and the X-rays above it. 

The advantage of a misaligned jet over one viewed on-axis is that it offers a natural explanation for a delayed light curve peak as found in the earlier model, without requiring the bulk LF for the entire flow to be significantly different from the standard model. In comparison to our on-axis uniform jet model, the optical light curve peak is not so delayed and occurs much earlier at $T-T_0\sim0.06$\,days. We find a misaligned jet with core angle $\theta_c\simeq1.54^\circ$ and our LOS outside of the core at $\theta_{\rm obs}\simeq4.5\theta_c\simeq6.9^\circ$. The jet has a marginally steep energy profile with $a\simeq2.4$ that yields a total kinetic energy, after integrating over the angular structure, of
\begin{equation}
    E_k = \int \frac{dE_k}{d\Omega}d\Omega = 9.5\times10^{50}\,{\rm erg}\,,
\end{equation}
which is consistent with the estimate of the total energy release obtained in Sec.\,\ref{sec:bulk-Gamma-constraints}. The isotropic-equivalent kinetic energy along the LOS is $E_{\rm k,iso}(\theta_{\rm obs})\simeq1.7\times10^{52}$\,erg. When comparing this energy to that of $\gamma$-rays during the prompt emission, where $E_{\rm iso}\simeq3.6\times10^{50}$\,erg, we find a $\gamma$-ray efficiency of $\eta_\gamma\equiv E_{\rm iso}/(E_{\rm iso} + E_{\rm k,iso})\simeq2\%$. This level of efficiency is too small when compared to the $\eta_\gamma\sim20\%$ inferred from a sample of GRBs \citep{Beniamini+16}. The angular profile of the initial bulk LF is not so steep with $b\simeq1.54$, and with a core value of $\Gamma_c=363$ the material along our LOS had a bulk LF of $\Gamma(\theta_{\rm obs})\simeq35.4$, which broadly agrees with that obtained for the uniform jet model. The bulk LF angular profile is inconsistent with what is typically obtained in numerical simulations of hydrodynamic jets in collapsars \citep[e.g.][]{Gottlieb+21} that find $3.5\lesssim b \lesssim 5$ \citep{Beniamini+22}. This model finds an external medium density profile much closer to a uniform ISM, which is in contrast with a $k\sim1$ profile found in the uniform jet model. The low density, however, is consistent with the GRB located at the outskirts of its host galaxy.

\subsubsection{Jet Angular Structure \& light curve Shape}
Unlike in a spherical flow in which the observer receives flux from a region of angular size $1/\Gamma$ around the LOS that grows as the blast wave slows down, the steep angular profile (with $a > 2$) in a structured jet dictates when the observer receives flux contribution from smaller angles $\theta<\theta_{\rm obs}$ \citep{Beniamini+20,Beniamini+22}. When the jet has angular structure, at any given time in an ultra-relativistic flow the observer can receive flux from material at the smallest angle $\theta_{\rm min}<\theta_{\rm obs}$ whose beaming cone includes the observer's LOS. This condition is given by $\Gamma(\theta_{\min})(\theta_{\rm obs}-\theta_{\min})=1$. The emission from all other angles smaller than $\theta_{\min}$ is beamed away. If more energy resides in the flow at smaller angles towards the jet symmetry axis, which is typically the case, then material at angle $\theta_{\min}<\theta_F\lesssim\theta_{\rm obs}$ will make the dominant contribution to the observed flux. To determine $\theta_F$ we calculate the angle at which the infinitesimal flux contribution, $[dF_\nu(\theta)/d\Omega]\theta\sin\theta$, is maximized. In an axisymmetric jet with steep angular structure the maximum of this differential flux will move over time to smaller angles along the line connecting the observer's LOS and the jet symmetry axis.

\begin{figure}
    \centering
    \includegraphics[width=0.48\textwidth]{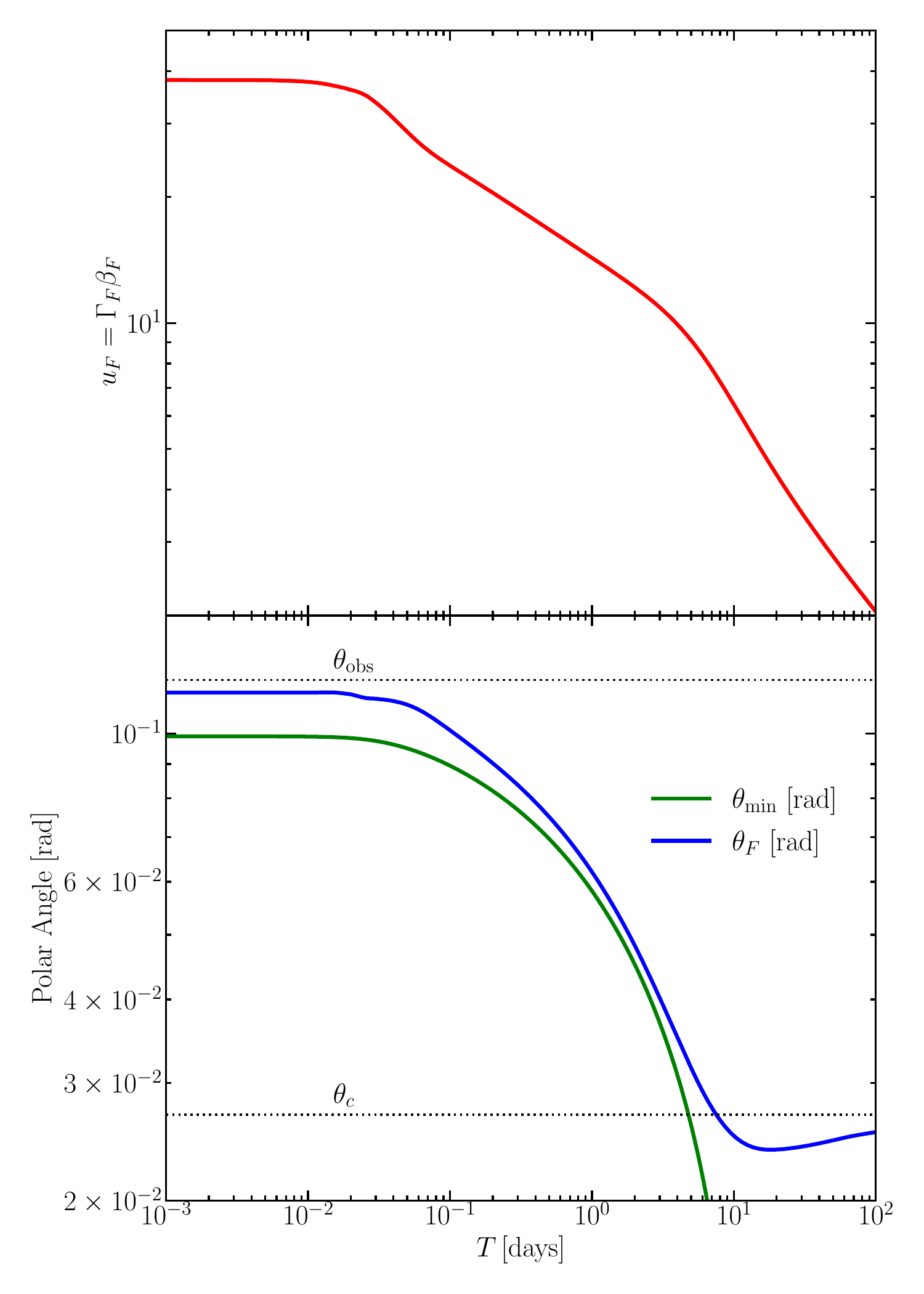}
    \caption{(\textbf{Top}) Proper speed, $\Gamma_F\beta_F$, of the material that makes the dominant contribution to the observed flux at any given time. 
    (\textbf{Bottom}) The angle that makes the dominant contribution is $\theta_F$ and the smallest angle from which the emission is beamed towards the observer is $\theta_{\min}$.
    }
    \label{fig:GammaF-thF-thmin}
\end{figure}

The lower panel of Fig.\,\ref{fig:GammaF-thF-thmin} shows the temporal evolution of both angles $\theta_{\min}$ and $\theta_F$. As $\theta_{\min}$ becomes progressively smaller over time, due to material at those angles decelerating sufficiently enough so that their beaming cones include our LOS, $\theta_F$ also moves to smaller angles due to the presence of more energy closer to the jet core. The angle $\theta_F$ continues to move to smaller angles over time until it reaches the core, after which time no additional energy can be added to our LOS. It is also at this moment the angular structure of the jet becomes unimportant, and the light curve shows a steeper decay that matches the decay trend for a spherical shell. The top panel of Fig.\,\ref{fig:GammaF-thF-thmin} shows the four speed, $u_F=\Gamma_F\beta_F$, where $\Gamma_F=\Gamma(\theta_F)$ and $\beta_F=(1-\Gamma_F^{-2})^{1/2}$, of the material making the dominant contribution to the observed flux. Initially, most of the emission is arising from $\theta_F\sim\theta_{\rm obs}$ until the material at that angle decelerates. That leads to the peak in both the optical and X-ray light curves as both energy bands are located at $\nu_m<\nu_o<\nu_X<\nu_c$ when the material emitting along our LOS decelerates. At $T\sim0.2$\,days $\nu_c$ crosses the X-ray energy band and the light curve correspondingly shows a steeper temporal decay as compared to that in the optical. When $\theta_F$ reaches $\theta_c$ at $T\sim8$\,days, the proper speed shows the same decay trend, i.e. $u_F\propto T^{-(3-k)/2(4-k)}$, as expected from a spherical flow after deceleration. The light curve then correspondingly shows the decay expected from a spherical shell. Over the timescales considered here the bulk LF of the material at $\theta_F$ remains relativistic ($u_F\gtrsim4$), which allows to ignore the effects of any lateral expansion, only becoming trans-Newtonian at $T\gtrsim50$\,days. 

\subsubsection{Afterglow Image Size \& Radio Flux Centroid Motion}
\label{sec:radiocentroid}
One critical test that serves to distinguish between an on-axis and misaligned jet is the offset of the radio flux centroid that evolves over time as the jet expands \citep{Sari-99,Granot-Loeb-03,Gill-Granot-18b}. No such motion of the flux centroid is expected if the jet is viewed on-axis (i.e. with $\theta_{\rm obs}\simeq0$) or when $\theta_{\rm obs}=\pi/2$, where in the latter case both the jet and counterjet are expected to contribute equally to the observed flux causing the flux centroid to remain at the source location. This offset can be calculated by projecting the emission from the jet onto the plane of the sky that has coodinates $(\tilde x, \tilde y)$, where $\tilde x$ is made to align with the line connecting the observer's LOS and the jet symmetry axis. Then, for an axisymmetric flow, it is expected that the flux centroid will move along the $\tilde x$ axis away from the origin that coincides with the GRB location. The flux centroid will continue to move away from the origin until emission from the counterjet starts to become visible. At that moment the offset will reach its farthest distance from the origin and start to move back towards the origin thereafter.

The position of the flux centroid is calculated from $\tilde x=\tilde\rho\cos\tilde\varphi=R\sqrt{1-\tilde\mu^2}\cos\tilde\varphi$, where $\tilde\rho$ and $\tilde\varphi$ are the polar coordinates of the GRB image on the sky plane, $R$ is the radial distance of the blastwave, and $\tilde\mu=\cos\tilde\theta$ with $\tilde\theta$ being the polar angle measured from our LOS in a spherical coordinate system centered at the GRB and in which the blast wave is propagating. Finally, each point on the image needs to be weighted by the differential flux density that it contributes to get the location of the flux centroid \citep{Gill-Granot-18b},
\begin{equation}
    \tilde x_{\rm fc} = \frac{\int dF_\nu\,\tilde x}{\int dF_\nu} 
    \quad\quad {\rm and} \quad\quad 
    \theta_{\rm fc}\equiv\frac{\tilde x_{\rm fc}}{d_A} = (1+z)^2\frac{\tilde x_{\rm fc}}{d_L}\,, 
\end{equation}
where $\theta_{\rm fc}$ is the angular displacement of the centroid, $d_A = (1+z)^{-2}d_L$ is the angular distance and $d_L$ is the luminosity distance.

\begin{figure}
    \centering
    \includegraphics[width=0.48\textwidth]{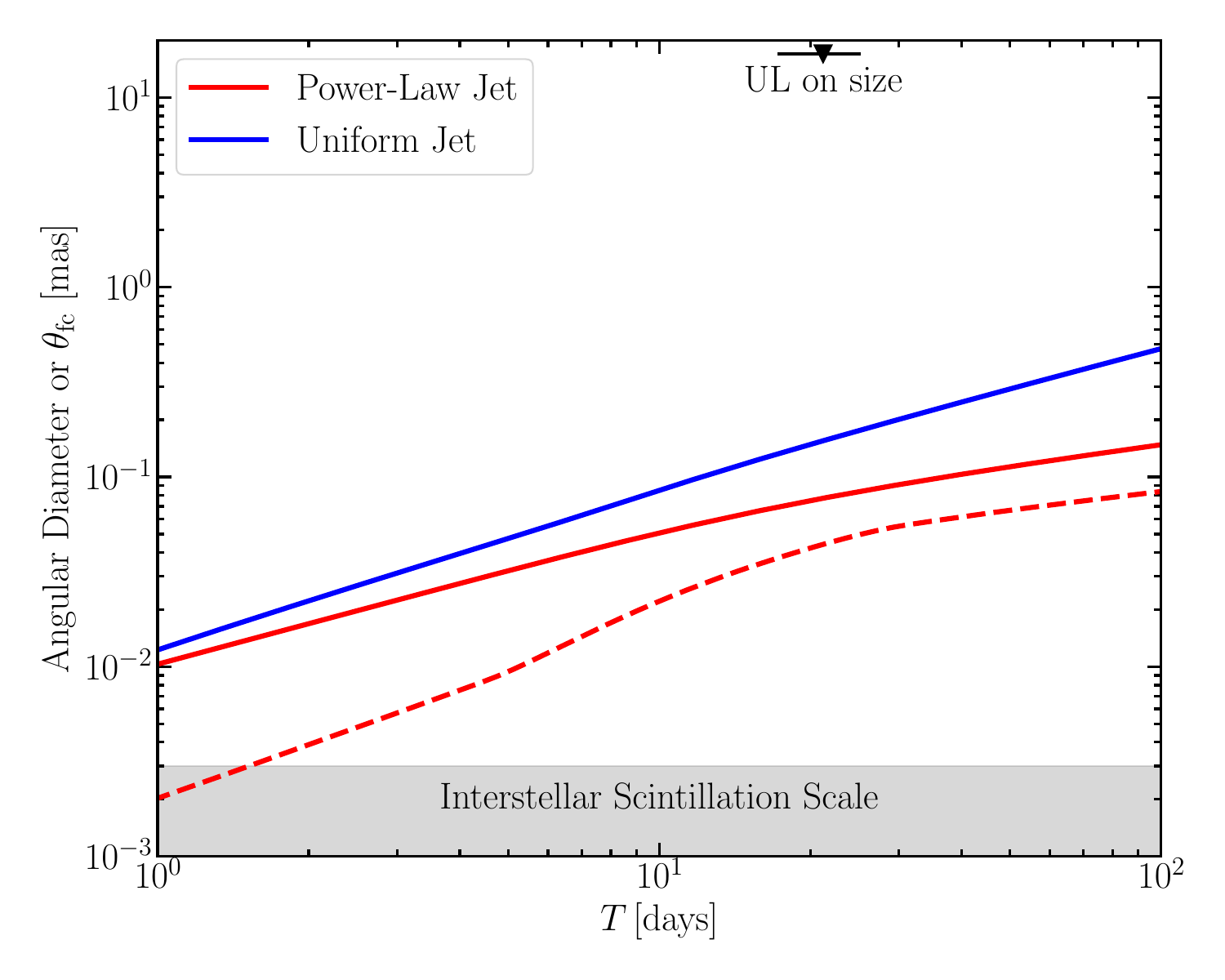}
    \caption{Radio flux centroid distance ($\theta_{\rm fc}$; dashed) away from the source location on the plane of the sky at 45\,GHz for a misaligned power-law structured jet. An on-axis jet would not produce any offset in the flux centroid. The solid curves show the angular diameter of the image of the afterglow for the two jet models. The upside down triangles mark the upper limits from radio observations from $T=(17.2-25.2)$\,days that show that the source is unresolved. The image sizes are larger than the typical interstellar scintillation scale, producing insignificant modulation of the radio light curve.}
    \label{fig:FC}
\end{figure}

Fig.~\ref{fig:FC} shows the angular offset of the flux centroid over time for the model adopted here. The offset from the GRB location is less than 100\,micro-arcsecond over the timescales modeled in this work. Since we do not have an earlier radio observation in our dataset for comparison, this offset cannot be confirmed. Also, given that the radio light curve starts to decline at $T\gtrsim20$\,days it may not be possible to measure relative offsets at later times since the radio flux will be too dim to register any useful measurement.

The image of the afterglow on the plane of the sky is limited by the EATS, that defines a maximum angle away from the LOS, $\tilde\theta_{\max}$ (in a coordinate system centered on the source), beyond which the image is terminated. For a spherical flow or uniform jet viewed sufficiently away from its edges, the image is a limb-brightened disk \citep{Granot+99}. In a spherical flow, the maximum angle is obtained from the condition $\mu_{\max}=\cos\theta_{\max}=\beta$, which simplifies to $\theta_{\max}\simeq1/\Gamma$ for a point-like ($\theta_{\max}\ll1$) ultra-relativistic ($\Gamma\gg1$) source. The radial size of the image at any time $T$ is given by $\tilde\rho=R(\theta_{\rm max})\sin\theta_{\rm max}\simeq R\theta_{\max}\simeq R/\Gamma$, where $\tilde\rho\propto T$ pre-deceleration of the blast wave and $\tilde\rho\propto T^{(5-k)/2(4-k)}$ post-deceleration. In a misaligned angular structured jet the image is not circular, and if the jet is axisymmetric the image will be slightly elongated along the line connecting the viewing angle and the jet symmetry axis. To determine the size of the image we fit an elliptical Gaussian to the projection of the EATS on the sky and report the angular diameter as twice the semi-major axis. The solid curves in Fig.\,\ref{fig:FC} show the size of the image (angular diameter) for the two jet models explored in this work. These are much smaller than the upper-limit imposed by radio observations and therefore the images are simply point-like and unresolved.

Radio light curves will show variability due to interstellar scintillation \citep{Goodman-97} when the size of the image is smaller than the typical scale given by the first Fresnel zone,  
\begin{equation}
    \theta_{\rm Fr}=2.93\,\nu_{10}^{-11/5}\left(\frac{SM}{10^{-3.5}m^{-20/3}\,{\rm kpc}}\right)^{3/5}\,\mu{\rm as}\,,
\end{equation}
where $SM$ is the scattering measure and $\nu = 10\nu_{10}$\,GHz is the transition frequency. In general, the $SM$ and transition frequency vary in different source directions but the above assumes a typical value. The image sizes from the two models, at 45\,GHz and therefore in the weak scattering regime ($\nu>\nu_0$; \citealt{Granot-van-der-Horst-14}), are larger than this scale, and therefore no significant scintillation is expected in the radio light curve.

\subsection{Late-Time Optical \& X-ray Excesses}
Both the optical and X-ray light curves show clear excesses at $T-T_0\gtrsim20$\,days. These could be produced in two distinct shocked media, i.e. from the relativistic collimated blast wave or from the non-relativistic spherical SN blast wave. To calculate the radius at which this emission is produced we first consider the simpler uniform jet model. The dynamical evolution of the blast wave is shown in Fig.\,\ref{fig:Gamma-R-Sph-Einj}, according to which the apparent time of the rebrightening occurs at a radius of $R\simeq8\times10^{18}\,{\rm cm}\simeq2.6$\,pc. The dynamical evolution of the misaligned jet is shown in Fig.\,\ref{fig:GammaF-thF-thmin} that shows that at $T-T_0>20$\,days the flux is dominated by the core. At the time of the rebrightening the jet core has reached a distance of $R\simeq3.8\times10^{18}\,{\rm cm}=1.23$\,pc. These radii are too large for the engine to directly inject energy into the blast wave and therefore it cannot arise from long-lasting engine activity. Alternatively, a local disturbance, e.g., a sharp over-density in the circumburst environment \citep{Nakar-Granot-07} could potentially produce a factor of $\sim2$ flux variation. 

The second scenario involves an over dense region in the external medium encountered by the SN blast wave, the dynamical evolution of which is shown in Fig.\,\ref{fig:sn-vel}. Taking a mean expansion velocity of $v_{\rm sn}\sim2\times10^4\,{\rm km\,s^{-1}}$, the SN blast wave would have propagated to $R=v_{\rm sn}T_{\rm bright}/(1+z)\simeq4\times10^{15}$\,cm by the time of the rebrightening.

Both possibilities require further detailed modeling that is outside the scope of the present work and will be the subject of a future study.

\section{Discussion}
\label{sec:discussion}

\subsection{Two Distinct Jet Models}
In this paper, we compare two scenarios to explain the observed multi-wavelength behavior: (1) an on-axis uniform jet with energy injection, and (2) a misaligned structured jet. While both models can broadly reproduce the overall light-curve evolution, they differ in their physical interpretation of the shallow temporal decay of the optical light curve post-peak. In the structured jet scenario, the light-curve evolution is primarily governed by the energy angular structure and geometric effects, as emission from progressively smaller angles below the viewing angle enters the line of sight. In contrast, the uniform jet model achieves the same via injection of energy into the blast wave. By doing so it also naturally explains the shallow decline of the light curve post-peak in the optical and X-ray light curves, including the brief plateau, and the power-law decay thereafter. The misaligned structured jet, on the other hand, requires some form of energy injection to fully explain the observed excesses near the plateau. While the misaligned structured jet may offer a more natural explanation for the entire afterglow evolution, the on-axis heavily baryon-loaded jet scenario demands special explosion conditions that makes it a rare event.

\subsection{Chromatic Evolution During Rebrightening}
The observed color evolution during the rebrightening phase provides an additional constraint on the underlying physics. The systematic decrease of $r-z$ and $g-z$, together with a nearly constant $g-r$, indicates a frequency-dependent spectral evolution. In a simplified interpretation, where we consider an additional overlapping emission component, this behavior is qualitatively consistent with a spectral break sweeping through the optical bands in this component, as would occur if the characteristic synchrotron frequency $\tilde\nu_m$ evolves across the observed frequency range. As discussed in the energy-injection framework, this can be realized when the observed flux may in general include contributions from multiple emitting regions, such as a possible reverse shock, or from spatially distinct shocked material with different spectral properties. In such a case, the measured colors would reflect a combination of components whose relative flux contributions evolve with time. 

The presence of an additional emission component is further strengthened by the mismatch between the model predictions and the radio observations which appears to be independent of the assumed geometry. Both the spherical energy injection and structured jet models rely on forward shock emission and therefore fail to reproduce the observed radio flux level, indicating the presence of an additional emission component, most likely a reverse shock.

At later times, however, the color evolution changes qualitatively. Beyond about 18 days, the optical colors evolve toward the red and a strong excess develops preferentially in the redder bands. Unlike the earlier blueing trend, this late-time behaviour cannot be straightforwardly explained by synchrotron spectral evolution alone and instead indicates the emergence of thermal emission associated with the accompanying supernova. The simultaneous appearance of a late X-ray rebrightening suggests that the observed optical bump may itself be composed of multiple contributions, namely an achromatic refreshed-shock component superposed on the evolving SN emission. After accounting for this additional afterglow component, the residual optical light curves and color evolution are well reproduced by a faint and fast SN1998bw-like event.

\subsection{Comparison with other long GRBs}
\label{sec:comparison}

\begin{figure*}
    \centering
    \includegraphics[width=0.45\textwidth]{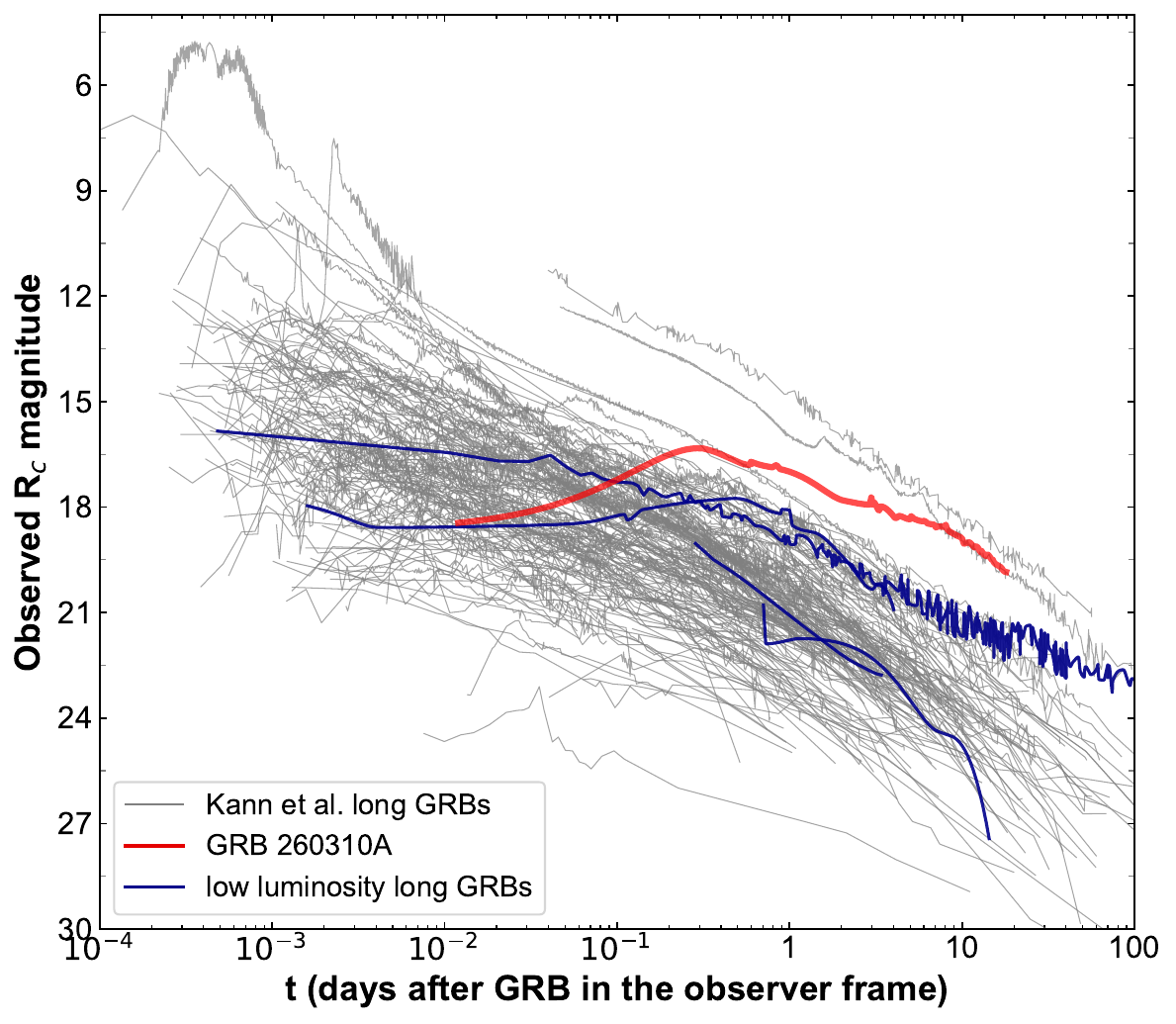}
    \includegraphics[width=0.45\textwidth]{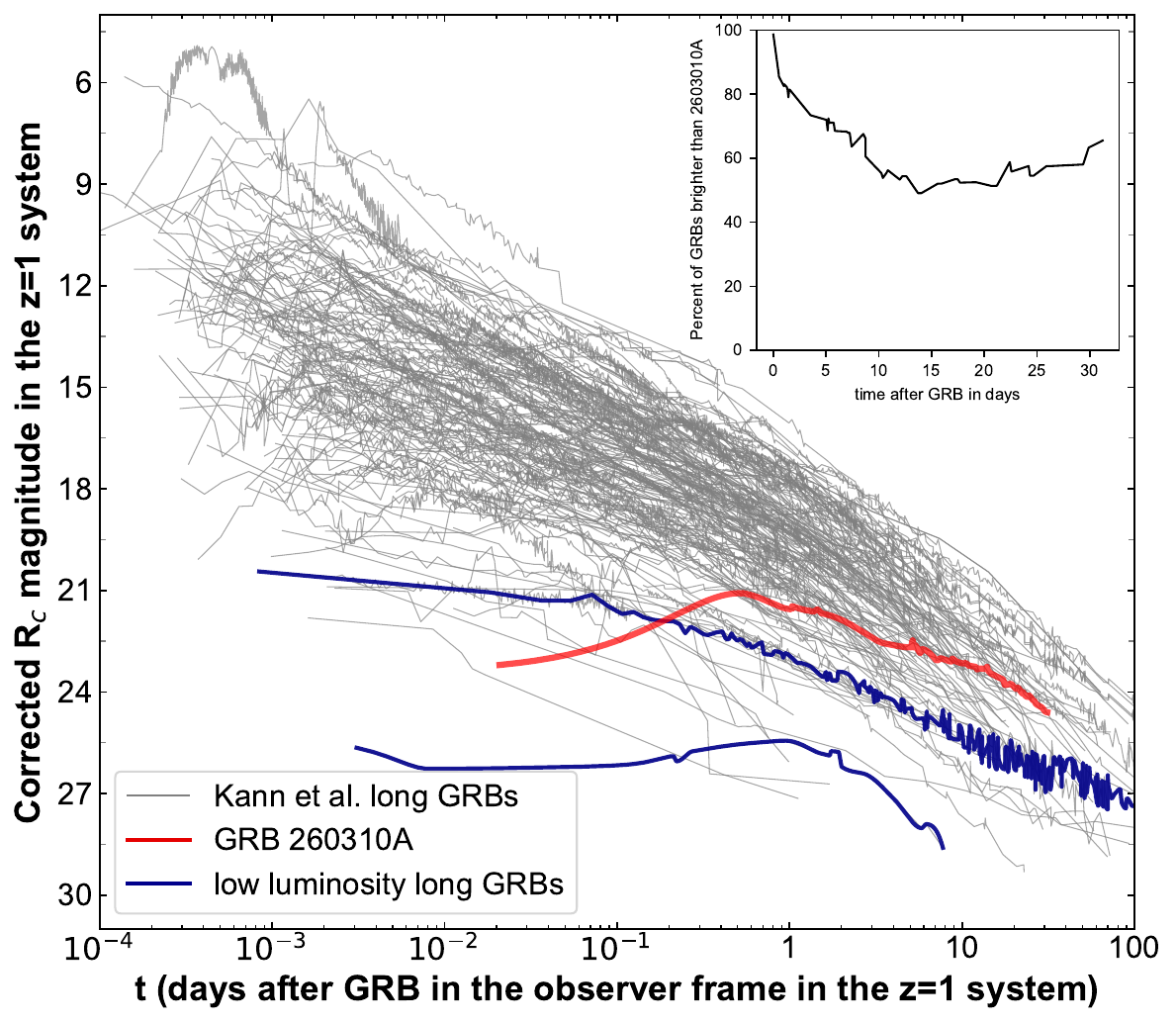}
     \caption{Kann plot for long GRBs in observer frame (left) and in the z$=$1 (right). Data are taken from \protect\cite{Kann+06, Kann10, Kann+11} and the latest update from \protect\cite{Kann24}. The inset in the z$=$1 plot shows the percentile of luminosity of the afterglow of GRB\,260310A within the sample as time evolves. We only plot its LC until day 17 when the SN emission on top of the afterglow decay becomes evident. For GRB\,260310A we only correct for Galactic but not intrinsic extinction for the transformation to the z=1 system since the afterglow extinction is consistent with zero. For comparison, we also plot available afterglow lightcurves of other low redshift low luminosity GRBs in the \protect\cite{Kann24} sample (blue lines): GRB\,060505, GRB\,161219B, GRB\,171205A and GRB\,190829A (the latter can be considered priv. comm.). 
    }
    \label{fig:Kannplot}
\end{figure*}

Figure~\ref{fig:Kannplot} compares the optical afterglow of GRB\,260310A with the long-GRB afterglow sample of \citet{Kann+06,Kann10, Kann+11, Kann24} in both the observer frame (left panel) and after shifting the light curves to a common redshift of $z=1$ (right panel). Although in the observer frame GRB\,260310A stands out because of its apparent brightness (also noticed by \citealt{OConnor2026}), in the common-redshift frame the burst lies among the faintest optical afterglows, particularly at very early times (within the faintest $\sim2\%$ of events in the comparison sample), as inferred from the GOTO observation. Pending confirmation from early-time observations from other facilities, this behavior may be consistent with the delayed rise as inferred from our afterglow model fit to an on-axis uniform jet. In this scenario, GRB\,260310A evolves from the low-luminosity tail of the long-GRB optical afterglow distribution to an average afterglow at late times. Although the sample is still small, other long GRBs with intrinsically low $\gamma$ luminosity seem to also have low luminosity afterglows. A special case is GRB\,171205A, where the emission of the cocoon contributed to a flat evolution of the early afterglow phase \citep{Izzo2019}.

\subsection{Connection with Orphan Afterglows and Gamma-Ray Quiet FXT}

The multi-wavelength properties of GRB\,260310A provide an interesting point of comparison with recently identified orphan afterglow candidates\footnote{Orphan afterglows have been conventionally referred to afterglows from off-axis jets in which the $\gamma$-rays are beamed away from the observer and therefore the GRB itself is not observed \citep{Rhoads-97}. Here we refer to any afterglow that is missing a GRB counterpart as an orphan afterglow, including afterglows of dirty fireballs \citep{Rhoads-03}.}, like AT2019pim \citep{Perley+25}, AT2023lcr \citep{Li+25}, AT2021lfa \citep{Ye+24} and AT2021any \citep{Xu+23}.
These events are characterized by luminous, rapidly rising optical transients, accompanied by bright X-ray and radio emission, but with tight upper limits on any coincident high-energy signal from \textit{Fermi}/GBM and \textit{Konus-Wind}. Their afterglow luminosities are comparable to those of classical long GRBs \citep[compared with the samples from][]{Kann+11,Cenko2009,Chandra2012,Evans2007}, despite the absence of detected prompt emission. This suggests that a population of relativistic explosions may exist in which the prompt emission is either intrinsically faint or observationally suppressed.

Similarly, recent \textit{Einstein Probe} discoveries fast X-ray transients (FXT), such as EP241021a \citep{Yadav2025}, exhibit GRB-like X-ray and radio properties, but without gamma-ray detection (we refer to as gamma-ray quiet), raising the possibility that they represent a related population of relativistic transients identified through soft X-ray surveys rather than high-energy triggers (\textit{Fermi}, \textit{Swift}, \textit{Konus-Wind}). In these cases, the absence of gamma-ray emission may result from observational limitations (e.g. detector coverage or sensitivity) or from intrinsic properties such as lower Lorentz factors, as found in dirty fireball models, and/or off-axis viewing angles in angular structured outflows. We collect the information on gamma-ray quiet fast X-rays transients from \citet{Yadav2025} and use the approximate duration reported in GCNs to estimate the upper limit of the gamma-ray fluence of each event.

This comparison highlights an important consideration: an event intrinsically similar to GRB\,260310A could, under different conditions (such as higher redshift, lower gamma-ray efficiency, or off-axis viewing) appear observationally similar to an orphan afterglow or a gamma-ray quiet FXT. In such a scenario, the afterglow emission would remain detectable at optical, X-ray, and radio wavelengths, while the prompt gamma-ray emission could fall below the detection threshold, thereby imitating the observational properties of orphan afterglows.

\begin{figure}
    \centering
    \includegraphics[width=0.48\textwidth]{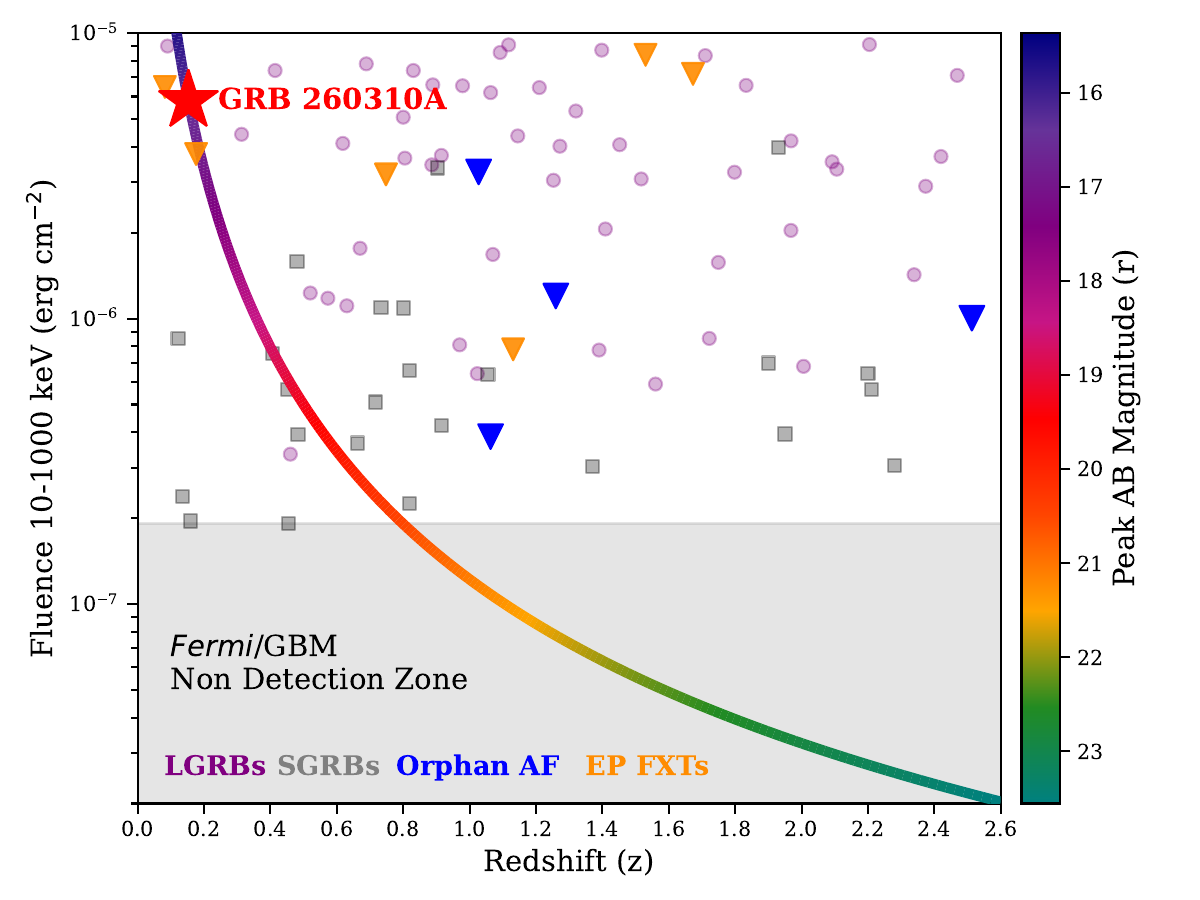}
    \caption{Detectability of the gamma-ray emission from GRB~260310A / SN\,2026fgk by \textit{Fermi}/GBM as a function of redshift. The red star marks the observed properties of GRB\,260310A at $z=0.153$, while the coloured curve shows the expected evolution of its observed fluence in the 10--1000~keV band when shifted to higher redshifts, including the corresponding $\gamma$-ray $k$-correction derived from the cutoff power-law spectrum reported by \citet{GCN43975}. The colour along the curve represents the expected apparent peak $r$-band magnitude. 
    The gray shaded region indicates the approximate \textit{Fermi}/GBM non-detection zone. Long (type II) and short (type I) \textit{Fermi}/GRBs from the comparison sample are shown as purple circles and gray squares, respectively. Upper limits or inferred fluences for orphan afterglow candidates are displayed as blue inverted triangles, while EP FXTs without gamma-ray emission are shown in orange. An event intrinsically similar to GRB\,260310A would rapidly move below the \textit{Fermi}/GBM sensitivity threshold at moderate redshift ($z\approx0.8$), while still remaining detectable by COLIBRÍ (assuming a detecting magnitude of $r\approx 24.5$).}
    \label{fig:orphan_GRBs}
\end{figure}

To explore this possibility, we performed a simple test by placing GRB\,260310A ($z=0.153$; see Section~\ref{sec:hg}) at different redshifts (continuous line in Figure~\ref{fig:orphan_GRBs}), scaling the gamma-ray signal according to luminosity distance and cosmological time dilation, while assuming the same intrinsic emission and instrumental background. Figure~\ref{fig:orphan_GRBs} also shows the distribution of long and short GRBs detected by \textit{Fermi}/GBM, together with orphan afterglow candidates and gamma-ray quiet FXTs discovered by \textit{Einstein Probe}.

Under these assumptions, the expected gamma-ray signal drops to only a few counts per bin, comparable to background fluctuations, and would likely not trigger \textit{Fermi}/GBM. At the same time, the optical afterglow would remain sufficiently bright to be detected by current wide-field optical surveys, even at moderately high redshifts. The colour gradient along the GRB\,260310A placed at different distances in Figure~\ref{fig:orphan_GRBs} illustrates how the apparent peak optical brightness evolves with redshift, assuming $\beta=-0.9$ and that the spectral regime $\nu_m<\nu_o<\nu_c$ remains valid. The details of this are described in Appendix~\ref{sec:kcorrection}.

This result indicates that even a standard, on-axis GRB such as GRB\,260310A could remain undetected in gamma rays if observed at a higher redshift, reinforcing the idea that some percentage of the orphan afterglow and gamma-ray quiet FXTs populations may arise from an observational bias in a flux-limited detector rather than intrinsically distinct explosion mechanisms. On the other hand, the optical afterglow would have been detected by COLIBRÍ even out to a moderately high redshift of $z\gtrsim2.4$.

\subsection{Radio Polarization}
\label{sec:rad_pol_scintillation}
Several GRBs show optical afterglow linear polarization of $\Pi\sim(1-5)$ per cent at late-times when the forward shock emission dominates \citep[e.g.][]{Covino-Gotz-16,Agui-Fernandez+24}. It can be much higher ($\Pi\sim5-28$\,per cent) at early times when the emission is dominated by the reverse shock instead \citep[e.g.][]{Mundell+13,Steele+17}. In contrast, radio afterglow polarization measurements have always been at much lower level, with $\Pi\lesssim1$\,per\,cent \citep{Urata2019,Laskar+19,Urata2023}. 

GRB\,260310A provides the first detection of significant late-time afterglow polarization at centimeter wavelengths at $T_0+19.2$\,days \citep{Christy+2026}. The polarization degree showed a monotonic decline from $(3.18 \pm 0.18)$\,per\,cent at 25~GHz to $(0.69\pm0.22)$\,per cent at 11~GHz along with an unprecedented measurement of Faraday rotation in a GRB ever. These observations yielded the first-ever signs of depolarization that may have been caused by either the location of the synchrotron self-absorption frequency ($\nu_a$) in the middle of their observing band or Faraday depolarization due to an external Faraday screen. Alternatively, Faraday depolarization above $\nu_a$ can also be caused by cooler thermal electrons in partly ordered magnetic fields \citep{Toma2008}. A further measurement of radio polarization at 15\,GHz was made by \citet{Yang+2026} at $T_0+55$\,days that yielded $\Pi=1.7\pm0.4$\,per\,cent, with only $3\sigma$ upper limits ($\Pi_{\rm 10\,GHz}<2.5$\,per\,cent and $\Pi_{\rm 6\,GHz}<1.7$\,per\,cent) obtained at lower frequencies.

The saturation of the radio polarization at $\Pi\sim3$\,per cent at $T_0+19.2$\,days in GRB\,260310A towards higher frequencies yields the intrinsic degree of polarization that matches the level in optical observations. However, since the radio emission here may have an origin different from the forward-shock-produced optical emission, e.g. a reverse shock, the properties of the magnetic field can be very different. To qualitatively explain this intrinsic polarization we consider the two jet geometries used in this work and the structure of the magnetic field in the shocked media. If the upstream medium is weakly magnetized or unmagnetized, collisionless relativistic shocks generate microscopic scale magnetic fields via the two-stream and/or Weibel (filamentation) instabilities \citep{Weibel-59,Gruzinov-Waxman-99,Medvedev-Loeb-99}. This magnetic field is tangled, confined to the shock plane, and also axisymmetric around the local shock normal. Consequently, it generates negligible net polarization in a uniform flow, unless the symmetry of the afterglow image is broken when the beaming cone of emission includes the jet edge \citep{Ghisellini-Lazzati-99,Sari-99}. There is no clear indication of a jet break in the afterglow light curve, and so this possibility may not be realized. Alternatively, the symmetry of the image can be broken in a misaligned jet with angular structure, yielding net non-zero polarization \citep{Rossi+04,Gill-Granot-18b,Gill-Granot-20,Birenbaum+24}. This opens up the possibility of explaining the radio polarization in our misaligned power-law jet model if the magnetic field in the shocked material in indeed microscopic.

A net non-zero polarization can be obtained in a uniform flow if the shocked material contained an (partially) ordered field in addition to the shock-produced microscopic field \citep{Granot-Konigl-03}. Any preexistent weak ordered field in the ejecta may be amplified by a magnetohydrodynamic instability, e.g. turbulence, so that it forms several mutually incoherent patches within the observable region of the shocked shell \citep{Gruzinov-Waxman-99,Kuwata2023,Kuwata2024}. The angular size of these regions is $\theta_B<1/\Gamma$ so that the observable region contains $N\sim(\Gamma\theta_B)^{-2}$ patches that dilutes the net polarization to $\Pi\sim\Pi_{\max}/\sqrt{N}$. Here $\Pi_{\rm max}$ is the maximum polarization produced locally by an ordered field and which depends on the spectral index, such that $\Pi_{\max} = (\alpha+1)/(\alpha+5/3) = (p+1)/(p+7/3)$ where $\alpha=-d\ln F_\nu/d\ln\nu=(p-1)/2$. If $p\sim2.5$ then $\Pi_{\rm max}=72$\,per cent that yields $N\sim575$ such patches of angular size $\theta_B\sim5.2\times10^{-3}$\,rad for $\Gamma\simeq8$ at $T\simeq19$\,days  \citep{Christy+2026}. If the same emission region with a patchy ordered field continues to dominate the polarization at $T_0+55$\,days, when keeping the same $\theta_B$ an estimate of $\Gamma\simeq4.5$ is obtained, which is consistent with that obtained in Fig.\,\ref{fig:Gamma-R-Sph-Einj}.

The distinguishing feature between the ordered and microscopic field is that the polarization position angle should show random variations over time in the former case and either remain constant or jump by $90^\circ$ around the jet break time in the latter. Radio polarization measurements at different epochs should help in discriminating between the two scenarios. A more detailed investigation into the afterglow radio emission and its polarization in GRB\,260310A  will be the subject of a future work.

\section{Conclusions}
\label{sec:summary}

We have presented a detailed multi-waveband analysis of GRB\,260310A/\,SN\,2026fgk, combining gamma-ray, X-ray, optical/NIR, and radio observations spanning from minutes to several weeks after the burst. Despite its nearby distance ($z=0.153$), the event produced comparatively faint prompt $\gamma$-ray emission accompanied by an exceptionally bright and long-lived afterglow.

The afterglow evolution reveals a complex temporal and spectral behaviour with multiple mild flares. Both the X-ray and optical light curves exhibit a shallow phase around $\sim4$--$8$ days that is inconsistent with a simple afterglow model of a uniform jet. This is followed by a steeper decay and late-time flattening; while the optical colour evolution and spectroscopy during the latter reveal the emergence of an associated broad-lined supernova. In parallel, the host-galaxy analysis shows that the transient is associated with a relatively luminous star-forming galaxy at $z=0.153$, although with a comparatively large projected offset of $\sim15$~kpc from its center. Similar large-offset GRB-SN systems have been reported in a small number of nearby low- and intermediate-luminosity GRBs, suggesting that relativistic explosions with weak prompt $\gamma$-ray emission may arise in a broader range of local environments than classical cosmological GRBs.  While the optical-to-X-ray SED is broadly consistent with forward-shock synchrotron emission, the radio data require an additional component, most likely a reverse shock during the energy injection/refreshed shock process. The detection of centimetre-band radio polarization further demonstrates that ordered magnetic fields or geometric asymmetries persist in the emitting region at late times.

Interestingly, the optical afterglow of GRB\,260310A evolves from the faint tail of the long-GRB afterglow population to a more typical luminosity regime in the comparison sample (see Fig.~\ref{fig:Kannplot}), further emphasizing the unusual observational properties of the event that remains to be verified by early time observations. 

We modelled the multi-waveband afterglow of GRB\,260310A using an on-axis uniform jet with an episode of energy injection and a misaligned jet with a power-law angular structure. Both models are able to describe the optical and X-ray light curves well, but underproduce the radio emission, indicating the presence of an additional component. Whether the shallow afterglow light curves are connected to our viewing geometry or an intrinsically rare and different nature of the relativistic outflow remains unclear, while the observed colour evolution provides additional constraints on the physics of energy injection.

In both jet models considered here, the material along our LOS had kinetic energy $E_{\rm k,iso}\sim10^{52.2-52.9}$\,erg and must have coasted at an atypically low bulk $\Gamma\sim20-30$ before decelerating. This directly led to the low $\gamma$-ray efficiency and Comptonized spectrum of the prompt GRB. Current radio observations, including the flux centroid motion and image size, cannot distinguish between the two jet geometries. Further radio polarimetric measurements will help to break the degeneracy and constrain the magnetic field structure in the emission region.

GRB\,260310A / SN\,2026fgk demonstrates how relativistic explosions with underluminous prompt $\gamma$-ray emission, but otherwise ordinary afterglows, can easily evade detection by current gamma-ray detectors and instead appear as orphan afterglows or $\gamma$-ray quiet fast X-ray transients purely because of observational selection effects. We show that, had this event occurred at a redshift of $z\gtrsim0.5$, the prompt GRB would not have been detected by gamma-ray monitors like \textit{Fermi}/GBM. Instead, it would have been classified as an orphan afterglow since the peak optical magnitude can be easily detected up to $z\sim2$. It is not yet clear whether the outflow was a dirty fireball seen on-axis or a more standard structured outflow seen off-axis, it certainly reveals the underlying observational bias in detecting the GRB in these two scenarios. In addition, it highlights how one type of outflow can be easily mistaken for another due to the inherent degeneracy in the afterglow modelling.

Events such as GRB\,260310A may represent a direct connection between classical GRBs, orphan afterglow candidates, and the growing population of fast transients discovered by wide-field optical and X-ray surveys. As facilities such as \textit{Einstein Probe} and the Vera C. Rubin Observatory continue expanding the discovery space of fast extragalactic transients, systems like GRB\,260310A provide an important benchmark for identifying the continuous spectrum from moderately relativistic to ultra-relativistic jets and the diversity of transients that they power.

\section*{Acknowledgements}

We thank the anonymous referee for the careful reading of the manuscript and for the constructive comments and suggestions that helped improve the quality and clarity of this work.

We thank Stephane Favard, Jean Pierre Troncin, Yoann Degot-Longhi, Jean Balcaen, Jerome Schmitt, and Jean Claude Brunel (night operators, engineers, and technicians of MISTRAL), as well as the SOPHIE observers during the ToOs who allowed us to carry out the observations: Jannat Alazzawi, Lena Parc, and Remi Fahed.

COLIBRI received support from the French government under the France 2030 investment plan, as part of the Initiative d’Excellence d’Aix-Marseille Université-A*MIDEX through (ANR-11-LABX-0060 - OCEVU) and (AMX-19-IET-008 - IPhU), from LabEx FOCUS (ANR-11-LABX-0013), From Centre National d'Etudes Spatiale (CNES) and from CSAA-INSU-CNRS support program, and in Mexico from UNAM (Secretaria Administrativa, Coordinacion de la Investigacion Cientıfica, Instituto de Astronomıa and PAPIIT grant IN105921), and SECIHTI/CONACyT (277901, Ciencias de Frontera 1046632 and Laboratorios Nacionales).

The COLIBRÍ team thanks the staff of the Observatorio Astronómico Nacional at Sierra de San Pedro Mártir, as well as the technical and engineering teams at CEA, CPPM, IRAP, LAM, OHP, OSU Pytheas, and UNAM.

This research has made use of the MISTRAL database, based on observations made at Observatoire de Haute Provence (CNRS), France, with the MISTRAL spectro-imager, and operated at CeSAM (LAM), Marseille, France.

This work has made use of data from the Asteroid Terrestrial-impact Last Alert System (ATLAS) project. The Asteroid Terrestrial-impact Last Alert System (ATLAS) project is primarily funded to search for near earth asteroids through NASA grants NN12AR55G, 80NSSC18K0284, and 80NSSC18K1575; byproducts of the NEO search include images and catalogs from the survey area. This work was partially funded by Kepler/K2 grant J1944/80NSSC19K0112 and HST GO-15889, and STFC grants ST/T000198/1 and ST/S006109/1. The ATLAS science products have been made possible through the contributions of the University of Hawaii Institute for Astronomy, the Queen’s University Belfast, the Space Telescope Science Institute, the South African Astronomical Observatory, and The Millennium Institute of Astrophysics (MAS), Chile. This research used the facilities of the Canadian Astronomy Data Centre operated by the National Research Council of Canada with the support of the Canadian Space Agency. This research has made use of the NASA/IPAC Infrared Science Archive, which is funded by the National Aeronautics and Space Administration and operated by the California Institute of Technology.

The Caltech OVRO 40~m Telescope monitoring program is supported by NSF grants AST 2407603 and AST 2407604. P.V.d.l.P. acknowledges support from ANID Basal AFB-170002, Núcleo Milenio TITANs (NCN2023$\_$002), CATA BASAL FB210003 and UdeC-VRID 2025001479INV.

The National Radio Astronomy Observatory is a facility of the U.S. National Science Foundation operated under cooperative agreement by Associated Universities, Inc.

CAV acknowledges support from a SECIHTI fellowship. 

NG and LGG gratefully acknowledge the support of the Simons Foundation (MP-SCMPS-00001470, N.G.)

EAR acknowledges support from the UNAM/DGAPA Elisa Acuña and SECIHTI postdoctoral fellowships.

AMW is grateful for support from UNAM/PAPIIT project IN109224.

AdUP is supported by the Programme National Astro of CNRS/INSU with INP and IN2P3, co-funded by CEA and CNES through the Thematic Actions "Phénomènes Extrêmes et Multi-messagers" (PEM), “Physique et Chimie du Milieu Interstellaire” (PCMI) and "Cosmologie et Galaxies" (CG) of INSU Programme National “Astro".

DBM is funded by the European Union (ERC, HEAVYMETAL, 101071865). Views and opinions expressed are, however, those of the authors only and do not necessarily reflect those of the European Union or the European Research Council. Neither the European Union nor the granting authority can be held responsible for them. The Cosmic Dawn Center (DAWN) is funded by the Danish National Research Foundation under grant DNRF140.

AK is supported by the UNAM/DGAPA Elisa Acuña postdoctoral fellowship.

MAA acknowledges support from grants PID2021-127495NB-I00, PID2025-171322NB-C22 and CIPROM/2022/13.
\section*{Data Availability}

The data underlying this article will be shared on reasonable request to the corresponding author.



\bibliographystyle{mnras}
\bibliography{refs.bib} 





\clearpage
\appendix


\section{Host Galaxy}
\subsection{Spectral Energy Distribution}
\label{app:hgphoto}

We compiled archival photometry of the host galaxy, free from afterglow emission, from publicly available optical and infrared surveys to construct the broadband spectral energy distribution used for the fitting analysis. Optical measurements in the $g$, $r$, $i$, $z$, and $y$ bands were retrieved from the Pan-STARRS DR2 catalog \citep{Chambers2016}, while the $u$-band photometry was obtained from CFHT observations \citep{CADC_CFHT}. Mid-infrared photometry at 3.4, 4.6, 12, and $22\,\mu$m was extracted from the WISE AllWISE catalog \citep{Cutri2013}. The host galaxy photometry is provided in Table~\ref{tab:photometry_host}.

\begin{table}
\centering
\caption{Host galaxy photometry used for the SED fitting.}
\label{tab:photometry_host}
\begin{tabular}{lcc}
\hline
Instrument & Band & AB Magnitude \\
\hline
CFHT        & $u$                & $19.87 \pm 0.05$ \\
Pan-STARRS  & $g$                & $18.85 \pm 0.03$ \\
Pan-STARRS  & $r$                & $18.09 \pm 0.02$ \\
Pan-STARRS  & $i$                & $17.87 \pm 0.02$ \\
Pan-STARRS  & $z$                & $17.85 \pm 0.04$ \\
Pan-STARRS  & $y$                & $17.62 \pm 0.20$ \\
WISE1       & $3.4\,\mu$m        & $18.14 \pm 0.05$ \\
WISE2       & $4.6\,\mu$m        & $18.57 \pm 0.09$ \\
WISE3       & $12\,\mu$m         & $17.48 \pm 0.15$ \\
WISE4       & $22\,\mu$m         & $16.24 \pm 0.53$ \\
\hline
\end{tabular}
\end{table}

\section{Prompt Emission Spectral Models}
\label{sec:prompt-models}
We use two different spectral models to fit the prompt emission spectrum.

\begin{figure*}
    \centering
    \includegraphics[width=0.48\textwidth]{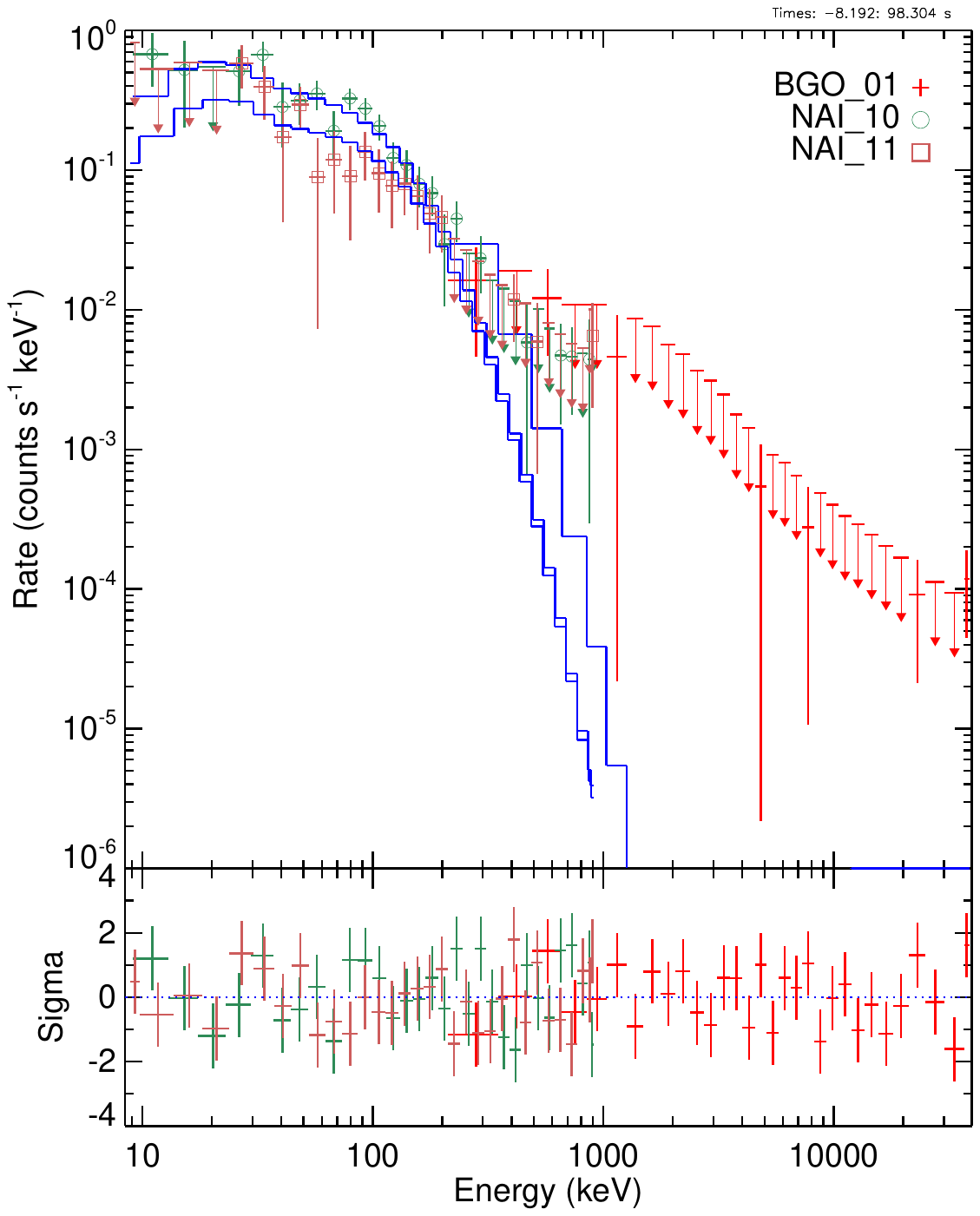}
    \includegraphics[width=0.48\textwidth]{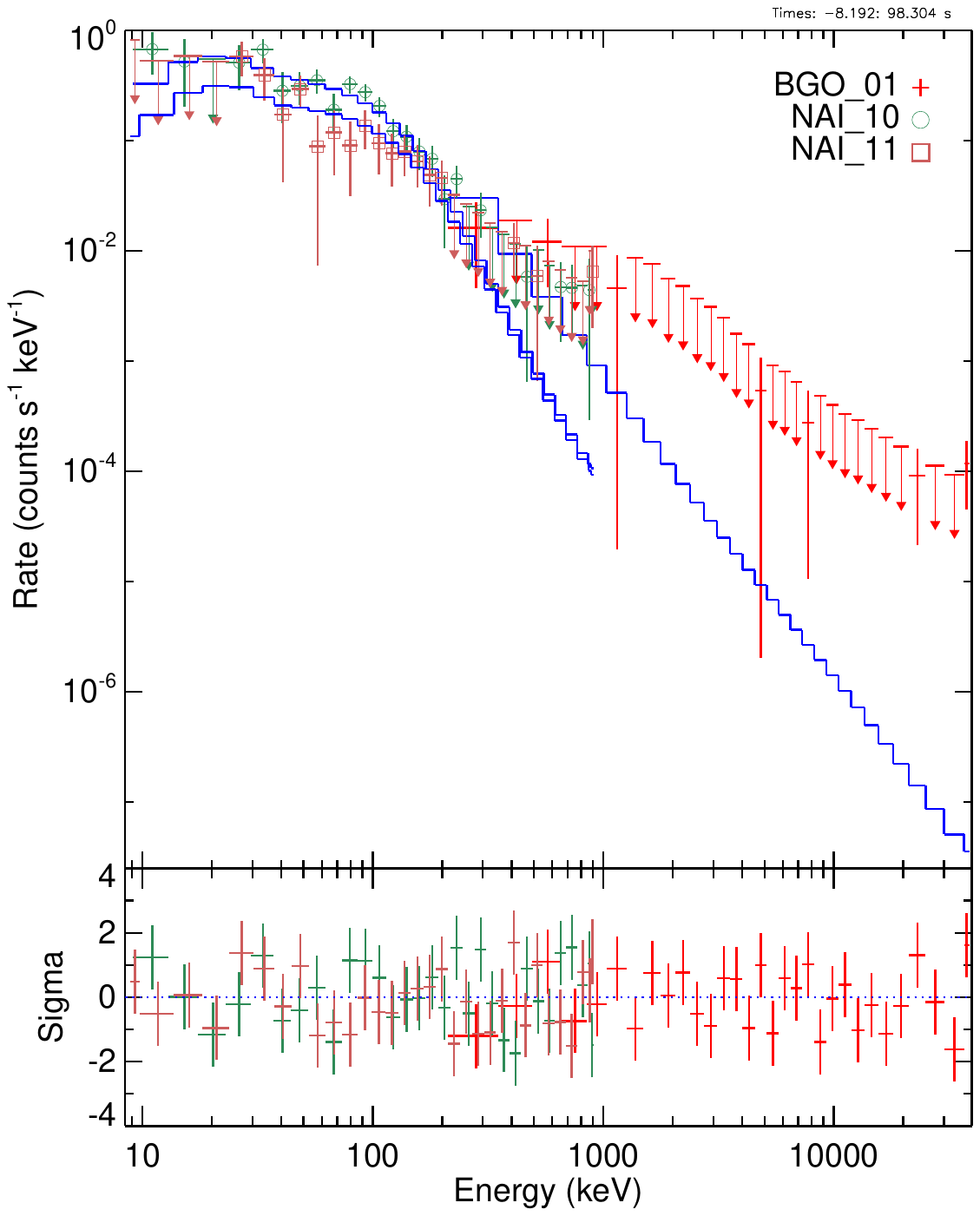}
    \caption{(\textbf{Left}) A cutoff power-law or Comptonized spectral model fit (blue curve) to the time-integrated prompt emission spectrum. (\textbf{Right}) The Band function fit (blue curve) to the same spectrum. In the bottom panels \emph{Sigma} shows the residuals between the model, folded through the detector responses (blue lines), and the observed count rates, normalized by the uncertainty.}
    \label{fig:prompt-spec-fit}
\end{figure*}

(i) \textit{Cutoff Power Law}: This model is described by a power law with an exponential cutoff \citep[see e.g.][]{Kaneko+06,Goldstein2012}. The photon spectrum is given by
\begin{equation}
  N(E) = A \left( \frac{E}{E_{\rm piv}} \right)^{\alpha_{\rm CPL}} \exp \left[ -\frac{(\alpha_{\rm CPL} + 2) E}{E_{\rm peak}}  \right] \label{eq:comp}\,,
\end{equation}
where $E$ is the photon energy in keV, $N(E)$ is the photon spectrum in units of photons\,${\rm cm^{-2}\,s^{-1}\,{keV}^{-1}}$, $\alpha_{\rm CPL}$ is the photon index, $A$ is the normalization, $E_{\rm peak}$ is the peak energy, and $E_{\rm piv}=100$\,keV is the pivot energy. The best-fit model parameters from this spectral model are shown in Table\,\ref{tab:prompt-spec-params} and the model fit is shown in the left panel of Fig.\,\ref{fig:prompt-spec-fit} (online supplementary material).

\begin{table*}
    \centering
\begin{tabular}{c| c c c c c}
Time &  $\alpha_{\rm CPL}$ &  $E_{\rm peak}$  & Flux & E$_{\rm iso}$ & Cstat/dof  \\
  s&      &  keV &     10$^{-8}$ erg cm$^{-2}$ s$^{-1}$ & 10$^{49}$ erg &   \\
\hline\hline
-8.192- 40.960  &    -0.06 $\pm$ 0.21  &  198.5  $\pm$   16.8 &  11.50 $\pm$ 0.77  & 31.2 $\pm$ 2.1 & 518.1/355  \\
40.960- 98.304  &  -0.49 $\pm$   0.71    & 74.5 $\pm$    18.5  & 1.79 $\pm$ 0.35   & 6.1 $\pm$ 1.2 & 525.7/355  \\
\hline
 -8.192- 98.304 & -0.39 $\pm$0.20   &  172.4 $\pm$18.7   & 6.07 $\pm$ 0.48 & 36.5$\pm$ 2.9 & 646.8/355  \\
  \end{tabular}
 \caption{Spectral parameters, for the two intervals separately, and analysed as a whole. The best fitting spectral model is a power law with exponential cutoff. 
 }
 \label{tab:prompt-spec-params}
\end{table*}

(ii) \textit{Band Function}: The Band function \citep{Band1993} is a smoothly broken power law given by
\begin{equation}
   N(E) = A \left\{
               \begin{array}{ll}
                 \left( \frac{E}{E_{\rm piv}} \right)^{\alpha_B} e^{\left[-\frac{(\alpha_B + 2) E}{E_{\rm peak}}  \right]}, & E \geq E_b    \\
                 \left( \frac{E}{E_{\rm piv}} \right)^{\beta_B} e^{\left(\beta_B - \alpha_B \right)}
                 \left[ \frac{(\alpha_B - \beta_B) E_{\rm peak}}{E_{\rm piv} (\alpha_B +2)}\right]^{\alpha_B - \beta_B}, & E < E_b
               \end{array}
             \right.\label{eq:band}
\end{equation}
where $E_b = [(\alpha_B - \beta_B) E_{\rm peak}]/(\alpha_B +2)$ is the break energy, and $\alpha_B$ and $\beta_B$ are the low and high energy photon indices. The model fit is shown in the right panel of Fig.\,\ref{fig:prompt-spec-fit}.

\section{MCMC Fit Posterior Distributions}
\label{app:mcmc}

\begin{figure*}
    \centering
    \includegraphics[width=0.9\textwidth]{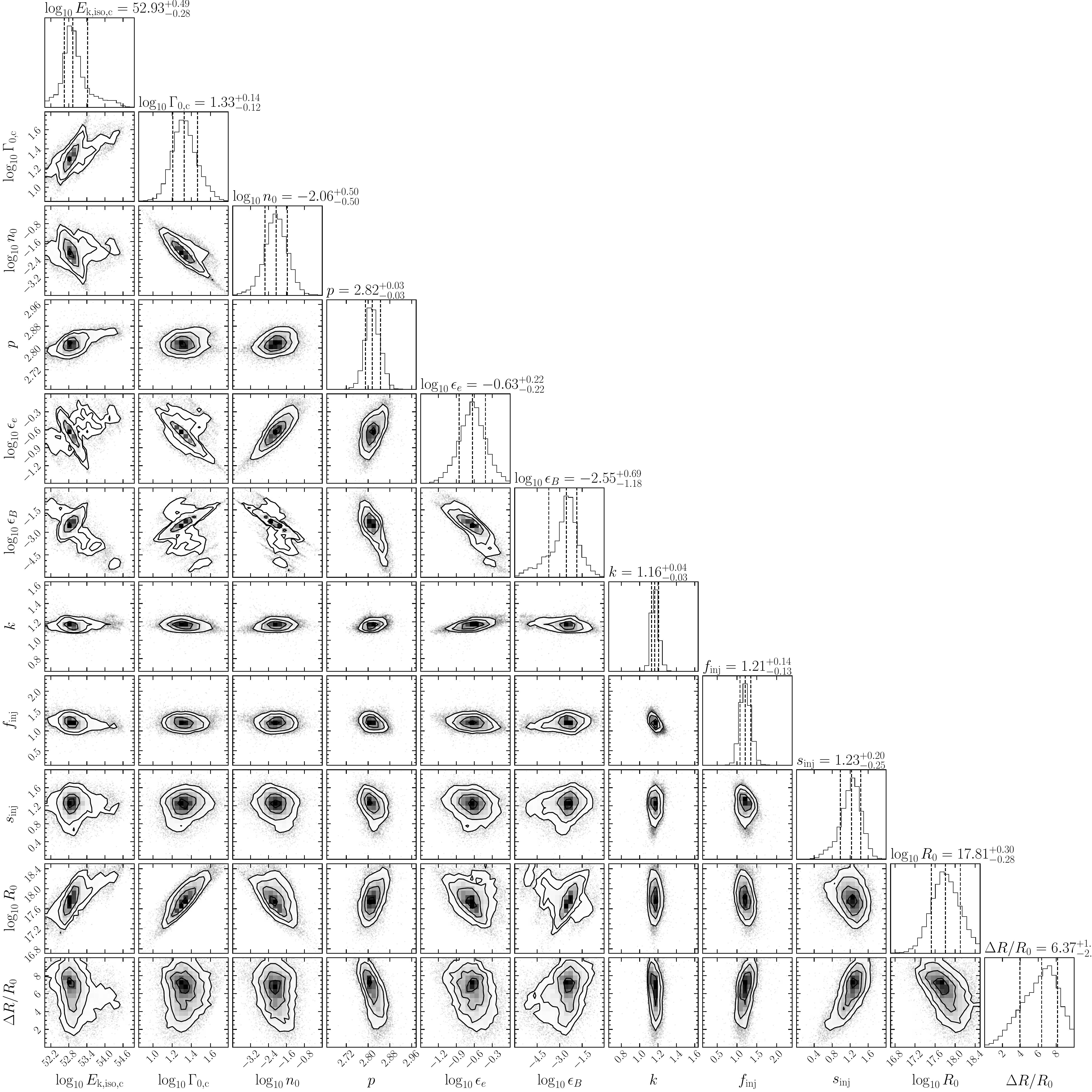}
    \caption{Afterglow model parameter posterior distributions obtained from the fit to the multi-waveband light curve from a uniform jet with energy injection.}
    \label{fig:fit-posteriors-Sph-Einj}
\end{figure*}

\begin{figure*}
    \centering
    \includegraphics[width=0.9\textwidth]{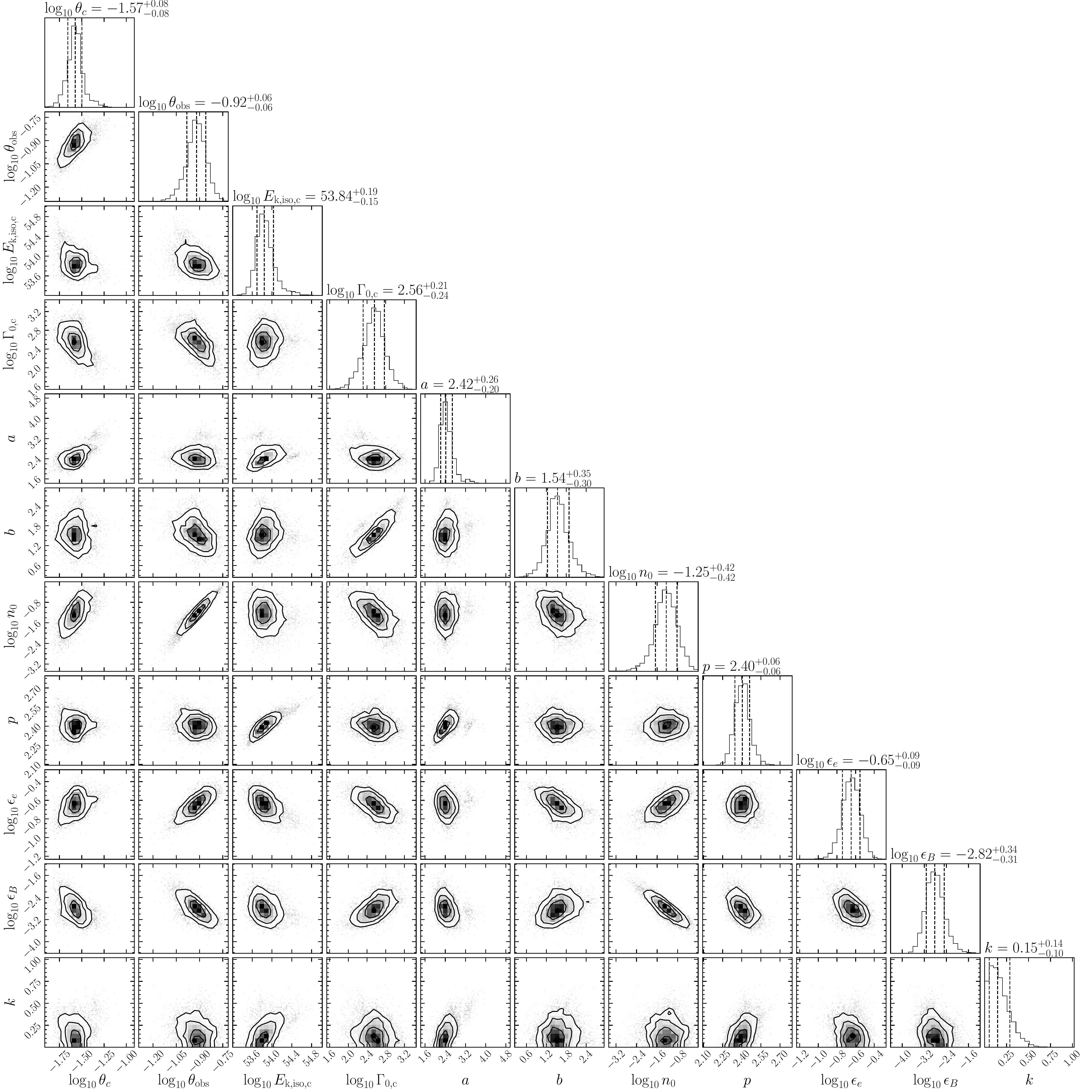}
    \caption{Afterglow model parameter posterior distributions obtained from the fit to the multi-waveband light curve from a misaligned power-law angular structured jet.}
    \label{fig:fit-posteriors-PLJ}
\end{figure*}

The parameter posterior distributions from the two jet models used in this work are shown in figures \ref{fig:fit-posteriors-Sph-Einj} and \ref{fig:fit-posteriors-PLJ} (online supplementary material).

\section{Photometry}
\label{appendix:photometry}
Table \ref{tab:xray_photometry} shows our new X-ray photometry from \textit{EP\/}/FXT.
Tables \ref{tab:new_optical_photometry}, \ref{tab:new_J_photometry}, and \ref{tab:new_extraphotometry} show the optical and infrared photometry used in this paper. Table \ref{tab:radio_photometry} shows radio photometry from GCNs (cited in the table) and our new OVRO observations.

Table \ref{tab:new_optical_photometry} shows optical photometry from GOTO, ATLAS, LAST, and COLIBRÍ. All of this photometry is after correcting for the host galaxy by subtracting a template image. The GOTO and LAST photometry is from GCNs (cited in the table), the ATLAS photometry is from the ATLAS forced-photometry service, and the COLIBRÍ photometry is new. We use the GOTO, ATLAS, LAST, and COLIBRÍ $r$ data from Table \ref{tab:new_optical_photometry} when fitting the model light curves (i.e., in Figures \ref{fig:fit-spherical-shell} and \ref{fig:fit-misaligned-jet}), is shown in Figure \ref{fig:multi-waveband-LC-Spectra}, and is used to derive the SEDs.

Table \ref{tab:new_J_photometry} shows $J$ photometry from GCNs (cited in the table) and new photometry from ALT100C. None of this photometry is after subtracting a template image. This photometry is used as an independent check of the model light curves (i.e., in Figures \ref{fig:fit-spherical-shell} and \ref{fig:fit-misaligned-jet}), is shown in Figure \ref{fig:multi-waveband-LC-Spectra}, and is used to derive the SEDs.

Table \ref{tab:new_extraphotometry} shows additional new optical and infrared photometry from NOT, TRT-SRO, and ALT100B and photometry published in GCNs (cited in the table). None of this photometry is after subtracting a template image. This photometry is not used in our model fitting, but it shown in Figure \ref{fig:multi-waveband-LC-Spectra} and is used to derive the SEDs.

None of the optical or infrared photometry in Tables \ref{tab:new_optical_photometry}, \ref{tab:new_J_photometry}, or \ref{tab:new_extraphotometry} has been corrected for Galactic extinction. However, prior to analysis, we apply corrections from Table \ref{tab:filters}, which shows our assumed filter wavelengths and values of the Galactic extinction from the model of \cite{Schlafly+2011}. For the GOTO $L$ photometry, with a band-pass from 420-690~nm, we assume $A_L \approx 0.5 (A_g + A_r)$ \citep{Steeghs+2022}, for the ATLAS $o$ photometry, with a band-pass from 560-820~nm, we assume $A_o \approx 0.5 (A_r + A_i)$, and for the LAST photometry we assume a similar band-pass and extinction to the DDOTI telescope and so $A_w \approx 0.23 A_g + 0.77 A_r$ \citep{Watson+2020}.

\clearpage
\onecolumn

\begin{table*}
\centering
\caption{X-ray photometry of GRB~260310A / AT~2026fgk from this work. Uncertainties are quoted at the $1\sigma$ confidence level. We list the first five rows of the table; the full version is available online and in Table~\ref{tab:xray_photometry_full}.}
\label{tab:xray_photometry}
\begin{tabular}{cccccc}
\hline
\hline
Time & Exposure & Telescope/Instrument & Band & Flux Density \\
(d) & (s) & & (keV) & ($\mu$Jy @1 keV) \\
\hline
2.520  & 3259 & \textit{EP\/}/FXT & 0.5--10 & 0.742$\pm$0.030 \\
4.310  & 3893 & \textit{EP\/}/FXT & 0.5--10 & 0.480$\pm$0.021 \\
5.780  & 3018 & \textit{EP\/}/FXT & 0.5--10 & 0.455$\pm$0.020 \\
7.640  & 1975 & \textit{EP\/}/FXT & 0.5--10 & 0.361$\pm$0.023 \\
9.440  & 2777 & \textit{EP\/}/FXT & 0.5--10 & 0.267$\pm$0.017 \\
\hline
\end{tabular}
\end{table*}

\begin{longtable}{cccccl}
\caption{Optical photometry of GRB~260310A / AT~2026fgk. Uncertainties are quoted at the $1\sigma$ confidence level. The magnitudes are in the AB system and are not corrected for Galactic extinction but are corrected for the host contribution. Times are relative to $T_0$. Where one time is given it is the mid time. Where two times are given, they are the start and end times. We list the first five rows of the table; the full version is available online and in Table~\ref{tab:new_optical_photometry_full}. For COLIBRÍ photometry, the quoted uncertainties include the statistical measurement uncertainty and a 1\% calibration systematic added in quadrature, following the standard COLIBRÍ reduction procedure. For all other instruments, the photometric measurements and their associated uncertainties are adopted directly from the corresponding references.}
\label{tab:new_optical_photometry}\\
\hline
\hline
\multicolumn{1}{c}{Time} & Exposure & Telescope/ Instrument & Filter & Magnitude & Reference \\
\multicolumn{1}{c}{(d)} & (s) & & & & \\
\hline
\endfirsthead

\multicolumn{6}{c}{\tablename\ \thetable\ -- continued from previous page}\\
\hline
\hline
Time & Exposure & Telescope/ Instrument & Filter & Magnitude & Reference \\
\multicolumn{1}{c}{(d)} & (s) & & & & \\
\hline
\endhead

\hline
\multicolumn{6}{r}{Continued on next page}\\
\endfoot

\hline
\hline
\endlastfoot

0.012&---&GOTO&$L$&$18.84 \pm 0.10$&\citet{ONeill2026}\\

0.297&---&ATLAS&$o$&$16.70\pm0.02$&\cite{Smith2020}\\
0.301&---&ATLAS&$o$&$16.71\pm0.02$&\cite{Smith2020}\\
0.307&---&ATLAS&$o$&$16.70\pm0.02$&\cite{Smith2020}\\
0.317&---&ATLAS&$o$&$16.71\pm0.02$&\cite{Smith2020}\\

0.585&---&LAST&---&$17.18\pm0.04$&\cite{GCN43974}\\
\end{longtable}

\begin{longtable}{cccccl}
\caption{Photometry in $J$ of GRB~260310A / AT~2026fgk. Uncertainties are quoted at the $1\sigma$ confidence level. The magnitudes are in the AB system and are not corrected for Galactic extinction or the host contribution. Times are relative to $T_0$ and are mid times. We list the first five rows of the table; the full version is available online and in Table~\ref{tab:new_J_photometry_full}.}

\label{tab:new_J_photometry}\\
\hline
\hline
\multicolumn{1}{c}{Time} & Exposure & Telescope/ Instrument & Filter & Magnitude & Reference \\
\multicolumn{1}{c}{(d)} & (s) & & & & \\
\hline
\endfirsthead

\multicolumn{6}{c}{\tablename\ \thetable\ -- continued from previous page}\\
\hline
\hline
Time & Exposure & Telescope/ Instrument & Filter & Magnitude & Reference \\
\multicolumn{1}{c}{(d)} & (s) & & & & \\
\hline
\endhead

\hline
\multicolumn{6}{r}{Continued on next page}\\
\endfoot

\hline
\hline
\endlastfoot

2.208&--&Palomar/WINTER&$J$&$17.12\pm0.09$&\citet{GCN43996}\\
2.232&--&Palomar/WINTER&$J$&$17.24\pm0.09$&\citet{GCN43996}\\
2.995&--&Palomar/WINTER&$J$&$17.30\pm0.12$&\citet{GCN43996}\\
3.413&7560&SYSU/SYSU&$J$&$17.56\pm0.10$&\citet{GCN43993}\\
4.021&2400&TNG/NICS&$J$&$17.87\pm0.10$&\citet{GCN44004}\\
\end{longtable}

\begin{longtable}{cccccl}
\caption{Additional optical photometry of GRB~260310A / AT~2026fgk. Uncertainties are quoted at the $1\sigma$ confidence level. The magnitudes are in the AB system and are not corrected for Galactic extinction or the host contribution. Times are relative to $T_0$ and are mid times. We list the first five rows of the table; the full version is available online and in Table~\ref{tab:new_extraphotometry_full}.} \label{tab:new_extraphotometry}\\
\hline
\hline
\multicolumn{1}{c}{Time} & Exposure & Telescope/ Instrument & Filter & Magnitude & Reference \\
\multicolumn{1}{c}{(d)} & (s) & & & & \\
\hline
\endfirsthead

\multicolumn{6}{c}{\tablename\ \thetable\ -- continued from previous page}\\
\hline
\hline
Time & Exposure & Telescope/ Instrument & Filter & Magnitude & Reference \\
\multicolumn{1}{c}{(d)} & (s) & & & & \\
\hline
\endhead

\hline
\multicolumn{6}{r}{Continued on next page}\\
\endfoot

\hline
\hline
\endlastfoot
22.047&--& Palomar/SED Machine &$ r $&$ 18.081 \pm 0.040 $& \citet{GCN44096} \\
2.255&--& Palomar/WINTER & H$_s$ & $17.085 \pm 0.210 $& \citet{GCN43996} \\
2.890&--& NOT &$ r $&$ 18.150 \pm 0.010 $& This work\\
2.910&--& NOT &$ g $&$ 18.420 \pm 0.010 $& This work\\
2.910&--& NOT &$ i $&$ 17.850 \pm 0.010 $& This work\\

\end{longtable}

\begin{longtable}{ccccll}
\caption{Radio photometry of GRB~260310A / AT~2026fgk from GCNs (cited in the table) and this work. Uncertainties are quoted at the $1\sigma$ confidence level. We list the first five rows of the table; the full version is available online and in Table~\ref{tab:radio_photometry_full}.}
\label{tab:radio_photometry}\\

\hline
\hline
Time & Exposure & Telescope &  Band & Flux Density & Reference \\
(d) & (hr) & & (GHz) &\multicolumn{1}{c}{($\mu$Jy)} & \\
\hline
\endfirsthead

\multicolumn{6}{c}{\tablename\ \thetable\ -- continued from previous page}\\
\hline
\hline
Time & Exp. & Telescope & Band & Flux Density & Reference \\
(d) & (hr) & & (GHz) & ($\mu$Jy) & \\
\hline
\endhead

\hline
\multicolumn{6}{r}{Continued on next page}\\
\endfoot

\hline
\hline
\endlastfoot

4.260  & --   & VLA & \phantom{0}6.0  & $\phantom{0}3894\pm8$   & \citet{GCN44045} \\
4.260  & --   & VLA & 10.0 & $\phantom{0}6368\pm13$ & \citet{GCN44045} \\
4.260  & --   & VLA & 15.0 & $\phantom{0}9104\pm29$  & \citet{GCN44045} \\

7.344  & -- & OVRO & 15.0 & $18100\pm1600$ & This work\\
7.345  & -- & OVRO & 15.0 & $20600\pm1500$ & This work\\

\end{longtable}

\label{app:filters}
\begin{table}
    \centering
    \caption{Effective wavelengths and corresponding Milky Way extinction along the line of sight of GRB~260310A in each band following \citet{Schlafly+2011} used in this work. \label{tab:extinction}}
    \label{tab:filters}
    \begin{tabular}{lcc}
\toprule
Filter&$\lambda_\mathrm{eff}$ (nm)&$A_{\lambda}$ (mag)\\
\midrule
$L$&554&0.071\\
$w$&590&0.065\\
$o$&686&0.051\\
$g$&488&0.083\\
$r$&620&0.059\\
$i$&752&0.044\\
$z$&866&0.034\\
$y$&971&0.028\\
$J$ & 1250 & 0.021 \\
$Hs$ & 1600 & 0.015 \\
$H$ & 1650 & 0.014 \\
$K$ & 2150 & 0.009 \\

\bottomrule
\end{tabular}

\end{table}

\subsection{MISTRAL campaign}
\begin{figure}
\includegraphics[clip, width=1\linewidth]{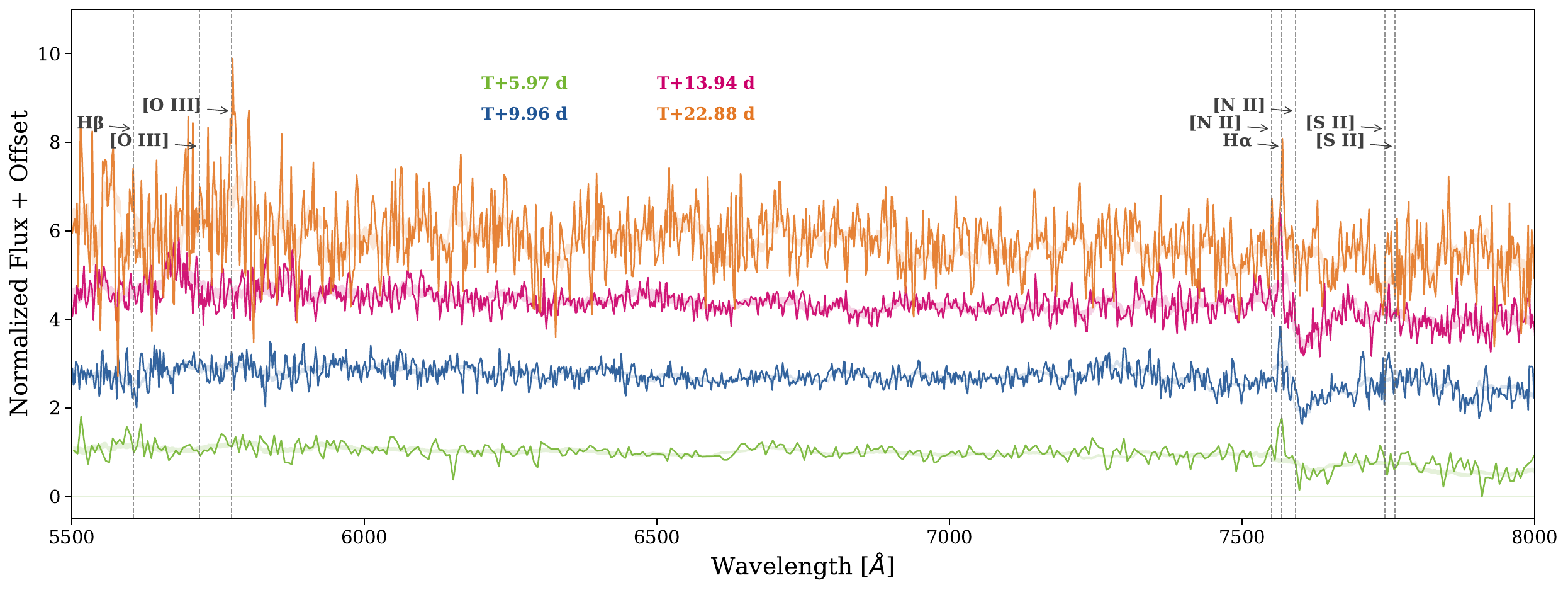}
    \caption{Spectroscopic campaign obtained with the MISTRAL spectro-imager at OHP T193~cm telescope at $T_0+5.97$, $T_0+9.96$, $T_0+13.94$, and $T_0+22.88$~days after the Fermi/GBM trigger. The spectra have been normalized and vertically offset for clarity. For the transient spectra, the solid curves correspond to the cleaned spectra binned in wavelength.
    Vertical dashed lines mark the expected observed-frame positions of prominent nebular emission lines at $z=0.153$, including \ion{O}{II}, \ion{H}{$\beta$}, \ion{O}{III}, \ion{H}{$\alpha$}, \ion{N}{II}, and \ion{S}{II}. Labels and arrows identify selected transitions. The host spectrum reveals clear narrow emission features consistent with active star formation in the proposed host galaxy.
\label{fig:ohp}}
\end{figure}

We also show the spectrum obtained by the MISTRAL team in Fig.~\ref{fig:ohp} (online supplementary material). The details of the campaign are described in Section~\ref{sec:mistral}

\subsection{Radio image}

\begin{figure}
    \centering
     \includegraphics[width=0.48\textwidth]{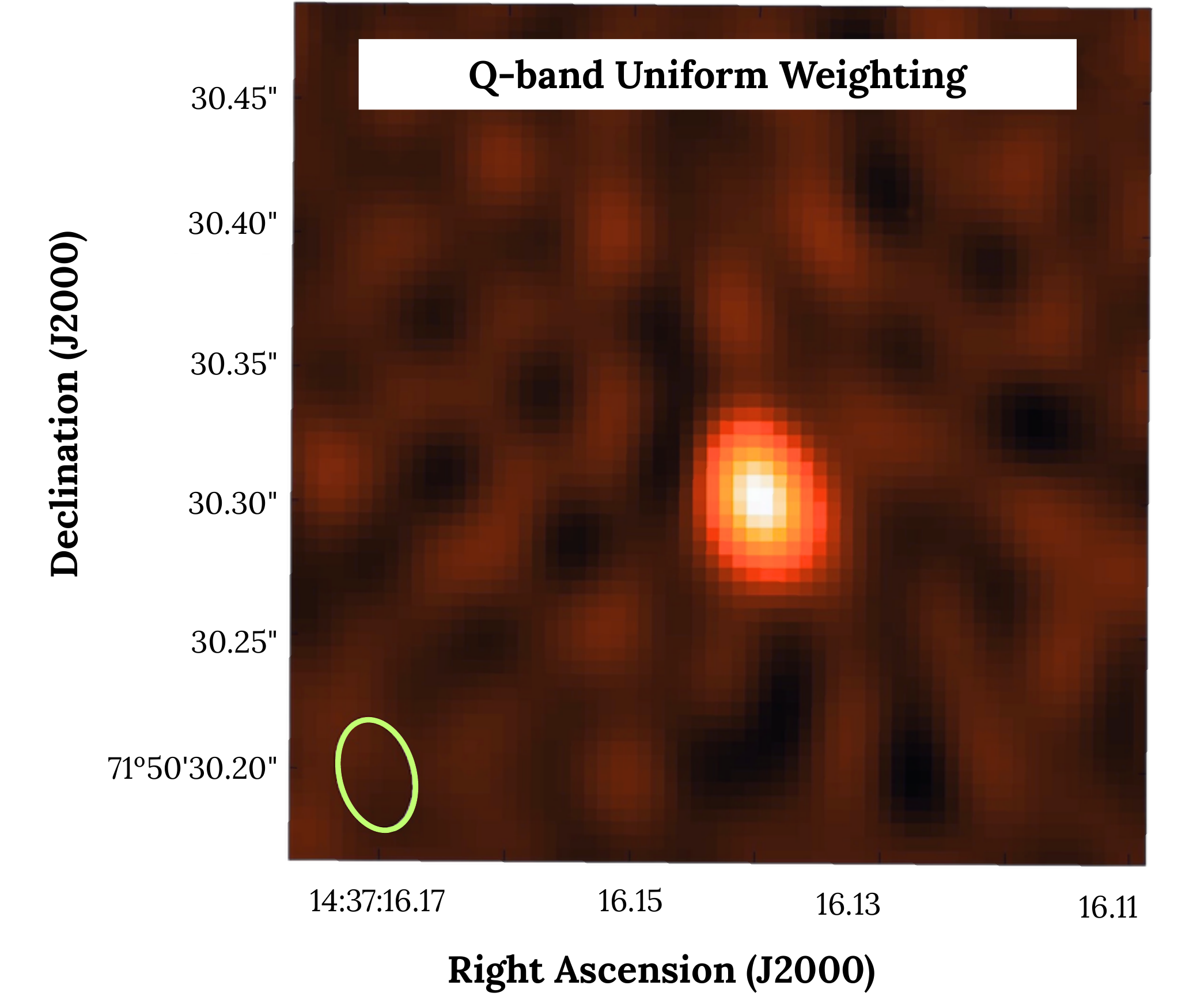}
     \caption{\textbf{45}\,GHz radio image of GRB\,260310A / SN\,2026fgk at $T=T_0+25.2$\,days. The beam size is shown in the bottom-left corner.}
    \label{fig:radio-image}
 \end{figure}

As it was mentioned in Section~\ref{sec:radiolog}. There is a third epoch of observations carried out with the VLA at $T_0+25.2$~d which spanned 1.5-45~GHz \citep{GCN44235}. We illustrate the image obtained at this epoch in Fig.~\ref{fig:radio-image} (online supplementary material).

\section{Fluence and Optical Magnitude Scaling with Redshift}
\label{sec:kcorrection}

The fluence of a source scales with redshift and luminosity distance ($d_L$) for a given $E_{\gamma,\rm iso}$, such that the detector fluence is 
\begin{equation}
    S_{\rm det} = (1+z)\frac{E_{\gamma,\rm iso}}{4\pi k(z)d_L^2}
\end{equation}
where 
\begin{equation}
    k(z) = \frac{\int_{E_1(1+z)^{-1}}^{E_2(1+z)^{-1}}EN(E)dE}{\int_{E_{1,\rm det}}^{E_{2,\rm det}}EN(E)dE}
\end{equation}
is the k-correction for a fixed energy bandwidth of $(E_1,E_2)=(1,10^4)$\,keV in the GRB rest-frame based on the detector bandwidth of $(E_{1,\rm det},E_{2,\rm det})=(10\,{\rm keV},1\,{\rm MeV})$ in the observer frame. The flux density at observer-frame frequency $\nu$ and time T is given by 
\begin{equation}
    F_\nu(\nu,T)=(1+z)\frac{L_{\nu_*}(\nu_*,t_*)}{4\pi d_L^2}
\end{equation}
where $t_* = T/(1+z)$ and $\nu_*=(1+z)\nu$ are the apparent time and frequency in the GRB rest frame, and $L_{\nu_*}\propto \nu_*^\beta$ is the power-law spectral luminosity that has spectral index $\beta$. Let the measured peak flux density be $F_0$ at $\nu=\nu_0$ for a source at the reference redshift of $z=z_0$, corresponding to a luminosity distance $d_L=d_0$. The peak flux density for the same source observed at $\nu=\nu_0$ from any $z>z_0$ is given by 
\begin{equation}
\label{eq:frequrestframe}
    F_\nu(z)=\left(\frac{1+z}{1+z_0}\right)^{1+\beta}\left(\frac{d_0}{d_L}\right)^2F_0
\end{equation}

Note that Eq.~\ref{eq:frequrestframe} does not require a cosmological correction for the peak time in the GRB rest frame since it only cares for the peak flux that is obtained at some redshift corrected rest frame time. This yields the AB magnitude of: 
\begin{equation}
    m_{\rm AB}=-2.5\log_{10}\left[\left(\frac{1+z}{1+z_0}\right)^{1+\beta}\left(\frac{d_0}{d_L}\right)^2\right]+m_0\,,
\end{equation}
where $m_0$ is the actual measured apparent magnitude.




\clearpage

\begin{center}
\Large \textbf{Online Supplementary Material}
\end{center}

\section{Figures}

\begin{figure}
	\includegraphics[clip, trim={0 0 1.3cm 1.3cm},width=\columnwidth]{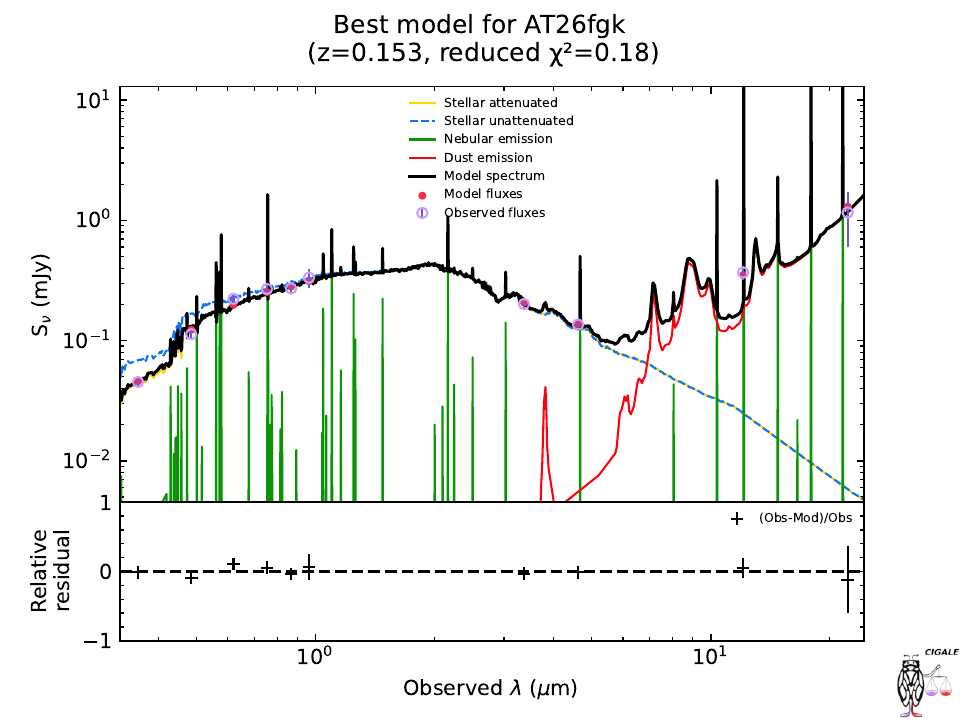}
    \caption{Spectral energy distribution of the host galaxy of GRB\,260310A, used to determine the integrated properties of the host. The solid black curve shows the best-fitting model using CIGALE \citep[see Section~\ref{sec:hg} for details of the fitting procedure and adopted model parameters;][]{Boquien2019}.}
\label{fig:host}
\end{figure}

\section{Tables}
\clearpage
\onecolumn

\begin{table*}
\centering
\caption{X-ray photometry of GRB~260310A / AT~2026fgk from this work. Uncertainties are quoted at the $1\sigma$ confidence level.}
\label{tab:xray_photometry_full}
\begin{tabular}{cccccc}
\hline
\hline
Time & Exposure & Telescope/Instrument & Band & Flux Density \\
(d) & (s) & & (keV) & ($\mu$Jy @1 keV) \\
\hline
2.520  & 3259 & \textit{EP\/}/FXT & 0.5--10 & 0.742$\pm$0.030 \\
4.310  & 3893 & \textit{EP\/}/FXT & 0.5--10 & 0.480$\pm$0.021 \\
5.780  & 3018 & \textit{EP\/}/FXT & 0.5--10 & 0.455$\pm$0.020 \\
7.640  & 1975 & \textit{EP\/}/FXT & 0.5--10 & 0.361$\pm$0.023 \\
9.440  & 2777 & \textit{EP\/}/FXT & 0.5--10 & 0.267$\pm$0.017 \\
11.040 & 2689 & \textit{EP\/}/FXT & 0.5--10 & 0.221$\pm$0.016 \\
15.160 & 4876 & \textit{EP\/}/FXT & 0.5--10 & 0.097$\pm$0.008 \\
17.290 & 2918 & \textit{EP\/}/FXT & 0.5--10 & 0.081$\pm$0.005 \\
18.290 & 9019 & \textit{EP\/}/FXT & 0.5--10 & 0.078$\pm$0.005 \\
23.680 & 3285 & \textit{EP\/}/FXT & 0.5--10 & 0.080$\pm$0.008 \\
25.340 & 3328 & \textit{EP\/}/FXT & 0.5--10 & 0.074$\pm$0.008 \\
26.940 & 961  & \textit{EP\/}/FXT & 0.5--10 & 0.089$\pm$0.009 \\
29.540 & 2649 & \textit{EP\/}/FXT & 0.5--10 & 0.107$\pm$0.011 \\
32.403 & 4159 & \textit{EP\/}/FXT & 0.5--10 & 0.070$\pm$0.007 \\
39.067 & 2704 & \textit{EP\/}/FXT & 0.5--10 & 0.051$\pm$0.007 \\
40.530 & 6244 & \textit{EP\/}/FXT & 0.5--10 & 0.062$\pm$0.009 \\
41.950 & 3245 & \textit{EP\/}/FXT & 0.5--10 & 0.052$\pm$0.006 \\
43.990 & 2864 & \textit{EP\/}/FXT & 0.5--10 & 0.049$\pm$0.007 \\
45.980 & 2995 & \textit{EP\/}/FXT & 0.5--10 & 0.045$\pm$0.006 \\
46.120 & 9177 & \textit{EP\/}/FXT & 0.5--10 & 0.044$\pm$0.003 \\
\hline
\end{tabular}
\end{table*}

\begin{longtable}{cccccl}
\caption{Optical photometry of GRB~260310A / AT~2026fgk. Uncertainties are quoted at the $1\sigma$ confidence level. The magnitudes are in the AB system and are not corrected for Galactic extinction but are corrected for the host contribution. Times are relative to $T_0$. Where one time is given it is the mid time. Where two times are given, they are the start and end times.}
\label{tab:new_optical_photometry_full}\\
\hline
\hline
\multicolumn{1}{c}{Time} & Exposure & Telescope/ Instrument & Filter & Magnitude & Reference \\
\multicolumn{1}{c}{(d)} & (s) & & & & \\
\hline
\endfirsthead

\multicolumn{6}{c}{\tablename\ \thetable\ -- continued from previous page}\\
\hline
\hline
Time & Exposure & Telescope/ Instrument & Filter & Magnitude & Reference \\
\multicolumn{1}{c}{(d)} & (s) & & & & \\
\hline
\endhead

\hline
\multicolumn{6}{r}{Continued on next page}\\
\endfoot

\hline
\hline
\endlastfoot

0.012&---&GOTO&$L$&$18.84 \pm 0.10$&\citet{ONeill2026}\\

0.297&---&ATLAS&$o$&$16.70\pm0.02$&\cite{Smith2020}\\
0.301&---&ATLAS&$o$&$16.71\pm0.02$&\cite{Smith2020}\\
0.307&---&ATLAS&$o$&$16.70\pm0.02$&\cite{Smith2020}\\
0.317&---&ATLAS&$o$&$16.71\pm0.02$&\cite{Smith2020}\\

0.585&---&LAST&---&$17.18\pm0.04$&\cite{GCN43974}\\
0.590&---&LAST&---&$17.17\pm0.04$&\cite{GCN43974}\\
0.596&---&LAST&---&$17.09\pm0.04$&\cite{GCN43974}\\
0.600&---&LAST&---&$17.16\pm0.04$&\cite{GCN43974}\\
0.606&---&LAST&---&$17.10\pm0.04$&\cite{GCN43974}\\
0.611&---&LAST&---&$17.11\pm0.04$&\cite{GCN43974}\\
0.749&---&LAST&---&$17.23\pm0.04$&\cite{GCN43974}\\
0.792&---&LAST&---&$17.25\pm0.04$&\cite{GCN43974}\\
0.835&---&LAST&---&$17.19\pm0.04$&\cite{GCN43974}\\
0.879&---&LAST&---&$17.30\pm0.04$&\cite{GCN43974}\\
0.917&---&LAST&---&$17.32\pm0.04$&\cite{GCN43974}\\

2.903--2.926&\phantom{0}600&COLIBRÍ/DDRAGO&$g$&$18.53\pm0.01$&This work\\
2.903--2.926&\phantom{0}600&COLIBRÍ/DDRAGO&$z$&$17.76\pm0.01$&This work\\
2.904--2.926&\phantom{0}600&COLIBRÍ/DDRAGO&$r$&$18.20\pm0.01$&This work\\
2.904--2.927&\phantom{}1200&COLIBRÍ/DDRAGO&$y$&$17.62\pm0.01$&This work\\
2.905--2.927&\phantom{0}600&COLIBRÍ/DDRAGO&$i$&$17.99\pm0.01$&This work\\
3.006--3.034&\phantom{0}960&COLIBRÍ/DDRAGO&$g$&$18.53\pm0.01$&This work\\
3.006--3.035&\phantom{}1920&COLIBRÍ/DDRAGO&$z$&$17.75\pm0.01$&This work\\
3.007--3.035&\phantom{0}960&COLIBRÍ/DDRAGO&$r$&$18.20\pm0.01$&This work\\
3.294--3.324&\phantom{}1020&COLIBRÍ/DDRAGO&$g$&$18.70\pm0.01$&This work\\
3.294--3.324&\phantom{}2040&COLIBRÍ/DDRAGO&$z$&$17.91\pm0.01$&This work\\
3.295--3.324&\phantom{}1020&COLIBRÍ/DDRAGO&$r$&$18.37\pm0.01$&This work\\
3.909--3.921&\phantom{0}480&COLIBRÍ/DDRAGO&$g$&$18.81\pm0.03$&This work\\
3.909--3.922&\phantom{0}960&COLIBRÍ/DDRAGO&$z$&$18.04\pm0.03$&This work\\
3.910--3.922&\phantom{0}480&COLIBRÍ/DDRAGO&$r$&$18.47\pm0.03$&This work\\
4.001--4.013&\phantom{0}480&COLIBRÍ/DDRAGO&$g$&$18.87\pm0.01$&This work\\
4.001--4.014&\phantom{0}960&COLIBRÍ/DDRAGO&$z$&$18.09\pm0.01$&This work\\
4.002--4.014&\phantom{0}480&COLIBRÍ/DDRAGO&$r$&$18.54\pm0.01$&This work\\
4.096--4.107&\phantom{0}480&COLIBRÍ/DDRAGO&$g$&$18.89\pm0.01$&This work\\
4.096--4.108&\phantom{0}480&COLIBRÍ/DDRAGO&$r$&$18.58\pm0.01$&This work\\
4.096--4.108&\phantom{0}960&COLIBRÍ/DDRAGO&$z$&$18.14\pm0.01$&This work\\
4.176--4.187&\phantom{0}480&COLIBRÍ/DDRAGO&$g$&$18.88\pm0.01$&This work\\
4.176--4.188&\phantom{0}480&COLIBRÍ/DDRAGO&$r$&$18.58\pm0.01$&This work\\
4.176--4.188&\phantom{0}960&COLIBRÍ/DDRAGO&$z$&$18.13\pm0.02$&This work\\
4.277--4.297&\phantom{0}480&COLIBRÍ/DDRAGO&$r$&$18.52\pm0.03$&This work\\
4.288--4.296&\phantom{0}360&COLIBRÍ/DDRAGO&$g$&$18.89\pm0.03$&This work\\
4.288--4.297&\phantom{0}720&COLIBRÍ/DDRAGO&$z$&$18.11\pm0.03$&This work\\
4.904--4.946&\phantom{}1440&COLIBRÍ/DDRAGO&$g$&$18.89\pm0.01$&This work\\
4.904--4.946&\phantom{}2760&COLIBRÍ/DDRAGO&$z$&$18.13\pm0.02$&This work\\
4.905--4.947&\phantom{}1440&COLIBRÍ/DDRAGO&$r$&$18.61\pm0.01$&This work\\
5.031--5.059&\phantom{0}960&COLIBRÍ/DDRAGO&$g$&$18.90\pm0.01$&This work\\
5.031--5.060&\phantom{}1920&COLIBRÍ/DDRAGO&$z$&$18.18\pm0.01$&This work\\
5.032--5.060&\phantom{0}960&COLIBRÍ/DDRAGO&$r$&$18.61\pm0.01$&This work\\
5.263--5.291&\phantom{0}960&COLIBRÍ/DDRAGO&$g$&$18.92\pm0.01$&This work\\
5.263--5.292&\phantom{}1920&COLIBRÍ/DDRAGO&$z$&$18.18\pm0.01$&This work\\
5.264--5.292&\phantom{0}960&COLIBRÍ/DDRAGO&$r$&$18.61\pm0.01$&This work\\
5.905--5.916&\phantom{0}480&COLIBRÍ/DDRAGO&$g$&$18.94\pm0.03$&This work\\
5.905--5.917&\phantom{0}840&COLIBRÍ/DDRAGO&$z$&$18.21\pm0.03$&This work\\
5.906--5.917&\phantom{0}420&COLIBRÍ/DDRAGO&$r$&$18.67\pm0.02$&This work\\
5.988--6.000&\phantom{0}480&COLIBRÍ/DDRAGO&$g$&$18.98\pm0.01$&This work\\
5.988--6.000&\phantom{0}960&COLIBRÍ/DDRAGO&$z$&$18.26\pm0.01$&This work\\
5.989--6.000&\phantom{0}480&COLIBRÍ/DDRAGO&$r$&$18.66\pm0.01$&This work\\
6.180--6.192&\phantom{0}480&COLIBRÍ/DDRAGO&$g$&$19.00\pm0.01$&This work\\
6.180--6.193&\phantom{0}960&COLIBRÍ/DDRAGO&$z$&$18.29\pm0.01$&This work\\
6.181--6.193&\phantom{0}480&COLIBRÍ/DDRAGO&$r$&$18.69\pm0.01$&This work\\
6.264--6.275&\phantom{0}480&COLIBRÍ/DDRAGO&$g$&$19.03\pm0.01$&This work\\
6.264--6.276&\phantom{0}480&COLIBRÍ/DDRAGO&$r$&$18.73\pm0.01$&This work\\
6.264--6.276&\phantom{0}960&COLIBRÍ/DDRAGO&$z$&$18.30\pm0.01$&This work\\
6.905--6.930&\phantom{0}960&COLIBRÍ/DDRAGO&$g$&$19.07\pm0.01$&This work\\
6.905--6.931&\phantom{}1920&COLIBRÍ/DDRAGO&$z$&$18.34\pm0.01$&This work\\
6.906--6.931&\phantom{0}960&COLIBRÍ/DDRAGO&$r$&$18.77\pm0.01$&This work\\
7.049--7.060&\phantom{0}480&COLIBRÍ/DDRAGO&$g$&$19.12\pm0.01$&This work\\
7.049--7.061&\phantom{0}960&COLIBRÍ/DDRAGO&$z$&$18.39\pm0.02$&This work\\
7.050--7.061&\phantom{0}480&COLIBRÍ/DDRAGO&$r$&$18.85\pm0.01$&This work\\
7.226--7.238&\phantom{0}480&COLIBRÍ/DDRAGO&$g$&$19.14\pm0.01$&This work\\
7.226--7.239&\phantom{0}960&COLIBRÍ/DDRAGO&$z$&$18.42\pm0.01$&This work\\
7.227--7.239&\phantom{0}480&COLIBRÍ/DDRAGO&$r$&$18.81\pm0.01$&This work\\
7.273--7.285&\phantom{0}480&COLIBRÍ/DDRAGO&$g$&$19.14\pm0.01$&This work\\
7.273--7.286&\phantom{0}960&COLIBRÍ/DDRAGO&$z$&$18.42\pm0.01$&This work\\
7.274--7.286&\phantom{0}480&COLIBRÍ/DDRAGO&$r$&$18.82\pm0.01$&This work\\
7.905--7.928&\phantom{0}600&COLIBRÍ/DDRAGO&$g$&$19.14\pm0.01$&This work\\
7.906--7.928&\phantom{0}600&COLIBRÍ/DDRAGO&$r$&$18.83\pm0.01$&This work\\
7.907--7.929&\phantom{0}600&COLIBRÍ/DDRAGO&$i$&$18.67\pm0.01$&This work\\
8.086--8.097&\phantom{0}480&COLIBRÍ/DDRAGO&$g$&$19.18\pm0.01$&This work\\
8.086--8.098&\phantom{0}960&COLIBRÍ/DDRAGO&$z$&$18.47\pm0.01$&This work\\
8.087--8.098&\phantom{0}480&COLIBRÍ/DDRAGO&$r$&$18.85\pm0.01$&This work\\
8.906--9.091&\phantom{}3960&COLIBRÍ/DDRAGO&$g$&$19.37\pm0.01$&This work\\
8.906--9.092&\phantom{}4320&COLIBRÍ/DDRAGO&$z$&$18.65\pm0.01$&This work\\
8.907--9.079&\phantom{}7200&COLIBRÍ/DDRAGO&$y$&$18.48\pm0.01$&This work\\
8.907--9.092&\phantom{}3960&COLIBRÍ/DDRAGO&$r$&$19.04\pm0.01$&This work\\
8.908--9.079&\phantom{}3600&COLIBRÍ/DDRAGO&$i$&$18.88\pm0.01$&This work\\
9.136--9.268&\phantom{}4320&COLIBRÍ/DDRAGO&$g$&$19.39\pm0.01$&This work\\
9.136--9.269&\phantom{}8640&COLIBRÍ/DDRAGO&$z$&$18.66\pm0.01$&This work\\
9.137--9.269&\phantom{}4320&COLIBRÍ/DDRAGO&$r$&$19.06\pm0.01$&This work\\
9.944--9.955&\phantom{0}480&COLIBRÍ/DDRAGO&$g$&$19.49\pm0.01$&This work\\
9.944--9.956&\phantom{0}480&COLIBRÍ/DDRAGO&$r$&$19.15\pm0.01$&This work\\
9.944--9.956&\phantom{0}960&COLIBRÍ/DDRAGO&$z$&$18.85\pm0.02$&This work\\
10.062--10.087&\phantom{0}960&COLIBRÍ/DDRAGO&$g$&$19.45\pm0.01$&This work\\
10.062--10.088&\phantom{}1920&COLIBRÍ/DDRAGO&$z$&$18.84\pm0.02$&This work\\
10.063--10.088&\phantom{0}960&COLIBRÍ/DDRAGO&$r$&$19.20\pm0.01$&This work\\
10.150--10.269&\phantom{}3900&COLIBRÍ/DDRAGO&$g$&$19.48\pm0.01$&This work\\
10.150--10.270&\phantom{}3840&COLIBRÍ/DDRAGO&$r$&$19.31\pm0.01$&This work\\
10.150--10.270&\phantom{}7740&COLIBRÍ/DDRAGO&$z$&$18.86\pm0.01$&This work\\
10.911--10.990&\phantom{}1800&COLIBRÍ/DDRAGO&$g$&$19.57\pm0.01$&This work\\
10.911--10.990&\phantom{}1800&COLIBRÍ/DDRAGO&$z$&$18.91\pm0.02$&This work\\
10.912--10.991&\phantom{}1800&COLIBRÍ/DDRAGO&$r$&$19.27\pm0.01$&This work\\
10.912--10.992&\phantom{}3600&COLIBRÍ/DDRAGO&$y$&$18.73\pm0.03$&This work\\
10.913--10.992&\phantom{}1800&COLIBRÍ/DDRAGO&$i$&$19.15\pm0.01$&This work\\
11.088--11.100&\phantom{0}480&COLIBRÍ/DDRAGO&$g$&$19.57\pm0.01$&This work\\
11.088--11.101&\phantom{0}960&COLIBRÍ/DDRAGO&$z$&$18.96\pm0.02$&This work\\
11.089--11.101&\phantom{0}480&COLIBRÍ/DDRAGO&$r$&$19.28\pm0.01$&This work\\
11.907--11.950&\phantom{}1440&COLIBRÍ/DDRAGO&$g$&$19.68\pm0.01$&This work\\
11.907--11.951&\phantom{}2880&COLIBRÍ/DDRAGO&$z$&$19.03\pm0.02$&This work\\
11.908--11.951&\phantom{}1440&COLIBRÍ/DDRAGO&$r$&$19.42\pm0.01$&This work\\
12.038--12.066&\phantom{0}960&COLIBRÍ/DDRAGO&$g$&$19.71\pm0.01$&This work\\
12.038--12.067&\phantom{}1920&COLIBRÍ/DDRAGO&$z$&$19.08\pm0.01$&This work\\
12.039--12.067&\phantom{0}960&COLIBRÍ/DDRAGO&$r$&$19.39\pm0.01$&This work\\
12.192--12.258&\phantom{}1500&COLIBRÍ/DDRAGO&$g$&$19.74\pm0.01$&This work\\
12.192--12.258&\phantom{}1500&COLIBRÍ/DDRAGO&$z$&$19.03\pm0.02$&This work\\
12.192--12.259&\phantom{}1500&COLIBRÍ/DDRAGO&$r$&$19.40\pm0.01$&This work\\
12.192--12.259&\phantom{}3000&COLIBRÍ/DDRAGO&$y$&$18.84\pm0.02$&This work\\
12.193--12.259&\phantom{}1500&COLIBRÍ/DDRAGO&$i$&$19.29\pm0.01$&This work\\
12.916--12.941&\phantom{0}960&COLIBRÍ/DDRAGO&$g$&$19.85\pm0.02$&This work\\
12.916--12.942&\phantom{0}960&COLIBRÍ/DDRAGO&$r$&$19.56\pm0.02$&This work\\
12.916--12.942&\phantom{}1920&COLIBRÍ/DDRAGO&$z$&$19.29\pm0.03$&This work\\
13.036--13.061&\phantom{0}960&COLIBRÍ/DDRAGO&$g$&$19.87\pm0.01$&This work\\
13.036--13.062&\phantom{}1920&COLIBRÍ/DDRAGO&$z$&$19.25\pm0.01$&This work\\
13.037--13.062&\phantom{0}960&COLIBRÍ/DDRAGO&$r$&$19.54\pm0.01$&This work\\
13.921--13.992&\phantom{}1440&COLIBRÍ/DDRAGO&$g$&$19.92\pm0.02$&This work\\
13.921--13.993&\phantom{}2880&COLIBRÍ/DDRAGO&$z$&$19.39\pm0.02$&This work\\
13.922--13.993&\phantom{}1440&COLIBRÍ/DDRAGO&$r$&$19.67\pm0.01$&This work\\
14.009--14.050&\phantom{0}960&COLIBRÍ/DDRAGO&$g$&$19.98\pm0.02$&This work\\
14.009--14.051&\phantom{}1920&COLIBRÍ/DDRAGO&$z$&$19.37\pm0.02$&This work\\
14.010--14.051&\phantom{0}960&COLIBRÍ/DDRAGO&$r$&$19.62\pm0.02$&This work\\
14.205--14.243&\phantom{0}900&COLIBRÍ/DDRAGO&$g$&$19.99\pm0.01$&This work\\
14.205--14.245&\phantom{}2700&COLIBRÍ/DDRAGO&$z$&$19.32\pm0.01$&This work\\
14.206--14.244&\phantom{0}900&COLIBRÍ/DDRAGO&$r$&$19.64\pm0.01$&This work\\
14.207--14.245&\phantom{0}900&COLIBRÍ/DDRAGO&$i$&$19.54\pm0.01$&This work\\
14.908--14.963&\phantom{}1920&COLIBRÍ/DDRAGO&$g$&$20.05\pm0.02$&This work\\
14.908--14.964&\phantom{}3840&COLIBRÍ/DDRAGO&$z$&$19.54\pm0.02$&This work\\
14.909--14.964&\phantom{}1920&COLIBRÍ/DDRAGO&$r$&$19.77\pm0.02$&This work\\
15.072--15.097&\phantom{0}960&COLIBRÍ/DDRAGO&$g$&$20.12\pm0.02$&This work\\
15.072--15.098&\phantom{}1920&COLIBRÍ/DDRAGO&$z$&$19.58\pm0.02$&This work\\
15.073--15.098&\phantom{0}960&COLIBRÍ/DDRAGO&$r$&$19.77\pm0.01$&This work\\
15.192--15.230&\phantom{0}900&COLIBRÍ/DDRAGO&$g$&$20.11\pm0.02$&This work\\
15.192--15.232&\phantom{}2700&COLIBRÍ/DDRAGO&$z$&$19.51\pm0.02$&This work\\
15.193--15.231&\phantom{0}900&COLIBRÍ/DDRAGO&$r$&$19.82\pm0.02$&This work\\
15.194--15.232&\phantom{0}900&COLIBRÍ/DDRAGO&$i$&$19.70\pm0.02$&This work\\
16.910--16.957&\phantom{}1080&COLIBRÍ/DDRAGO&$g$&$20.46\pm0.03$&This work\\
16.910--16.959&\phantom{}3240&COLIBRÍ/DDRAGO&$z$&$19.83\pm0.02$&This work\\
16.911--16.958&\phantom{}1080&COLIBRÍ/DDRAGO&$r$&$20.00\pm0.02$&This work\\
16.911--16.959&\phantom{}1080&COLIBRÍ/DDRAGO&$i$&$19.88\pm0.02$&This work\\
17.215--17.256&\phantom{0}900&COLIBRÍ/DDRAGO&$g$&$20.42\pm0.02$&This work\\
17.215--17.258&\phantom{}2700&COLIBRÍ/DDRAGO&$z$&$19.94\pm0.02$&This work\\
17.216--17.257&\phantom{0}900&COLIBRÍ/DDRAGO&$r$&$20.06\pm0.02$&This work\\
17.216--17.258&\phantom{0}900&COLIBRÍ/DDRAGO&$i$&$19.95\pm0.02$&This work\\
17.910--18.080&\phantom{}1560&COLIBRÍ/DDRAGO&$g$&$20.48\pm0.04$&This work\\
17.910--18.081&\phantom{}4680&COLIBRÍ/DDRAGO&$z$&$19.91\pm0.03$&This work\\
17.911--18.080&\phantom{}1560&COLIBRÍ/DDRAGO&$r$&$20.12\pm0.03$&This work\\
17.912--18.081&\phantom{}1560&COLIBRÍ/DDRAGO&$i$&$20.01\pm0.03$&This work\\
18.969--19.203&\phantom{}1380&COLIBRÍ/DDRAGO&$g$&$20.92\pm0.31$&This work\\
18.969--19.203&\phantom{}4140&COLIBRÍ/DDRAGO&$z$&$20.56\pm0.26$&This work\\
18.970--19.204&\phantom{}1440&COLIBRÍ/DDRAGO&$r$&$20.24\pm0.17$&This work\\
18.971--19.203&\phantom{}1440&COLIBRÍ/DDRAGO&$i$&$19.96\pm0.18$&This work\\
19.971--20.052&\phantom{}1800&COLIBRÍ/DDRAGO&$g$&$20.63\pm0.04$&This work\\
19.971--20.052&\phantom{}5340&COLIBRÍ/DDRAGO&$z$&$20.00\pm0.02$&This work\\
19.972--20.052&\phantom{}1800&COLIBRÍ/DDRAGO&$r$&$20.15\pm0.02$&This work\\
19.973--20.051&\phantom{}1740&COLIBRÍ/DDRAGO&$i$&$20.03\pm0.02$&This work\\
22.038--22.152&\phantom{}1500&COLIBRÍ/DDRAGO&$g$&$20.49\pm0.08$&This work\\
22.038--22.152&\phantom{}4440&COLIBRÍ/DDRAGO&$z$&$20.21\pm0.05$&This work\\
22.039--22.156&\phantom{}1500&COLIBRÍ/DDRAGO&$r$&$20.23\pm0.06$&This work\\
22.040--22.151&\phantom{}1560&COLIBRÍ/DDRAGO&$i$&$20.05\pm0.05$&This work\\
23.051--23.158&\phantom{}2220&COLIBRÍ/DDRAGO&$g$&$20.79\pm0.04$&This work\\
23.051--23.159&\phantom{}6840&COLIBRÍ/DDRAGO&$z$&$20.00\pm0.02$&This work\\
23.052--23.157&\phantom{}2280&COLIBRÍ/DDRAGO&$i$&$20.06\pm0.02$&This work\\
23.052--23.159&\phantom{}2340&COLIBRÍ/DDRAGO&$r$&$20.21\pm0.02$&This work\\
24.112--24.191&\phantom{}1800&COLIBRÍ/DDRAGO&$g$&$20.82\pm0.03$&This work\\
24.112--24.192&\phantom{}5340&COLIBRÍ/DDRAGO&$z$&$19.97\pm0.01$&This work\\
24.113--24.192&\phantom{}1800&COLIBRÍ/DDRAGO&$r$&$20.19\pm0.02$&This work\\
24.114--24.190&\phantom{}1740&COLIBRÍ/DDRAGO&$i$&$20.00\pm0.02$&This work\\
25.076--25.167&\phantom{}1980&COLIBRÍ/DDRAGO&$g$&$20.65\pm0.04$&This work\\
25.076--25.167&\phantom{}5760&COLIBRÍ/DDRAGO&$z$&$19.95\pm0.02$&This work\\
25.077--25.166&\phantom{}1920&COLIBRÍ/DDRAGO&$r$&$20.22\pm0.03$&This work\\
25.078--25.167&\phantom{}1860&COLIBRÍ/DDRAGO&$i$&$20.03\pm0.03$&This work\\
26.021--26.058&\phantom{0}900&COLIBRÍ/DDRAGO&$g$&$20.68\pm0.04$&This work\\
26.021--26.059&\phantom{0}900&COLIBRÍ/DDRAGO&$r$&$20.20\pm0.02$&This work\\
26.021--26.059&\phantom{}2640&COLIBRÍ/DDRAGO&$z$&$19.94\pm0.02$&This work\\
26.022--26.058&\phantom{0}840&COLIBRÍ/DDRAGO&$i$&$20.04\pm0.03$&This work\\
26.958--27.224&\phantom{}2220&COLIBRÍ/DDRAGO&$r$&$20.29\pm0.03$&This work\\
27.001--27.223&\phantom{}2100&COLIBRÍ/DDRAGO&$g$&$20.73\pm0.04$&This work\\
27.048--27.224&\phantom{}6000&COLIBRÍ/DDRAGO&$z$&$19.94\pm0.03$&This work\\
27.049--27.224&\phantom{}1980&COLIBRÍ/DDRAGO&$i$&$20.14\pm0.03$&This work\\
28.098--28.197&\phantom{}2220&COLIBRÍ/DDRAGO&$g$&$20.78\pm0.02$&This work\\
28.098--28.199&\phantom{}6660&COLIBRÍ/DDRAGO&$z$&$19.93\pm0.02$&This work\\
28.099--28.198&\phantom{}2220&COLIBRÍ/DDRAGO&$r$&$20.26\pm0.01$&This work\\
28.100--28.199&\phantom{}2220&COLIBRÍ/DDRAGO&$i$&$20.07\pm0.02$&This work\\
29.105--29.252&\phantom{}3240&COLIBRÍ/DDRAGO&$g$&$20.84\pm0.01$&This work\\
29.105--29.254&\phantom{}9720&COLIBRÍ/DDRAGO&$z$&$19.87\pm0.01$&This work\\
29.106--29.253&\phantom{}3240&COLIBRÍ/DDRAGO&$r$&$20.33\pm0.01$&This work\\
29.106--29.254&\phantom{}3240&COLIBRÍ/DDRAGO&$i$&$20.10\pm0.01$&This work\\
30.092--30.242&\phantom{}3300&COLIBRÍ/DDRAGO&$g$&$20.91\pm0.01$&This work\\
30.092--30.242&\phantom{}9780&COLIBRÍ/DDRAGO&$z$&$19.96\pm0.01$&This work\\
30.093--30.240&\phantom{}3240&COLIBRÍ/DDRAGO&$r$&$20.41\pm0.01$&This work\\
30.094--30.241&\phantom{}3240&COLIBRÍ/DDRAGO&$i$&$20.16\pm0.01$&This work\\
31.090--31.244&\phantom{0}420&COLIBRÍ/DDRAGO&$g$&$20.84\pm0.13$&This work\\
31.090--31.245&\phantom{}1260&COLIBRÍ/DDRAGO&$z$&$20.39\pm0.16$&This work\\
31.091--31.244&\phantom{0}540&COLIBRÍ/DDRAGO&$r$&$20.94\pm0.19$&This work\\
31.223--31.245&\phantom{0}420&COLIBRÍ/DDRAGO&$i$&$20.35\pm0.13$&This work\\
32.174--32.242&\phantom{}1500&COLIBRÍ/DDRAGO&$g$&$20.87\pm0.02$&This work\\
32.174--32.243&\phantom{}4440&COLIBRÍ/DDRAGO&$z$&$20.26\pm0.03$&This work\\
32.175--32.243&\phantom{}1500&COLIBRÍ/DDRAGO&$r$&$20.63\pm0.02$&This work\\
32.176--32.241&\phantom{}1440&COLIBRÍ/DDRAGO&$i$&$20.38\pm0.03$&This work\\
33.096--33.244&\phantom{}2820&COLIBRÍ/DDRAGO&$g$&$20.99\pm0.02$&This work\\
33.096--33.245&\phantom{}2820&COLIBRÍ/DDRAGO&$r$&$20.63\pm0.02$&This work\\
33.096--33.246&\phantom{}8400&COLIBRÍ/DDRAGO&$z$&$20.26\pm0.03$&This work\\
33.097--33.246&\phantom{}2700&COLIBRÍ/DDRAGO&$i$&$20.42\pm0.02$&This work\\
34.085--34.220&\phantom{}2580&COLIBRÍ/DDRAGO&$g$&$21.01\pm0.03$&This work\\
34.086--34.221&\phantom{}2760&COLIBRÍ/DDRAGO&$r$&$20.70\pm0.03$&This work\\
34.086--34.222&\phantom{}7020&COLIBRÍ/DDRAGO&$z$&$20.40\pm0.04$&This work\\
34.087--34.222&\phantom{}2400&COLIBRÍ/DDRAGO&$i$&$20.43\pm0.03$&This work\\
35.079--35.216&\phantom{}3000&COLIBRÍ/DDRAGO&$g$&$21.13\pm0.02$&This work\\
35.079--35.218&\phantom{}9000&COLIBRÍ/DDRAGO&$z$&$20.30\pm0.02$&This work\\
35.080--35.217&\phantom{}3000&COLIBRÍ/DDRAGO&$r$&$20.73\pm0.01$&This work\\
35.081--35.218&\phantom{}3000&COLIBRÍ/DDRAGO&$i$&$20.51\pm0.02$&This work\\
36.041--36.179&\phantom{}3000&COLIBRÍ/DDRAGO&$g$&$21.25\pm0.01$&This work\\
36.041--36.180&\phantom{}3000&COLIBRÍ/DDRAGO&$r$&$20.79\pm0.01$&This work\\
36.041--36.181&\phantom{}9000&COLIBRÍ/DDRAGO&$z$&$20.38\pm0.02$&This work\\
36.042--36.181&\phantom{}3000&COLIBRÍ/DDRAGO&$i$&$20.57\pm0.02$&This work\\
37.084--37.137&\phantom{}3120&COLIBRÍ/DDRAGO&$z$&$21.12\pm0.48$&This work\\
37.084--37.141&\phantom{}1140&COLIBRÍ/DDRAGO&$g$&$20.98\pm0.16$&This work\\
37.085--37.142&\phantom{}1080&COLIBRÍ/DDRAGO&$r$&$20.72\pm0.17$&This work\\
37.086--37.143&\phantom{}1140&COLIBRÍ/DDRAGO&$i$&$20.77\pm0.29$&This work\\
38.073--38.165&\phantom{}2040&COLIBRÍ/DDRAGO&$g$&$21.32\pm0.02$&This work\\
38.073--38.166&\phantom{}6120&COLIBRÍ/DDRAGO&$z$&$20.50\pm0.02$&This work\\
38.074--38.166&\phantom{}2040&COLIBRÍ/DDRAGO&$r$&$20.90\pm0.02$&This work\\
38.075--38.166&\phantom{}2040&COLIBRÍ/DDRAGO&$i$&$20.68\pm0.02$&This work\\
39.090--39.279&\phantom{}3540&COLIBRÍ/DDRAGO&$g$&$21.35\pm0.02$&This work\\
39.090--39.281&\phantom{}10620&COLIBRÍ/DDRAGO&$z$&$20.51\pm0.02$&This work\\
39.091--39.280&\phantom{}3540&COLIBRÍ/DDRAGO&$r$&$20.98\pm0.02$&This work\\
39.092--39.281&\phantom{}3540&COLIBRÍ/DDRAGO&$i$&$20.74\pm0.02$&This work\\
40.087--40.280&\phantom{}2280&COLIBRÍ/DDRAGO&$g$&$21.29\pm0.03$&This work\\
40.087--40.281&\phantom{}6180&COLIBRÍ/DDRAGO&$z$&$20.63\pm0.04$&This work\\
40.088--40.280&\phantom{}2220&COLIBRÍ/DDRAGO&$r$&$21.06\pm0.03$&This work\\
40.088--40.281&\phantom{}2280&COLIBRÍ/DDRAGO&$i$&$20.82\pm0.04$&This work\\
41.084--41.186&\phantom{}2220&COLIBRÍ/DDRAGO&$g$&$21.36\pm0.02$&This work\\
41.084--41.187&\phantom{}2220&COLIBRÍ/DDRAGO&$r$&$21.04\pm0.02$&This work\\
41.084--41.187&\phantom{}6600&COLIBRÍ/DDRAGO&$z$&$20.59\pm0.03$&This work\\
41.085--41.185&\phantom{}2160&COLIBRÍ/DDRAGO&$i$&$20.91\pm0.03$&This work\\
43.166--43.285&\phantom{}2640&COLIBRÍ/DDRAGO&$g$&$21.50\pm0.02$&This work\\
43.166--43.285&\phantom{}7800&COLIBRÍ/DDRAGO&$z$&$20.86\pm0.03$&This work\\
43.167--43.283&\phantom{}2580&COLIBRÍ/DDRAGO&$r$&$21.13\pm0.02$&This work\\
43.167--43.284&\phantom{}2580&COLIBRÍ/DDRAGO&$i$&$20.98\pm0.02$&This work\\
44.117--44.183&\phantom{}1500&COLIBRÍ/DDRAGO&$g$&$21.61\pm0.03$&This work\\
44.117--44.184&\phantom{}4440&COLIBRÍ/DDRAGO&$z$&$20.77\pm0.03$&This work\\
44.118--44.184&\phantom{}1500&COLIBRÍ/DDRAGO&$r$&$21.22\pm0.02$&This work\\
44.119--44.183&\phantom{}1440&COLIBRÍ/DDRAGO&$i$&$20.96\pm0.03$&This work\\
45.138--45.211&\phantom{}1500&COLIBRÍ/DDRAGO&$g$&$21.62\pm0.03$&This work\\
45.138--45.212&\phantom{}4440&COLIBRÍ/DDRAGO&$z$&$20.88\pm0.03$&This work\\
45.139--45.210&\phantom{}1440&COLIBRÍ/DDRAGO&$i$&$21.09\pm0.03$&This work\\
45.139--45.212&\phantom{}1500&COLIBRÍ/DDRAGO&$r$&$21.20\pm0.02$&This work\\
46.111--46.177&\phantom{}1500&COLIBRÍ/DDRAGO&$g$&$21.56\pm0.05$&This work\\
46.111--46.178&\phantom{}4440&COLIBRÍ/DDRAGO&$z$&$20.84\pm0.05$&This work\\
46.112--46.178&\phantom{}1500&COLIBRÍ/DDRAGO&$r$&$21.20\pm0.04$&This work\\
46.113--46.176&\phantom{}1440&COLIBRÍ/DDRAGO&$i$&$21.07\pm0.05$&This work\\
48.047--48.197&\phantom{}3300&COLIBRÍ/DDRAGO&$g$&$21.52\pm0.04$&This work\\
48.047--48.199&\phantom{}9900&COLIBRÍ/DDRAGO&$z$&$20.88\pm0.03$&This work\\
48.048--48.198&\phantom{}3300&COLIBRÍ/DDRAGO&$r$&$21.26\pm0.03$&This work\\
48.049--48.199&\phantom{}3300&COLIBRÍ/DDRAGO&$i$&$21.15\pm0.03$&This work\\
49.040--49.190&\phantom{}3300&COLIBRÍ/DDRAGO&$g$&$21.66\pm0.04$&This work\\
49.040--49.192&\phantom{}9900&COLIBRÍ/DDRAGO&$z$&$20.90\pm0.03$&This work\\
49.041--49.191&\phantom{}3300&COLIBRÍ/DDRAGO&$r$&$21.30\pm0.03$&This work\\
49.042--49.192&\phantom{}3300&COLIBRÍ/DDRAGO&$i$&$21.16\pm0.03$&This work\\
50.040--50.188&\phantom{0}960&COLIBRÍ/DDRAGO&$g$&$21.95\pm0.29$&This work\\
50.040--50.188&\phantom{}2340&COLIBRÍ/DDRAGO&$z$&$21.20\pm0.17$&This work\\
50.090--50.188&\phantom{}1020&COLIBRÍ/DDRAGO&$r$&$21.33\pm0.15$&This work\\
52.034--52.183&\phantom{}2280&COLIBRÍ/DDRAGO&$g$&$21.77\pm0.09$&This work\\
52.034--52.186&\phantom{}6780&COLIBRÍ/DDRAGO&$z$&$21.11\pm0.05$&This work\\
52.035--52.186&\phantom{}2280&COLIBRÍ/DDRAGO&$r$&$21.49\pm0.06$&This work\\
52.036--52.184&\phantom{}2280&COLIBRÍ/DDRAGO&$i$&$21.40\pm0.06$&This work\\
53.030--53.180&\phantom{}3300&COLIBRÍ/DDRAGO&$g$&$21.59\pm0.06$&This work\\
53.030--53.181&\phantom{}9900&COLIBRÍ/DDRAGO&$z$&$21.09\pm0.04$&This work\\
53.031--53.181&\phantom{}3300&COLIBRÍ/DDRAGO&$r$&$21.56\pm0.05$&This work\\
53.032--53.181&\phantom{}3300&COLIBRÍ/DDRAGO&$i$&$21.47\pm0.05$&This work\\
55.058--55.184&\phantom{}1560&COLIBRÍ/DDRAGO&$r$&$21.59\pm0.09$&This work\\
55.107--55.183&\phantom{}1620&COLIBRÍ/DDRAGO&$g$&$21.52\pm0.09$&This work\\
55.107--55.184&\phantom{}4620&COLIBRÍ/DDRAGO&$z$&$21.35\pm0.10$&This work\\
55.109--55.182&\phantom{}1560&COLIBRÍ/DDRAGO&$i$&$21.47\pm0.11$&This work\\
56.064--56.170&\phantom{}2340&COLIBRÍ/DDRAGO&$g$&$21.53\pm0.06$&This work\\
56.064--56.171&\phantom{}7020&COLIBRÍ/DDRAGO&$z$&$21.27\pm0.06$&This work\\
56.065--56.170&\phantom{}2340&COLIBRÍ/DDRAGO&$r$&$21.75\pm0.06$&This work\\
56.066--56.171&\phantom{}2340&COLIBRÍ/DDRAGO&$i$&$21.57\pm0.07$&This work\\
57.060--57.207&\phantom{}3240&COLIBRÍ/DDRAGO&$g$&$21.85\pm0.04$&This work\\
57.060--57.209&\phantom{}9720&COLIBRÍ/DDRAGO&$z$&$21.25\pm0.04$&This work\\
57.061--57.208&\phantom{}3240&COLIBRÍ/DDRAGO&$r$&$21.64\pm0.03$&This work\\
57.062--57.206&\phantom{}3180&COLIBRÍ/DDRAGO&$i$&$21.57\pm0.04$&This work\\
58.060--58.210&\phantom{}3300&COLIBRÍ/DDRAGO&$g$&$21.80\pm0.03$&This work\\
58.060--58.211&\phantom{}9900&COLIBRÍ/DDRAGO&$z$&$21.36\pm0.05$&This work\\
58.061--58.210&\phantom{}3300&COLIBRÍ/DDRAGO&$r$&$21.78\pm0.03$&This work\\
58.061--58.211&\phantom{}3300&COLIBRÍ/DDRAGO&$i$&$21.70\pm0.05$&This work\\
59.124--59.243&\phantom{}2640&COLIBRÍ/DDRAGO&$r$&$21.78\pm0.03$&This work\\
59.124--59.245&\phantom{}2700&COLIBRÍ/DDRAGO&$g$&$21.95\pm0.04$&This work\\
59.124--59.245&\phantom{}7980&COLIBRÍ/DDRAGO&$z$&$21.43\pm0.04$&This work\\
59.125--59.244&\phantom{}2640&COLIBRÍ/DDRAGO&$i$&$21.66\pm0.04$&This work\\
61.093--61.242&\phantom{}3300&COLIBRÍ/DDRAGO&$g$&$21.70\pm0.03$&This work\\
61.093--61.242&\phantom{}9780&COLIBRÍ/DDRAGO&$z$&$21.45\pm0.05$&This work\\
61.094--61.241&\phantom{}3240&COLIBRÍ/DDRAGO&$r$&$21.80\pm0.03$&This work\\
61.095--61.242&\phantom{}3240&COLIBRÍ/DDRAGO&$i$&$21.77\pm0.05$&This work\\

\end{longtable}

\begin{longtable}{cccccl}
\caption{Photometry in $J$ of GRB~260310A / AT~2026fgk. Uncertainties are quoted at the $1\sigma$ confidence level. The magnitudes are in the AB system and are not corrected for Galactic extinction or the host contribution. Times are relative to $T_0$ and are mid times.}

\label{tab:new_J_photometry_full}\\
\hline
\hline
\multicolumn{1}{c}{Time} & Exposure & Telescope/ Instrument & Filter & Magnitude & Reference \\
\multicolumn{1}{c}{(d)} & (s) & & & & \\
\hline
\endfirsthead

\multicolumn{6}{c}{\tablename\ \thetable\ -- continued from previous page}\\
\hline
\hline
Time & Exposure & Telescope/ Instrument & Filter & Magnitude & Reference \\
\multicolumn{1}{c}{(d)} & (s) & & & & \\
\hline
\endhead

\hline
\multicolumn{6}{r}{Continued on next page}\\
\endfoot

\hline
\hline
\endlastfoot

2.208&--&Palomar/WINTER&$J$&$17.16\pm0.09$&\citet{GCN43996}\\
2.232&--&Palomar/WINTER&$J$&$17.28\pm0.09$&\citet{GCN43996}\\
2.995&--&Palomar/WINTER&$J$&$17.34\pm0.12$&\citet{GCN43996}\\
3.413&7560&SYSU/SYSU&$J$&$17.61\pm0.10$&\citet{GCN43993}\\
4.021&2400&TNG/NICS&$J$&$17.91\pm0.10$&\citet{GCN44004}\\
6.200&--&LBT/LUCI&$J$&$18.03\pm0.05$&\citet{GCN44058}\\
10.620&--&ALT100C&$J$&$18.56\pm0.16$&This work\\
15.550&--&ALT100C&$J$&$19.23\pm0.19$&This work\\
18.469&--&ALT100C&$J$&$19.73\pm0.30$&This work\\

\end{longtable}

\begin{longtable}{cccccl}
\caption{Additional optical photometry of GRB~260310A / AT~2026fgk. Uncertainties are quoted at the $1\sigma$ confidence level. The magnitudes are in the AB system and are not corrected for Galactic extinction or the host contribution. Times are relative to $T_0$ and are mid times.} \label{tab:new_extraphotometry_full}\\
\hline
\hline
\multicolumn{1}{c}{Time} & Exposure & Telescope/ Instrument & Filter & Magnitude & Reference \\
\multicolumn{1}{c}{(d)} & (s) & & & & \\
\hline
\endfirsthead

\multicolumn{6}{c}{\tablename\ \thetable\ -- continued from previous page}\\
\hline
\hline
Time & Exposure & Telescope/ Instrument & Filter & Magnitude & Reference \\
\multicolumn{1}{c}{(d)} & (s) & & & & \\
\hline
\endhead

\hline
\multicolumn{6}{r}{Continued on next page}\\
\endfoot

\hline
\hline
\endlastfoot
22.047&--& Palomar/SED Machine &$ r $&$ 18.081 \pm 0.040 $& \citet{GCN44096} \\
2.255&--& Palomar/WINTER &$ H $&$ 17.085 \pm 0.210 $& \citet{GCN43996} \\
2.890&--& NOT &$ r $&$ 18.150 \pm 0.010 $& This work\\
2.910&--& NOT &$ g $&$ 18.420 \pm 0.010 $& This work\\
2.910&--& NOT &$ i $&$ 17.850 \pm 0.010 $& This work\\
2.920&--& NOT &$ z $&$ 17.770 \pm 0.020 $& This work\\
3.067&--& Palomar/SED Machine &$ r $&$ 18.131 \pm 0.030 $& \citet{GCN44096} \\
3.150&--& TRT-SRO &$ B $&$ 18.628 \pm 0.021 $& This work\\
3.160&--& TRT-SRO &$ V $&$ 18.266 \pm 0.014 $& This work\\
3.170&--& TRT-SRO &$ R $&$ 18.136 \pm 0.011 $& This work\\
3.180&--& TRT-SRO &$ I $&$ 17.889 \pm 0.020 $& This work\\
3.367&300& LOT &$ g $&$ 18.637 \pm 0.020 $& \citet{GCN44002} \\
3.367&300& LOT &$ g $&$ 18.480 \pm 0.020 $& \citet{GCN44002} \\
3.370&300& LOT &$ r $&$ 18.251 \pm 0.020 $& \citet{GCN44002} \\
3.370&300& LOT &$ r $&$ 18.361 \pm 0.020 $& \citet{GCN44002} \\
3.374&300& LOT &$ i $&$ 18.134 \pm 0.020 $& \citet{GCN44002} \\
3.374&300& LOT &$ i $&$ 18.216 \pm 0.020 $& \citet{GCN44002} \\
3.378&300& LOT &$ z $&$ 17.864 \pm 0.020 $& \citet{GCN44002} \\
3.378&300& LOT &$ z $&$ 17.926 \pm 0.050 $& \citet{GCN44002} \\
3.413&7560& SYSU &$ J $&$ 17.579 \pm 0.100 $& \citet{GCN43993} \\
4.021&2400& TNG/NICS &$ H $&$ 17.276 \pm 0.100 $& \citet{GCN44004} \\
4.021&2400& TNG/NICS &$ J $&$ 17.889 \pm 0.100 $& \citet{GCN44004} \\
4.222&--& TRT-SRO &$ B $&$ 18.970 \pm 0.086 $& This work\\
4.229&--& TRT-SRO &$ R $&$ 18.157 \pm 0.049 $& This work\\
4.237&--& TRT-SRO &$ I $&$ 18.328 \pm 0.046 $& This work\\
4.401&--& ALT100B &$ g $&$ 18.905 \pm 0.029 $& This work\\
4.413&--& ALT100B &$ r $&$ 18.571 \pm 0.034 $& This work\\
4.425&--& ALT100B &$ i $&$ 18.409 \pm 0.083 $& This work\\
5.050&--& Palomar/SED Machine &$ r $&$ 18.531 \pm 0.030 $& \citet{GCN44096} \\
5.157&--& TRT-SRO &$ B $&$ 18.835 \pm 0.044 $& This work\\
5.168&--& TRT-SRO &$ R $&$ 18.365 \pm 0.016 $& This work\\
5.179&--& TRT-SRO &$ I $&$ 18.205 \pm 0.028 $& This work\\
5.402&--& ALT100B &$ g $&$ 19.075 \pm 0.063 $& This work\\
5.410&--& ALT100B &$ r $&$ 18.770 \pm 0.065 $& This work\\
5.417&--& ALT100B &$ i $&$ 18.411 \pm 0.116 $& This work\\
6.157&--& TRT-SRO &$ B $&$ 19.132 \pm 0.072 $& This work\\
6.168&--& TRT-SRO &$ R $&$ 18.346 \pm 0.026 $& This work\\
6.179&--& TRT-SRO &$ I $&$ 18.361 \pm 0.039 $& This work\\
6.200&--& LBT/LUCI &$ J $&$ 18.009 \pm 0.050 $& \citet{GCN44058} \\
6.409&--& ALT100B &$ g $&$ 19.061 \pm 0.048 $& This work\\
6.416&--& ALT100B &$ r $&$ 18.715 \pm 0.053 $& This work\\
6.423&--& ALT100B &$ i $&$ 18.635 \pm 0.132 $& This work\\
7.377&--& ALT100B &$ g $&$ 19.337 \pm 0.097 $& This work\\
7.384&--& ALT100B &$ r $&$ 18.809 \pm 0.064 $& This work\\
7.391&--& ALT100B &$ i $&$ 18.680 \pm 0.151 $& This work\\
8.045&--& Palomar/SED Machine &$ r $&$ 18.771 \pm 0.030 $& \citet{GCN44096} \\
8.346&--& ALT100B &$ g $&$ 19.116 \pm 0.117 $& This work\\
8.357&--& ALT100B &$ r $&$ 18.919 \pm 0.049 $& This work\\
8.389&--& ALT100B &$ i $&$ 19.425 \pm 0.261 $& This work\\
9.353&--& ALT100B &$ r $&$ 19.140 \pm 0.070 $& This work\\
9.369&--& ALT100B &$ i $&$ 19.140 \pm 0.160 $& This work\\
9.413&--& ALT100B &$ g $&$ 19.740 \pm 0.180 $& This work\\
10.362&--& ALT100B &$ g $&$ 19.620 \pm 0.050 $& This work\\
10.383&--& ALT100B &$ r $&$ 19.270 \pm 0.040 $& This work\\
10.405&--& ALT100B &$ i $&$ 19.130 \pm 0.120 $& This work\\
11.461&--& ALT100B &$ g $&$ 19.559 \pm 0.058 $& This work\\
11.518&--& ALT100B &$ r $&$ 19.127 \pm 0.071 $& This work\\
11.539&--& ALT100B &$ i $&$ 19.159 \pm 0.285 $& This work\\
12.162&600& MDM/MIRAGE &$ h $&$ 18.446 \pm 0.160 $& \citet{GCN44122} \\
12.162&600& MDM/MIRAGE &$ j $&$ 19.289 \pm 0.140 $& \citet{GCN44122} \\
12.396&--& ALT100B &$ g $&$ 19.866 \pm 0.075 $& This work\\
12.417&--& ALT100B &$ r $&$ 19.442 \pm 0.072 $& This work\\
12.438&--& ALT100B &$ i $&$ 19.679 \pm 0.235 $& This work\\
13.016&--& Palomar/SED Machine &$ r $&$ 19.521 \pm 0.040 $& \citet{GCN44096} \\
13.433&--& ALT100B &$ g $&$ 20.257 \pm 0.318 $& This work\\
13.454&--& ALT100B &$ r $&$ 19.401 \pm 0.122 $& This work\\
13.486&--& ALT100B &$ i $&$ 18.973 \pm 0.104 $& This work\\
15.525&--& ALT100B &$ g $&$ 20.587 \pm 0.127 $& This work\\
15.571&--& ALT100B &$ r $&$ 19.769 \pm 0.054 $& This work\\
15.614&--& ALT100B &$ i $&$ 19.743 \pm 0.119 $& This work\\
16.486&--& ALT100B &$ g $&$ 20.553 \pm 0.082 $& This work\\
16.528&--& ALT100B &$ r $&$ 20.204 \pm 0.084 $& This work\\
16.571&--& ALT100B &$ i $&$ 20.065 \pm 0.187 $& This work\\
18.443&--& ALT100B &$ g $&$ 20.537 \pm 0.133 $& This work\\
18.528&--& ALT100B &$ r $&$ 20.063 \pm 0.062 $& This work\\
18.613&--& ALT100B &$ i $&$ 20.133 \pm 0.213 $& This work\\
19.562&--& ALT100B &$ g $&$ 20.754 \pm 0.191 $& This work\\
19.605&--& ALT100B &$ r $&$ 20.219 \pm 0.122 $& This work\\
19.647&--& ALT100B &$ i $&$ >19.500 $& This work\\
21.495&--& ALT100B &$ g $&$ 20.320 \pm 0.092 $& This work\\
21.538&--& ALT100B &$ r $&$ 20.254 \pm 0.114 $& This work\\
21.584&--& ALT100B &$ i $&$ 20.237 \pm 0.154 $& This work\\
22.558&--& ALT100B &$ g $&$ 20.821 \pm 0.248 $& This work\\
22.601&--& ALT100B &$ r $&$ 20.040 \pm 0.174 $& This work\\
22.643&--& ALT100B &$ i $&$ >20.200 $& This work\\
23.595&--& ALT100B &$ g $&$ 20.948 \pm 0.210 $& This work\\
23.638&--& ALT100B &$ i $&$ 20.480 \pm 0.358 $& This work\\
23.680&--& ALT100B &$ r $&$ 20.346 \pm 0.101 $& This work\\
24.501&--& ALT100B &$ g $&$ 20.567 \pm 0.119 $& This work\\
24.558&--& ALT100B &$ i $&$ 20.185 \pm 0.209 $& This work\\
24.615&--& ALT100B &$ r $&$ 20.288 \pm 0.135 $& This work\\
25.524&--& ALT100B &$ g $&$ 20.458 \pm 0.097 $& This work\\
25.577&--& ALT100B &$ i $&$ 20.646 \pm 0.328 $& This work\\
25.630&--& ALT100B &$ r $&$ 20.181 \pm 0.107 $& This work\\
26.596&--& ALT100B &$ g $&$ 20.902 \pm 0.278 $& This work\\
26.649&--& ALT100B &$ i $&$ >20.300 $& This work\\
26.702&--& ALT100B &$ r $&$ 20.205 \pm 0.132 $& This work\\
28.528&--& ALT100B &$ g $&$ 21.224 \pm 0.183 $& This work\\
28.597&--& ALT100B &$ i $&$ 20.533 \pm 0.314 $& This work\\
28.659&--& ALT100B &$ r $&$ 20.092 \pm 0.088 $& This work\\
29.524&--& ALT100B &$ g $&$ 20.976 \pm 0.129 $& This work\\
29.583&--& ALT100B &$ i $&$ 19.534 \pm 0.155 $& This work\\
29.636&--& ALT100B &$ r $&$ 20.442 \pm 0.113 $& This work\\
30.566&--& ALT100B &$ g $&$ > 21.198 $& This work\\
30.624&--& ALT100B &$ i $&$ > 20.039 $& This work\\
30.677&--& ALT100B &$ r $&$ > 20.784 $& This work\\

\end{longtable}

\begin{longtable}{ccccll}
\caption{Radio photometry of GRB~260310A / AT~2026fgk from GCNs (cited in the table) and this work. Uncertainties are quoted at the $1\sigma$ confidence level. }
\label{tab:radio_photometry_full}\\

\hline
\hline
Time & Exposure & Telescope &  Band & Flux Density & Reference \\
(d) & (hr) & & (GHz) &\multicolumn{1}{c}{($\mu$Jy)} & \\
\hline
\endfirsthead

\multicolumn{6}{c}{\tablename\ \thetable\ -- continued from previous page}\\
\hline
\hline
Time & Exp. & Telescope & Band & Flux Density & Reference \\
(d) & (hr) & & (GHz) & ($\mu$Jy) & \\
\hline
\endhead

\hline
\multicolumn{6}{r}{Continued on next page}\\
\endfoot

\hline
\hline
\endlastfoot

4.260  & --   & VLA & \phantom{0}6.0  & $\phantom{0}3894\pm8$   & \citet{GCN44045} \\
4.260  & --   & VLA & 10.0 & $\phantom{0}6368\pm13$ & \citet{GCN44045} \\
4.260  & --   & VLA & 15.0 & $\phantom{0}9104\pm29$  & \citet{GCN44045} \\

7.344  & -- & OVRO & 15.0 & $18100\pm1600$ & This work\\
7.345  & -- & OVRO & 15.0 & $20600\pm1500$ & This work\\
10.090 & -- & OVRO & 15.0 & $23000\pm1500$ & This work\\
14.397 & -- & OVRO & 15.0 & $21700\pm1600$ & This work\\
16.090 & -- & OVRO & 15.0 & $12700\pm1700$ & This work\\

17.200 & 3.35 & VLA & \phantom{0}3.0  & $\phantom{0}2900\pm40$  & \citet{GCN44160} \\
17.200 & 3.35 & VLA & \phantom{0}6.0  & $\phantom{0}7520\pm50$  & \citet{GCN44160} \\
17.200 & 3.35 & VLA & 10.0 & $11110\pm110$ & \citet{GCN44160} \\
17.200 & 3.35 & VLA & 15.0 & $11730\pm160$ & \citet{GCN44160} \\
17.200 & 3.35 & VLA & 33.0 & $\phantom{0}9580\pm390$  & \citet{GCN44160} \\
17.200 & 3.35 & VLA & 45.0 & $\phantom{0}7980\pm270$  & \citet{GCN44160} \\

20.145 & -- & OVRO & 15.0 & $13900\pm1500$ & This work\\
23.092 & -- & OVRO & 15.0 & $13300\pm1500$ & This work\\

25.200 & 3.62 & VLA & \phantom{0}1.5  & $\phantom{0}1740\pm60$   & \citet{GCN44235} \\
25.200 & 3.62 & VLA & \phantom{0}3.0  & $\phantom{0}5000\pm40$   & \citet{GCN44235} \\
25.200 & 3.62 & VLA & \phantom{0}6.0  & $\phantom{0}8390\pm50$   & \citet{GCN44235} \\
25.200 & 3.62 & VLA & 10.0 & $\phantom{0}8930\pm120$  & \citet{GCN44235} \\
25.200 & 3.62 & VLA & 15.0 & $\phantom{0}8160\pm220$  & \citet{GCN44235} \\
25.200 & 3.62 & VLA & 22.0 & $\phantom{0}7340\pm230$  & \citet{GCN44235} \\
25.200 & 3.62 & VLA & 33.0 & $\phantom{0}6140\pm190$  & \citet{GCN44235} \\
25.200 & 3.62 & VLA & 45.0 & $\phantom{0}5250\pm190$  & \citet{GCN44235} \\

30.054 & -- & OVRO & 15.0 & $11100\pm1300$ & This work\\
36.075 & -- & OVRO & 15.0 & $\phantom{0}6800\pm1500$  & This work\\

\end{longtable}

\bsp	
\label{lastpage}
\end{document}